\documentclass[useAMS,usenatbib,usegraphicx]{mn2e}

\usepackage{times}

\newcommand{\iraf}  {{\sc iraf}}
\newcommand{\midas}  {{\sc midas}}

\newcommand{\kms}{km\,s$^{-1}$}

\newcommand{\CII}  {C\,{\sc ii}}
\newcommand{\CIII} {C\,{\sc iii}}

\newcommand{\HeI}  {He\,{\sc i}}
\newcommand{\HeII} {He\,{\sc ii}}

\newcommand{\MgI} {Mg\,{\sc i}}
\newcommand{\MgII} {Mg\,{\sc ii}}
\newcommand{\NIII} {N\,{\sc iii}}

\newcommand{\Halpha} {H$\alpha$}
\newcommand{\Hbeta}  {H$\beta$}
\newcommand{\Hgamma} {H$\gamma$}
\newcommand{\Hdelta} {H$\delta$}

\newcommand{\lam}{$\lambda$}

\title[Spectroscopy of the cataclysmic variable UX UMa]
{Dark spot, Spiral waves and the SW Sex behaviour: it is all about UX Ursae Majoris}

\author[V.\,V.\,Neustroev et al.]
{V.\,V. Neustroev$^{1}$\thanks{E-mail:vitaly@neustroev.net}, V.\,F.\,Suleimanov$^{2, 3}$, N.\,V.\,Borisov$^4$,
K.\,V.\,Belyakov$^{3}$, and A.\,Shearer$^{1}$\\
$^{1}$Centre for Astronomy, National University of Ireland, Galway, Newcastle Rd., Galway, Ireland\\
$^{2}$Institute for Astronomy and Astrophysics, Kepler Center for Astro and Particle Physics, Eberhard Karls University, Sand 1, 72076 T\"{u}bingen, Germany\\
$^{3}$Kazan State University, Kremlevskaja str. 18, Kazan 420008, Russia\\
$^{4}$Special Astrophysical Observatory of the Russian AS, Nizhnij Arkhyz, Karachaevo-Cherkesia 369167, Russia\\
}
\begin{document}

\date{Accepted ???. Received ???; in original form 2009 December 7}

\pagerange{\pageref{firstpage}--\pageref{lastpage}} \pubyear{2010}

\maketitle

\label{firstpage}

\begin{abstract}
  We present an analysis of time-resolved,
  medium resolution optical spectroscopic observations of UX~UMa in the blue (3920--5250~{\AA}) and red
  (6100--7200~{\AA}) wavelength ranges, that were obtained in April 1999 and March 2008 respectively.
  The observed characteristics of our spectra indicate that UX~UMa has been in different states during
  those observations. The blue spectra are very complex. They are dominated by strong and broad single-peaked
  emission lines of hydrogen. The high-excitation lines of \HeII\ $\lambda4686$ and the Bowen blend are quite
  strong as well. All the lines consist of a mixture of absorption and emission components. Using Doppler
  tomography we have identified four distinct components of the system: the accretion disc, the secondary star,
  the bright spot from the gas stream/disc impact region, and the unique compact area of absorption
  in the accretion disc seen as a dark spot in the lower-left quadrant
  of the tomograms. In the red wavelength range, both the hydrogen (\Halpha) and neutral helium
  (\HeI\ $\lambda6678$ and \HeI\ $\lambda7065$) lines were observed in emission and both exhibited
  double-peaked profiles. Doppler tomography of these lines reveals spiral structure in the accretion disc,
  but in contrast to the blue wavelength range, there is no evidence for either the dark spot or
  the gas stream/disc impact region emission, while the emission from the secondary star is weak.
  During the observations in 1999, UX~UMa showed many of the defining properties of the SW~Sex stars.
  However, all these features almost completely disappeared in 2008.
  We have also estimated the radial velocity semi-amplitudes $K_1$ and $K_2$
  and evaluated the system parameters of UX~UMa. These estimates are inconsistent with previous values
  derived by means of analysis of WD eclipse features in the light curve in the different wavelength ranges.

\end{abstract}


\section{Introduction}

  Cataclysmic Variables (CVs) are close interacting binaries that contain a white dwarf (WD) accreting
  material from a companion, usually a late main-sequence star. CVs are very active photometrically,
  exhibiting variability on time scales from seconds to centuries (see \citealt{Warner} for a general
  review on CVs).

  Nova-like variables (NLs) are an important subset of CVs. By definition, NLs should not display any
  dwarf nova (DN) outbursts. Their almost steady brightness is thought to be due to their mass transfer
  rate $\dot{M}$ exceeding the upper stability limit $\dot{M}_{crit}$. Their discs are thus thermally
  and tidally stable. As a consequence of such a nebulous definition, the NL class is a very
  heterogeneous group of stars, and the definition of the subclasses within NLs is mostly based on
  observational features of objects. According to \citet{Warner}, the UX~Ursae~Majoris stars show
  persistent broad Balmer absorption lines, and this terminology has occasionally been used for all
  NLs. The RW~Triangulum stars, per contra, have pure emission-line spectra. A significant fraction
  of NLs show states of low luminosity in their long-term optical light curves at irregular intervals of
  weeks to months. Those of them having drops exceeding 1 mag are known as the ``anti-dwarf novae''
  or VY Sculptoris stars (\citealt{LaDous1993,Warner,Dhillon1996}).

  SW~Sextantis stars are another relatively large group of NLs, which was initially populated by eclipsing
  systems only \citep{Thorstensen1991}. These systems, largely occupying the narrow orbital period stripe
  between 3 and 4.5 hours, show many unusual yet consistent properties, including
  single-peaked emission lines despite the high inclination, strong high excitation spectral features,
  central absorption dips in the emission lines around phase  0.4 {--} 0.7, and high-velocity
  emission S-waves with maximum blueshift near phase $\sim0.5$. The unusual spectroscopic behaviour of
  the SW~Sex systems has led to their intensive studies and many NLs above the
  3 {--} 4.5 hr period interval, and even some Low Mass X-ray Binaries (LMXBs) have been found to exhibit
  distinctive SW Sex behaviour \footnote{See D.~W.\ Hoard's Big List of SW Sextantis Stars at\\
  http://web.ipac.caltech.edu/staff/hoard/cvtools/swsex/biglist.html \citep{Hoard2003}.}
  \citep{Rodriguez2007a,Rodriguez2007b}.

  In this paper we report on our observations of UX~UMa, a prototype of the NLs, and show that
  this system also exhibits some of the key features of the SW~Sex stars. UX UMa has been discovered by
  \citet{Beljavsky} and extensively studied in the past \citep{WalkerHerbig1954,WarnerNather1972}, and is
  considered to be one of the well known CVs. UX~UMa has an orbital period of 4.72~h \citep{Kukarkin1977}
  and is the brightest eclipsing NL.
  \citet{WarnerNather1972} found low-amplitude 29-s oscillations in the light curve of UX~UMa confirmed
  later by \citet{Knigge1998}. It could be interpreted in the context of a low-inertia magnetic accretor,
  in which accretion on to an equatorial belt of the WD primary causes the belt to vary its angular
  velocity \citep{WarnerWoudt2002}.
  \citet{Rutten1993, Rutten1994} presented spectrally-resolved eclipse maps of the UX~UMa accretion disc,
  obtained from low-resolution spectra spanning \lam\lam3600--9800 \AA.
  The system has been also well studied in ultraviolet
  \citep{Baptista95,Knigge1998,Froning2003, Linnell2008} and X-rays \citep{Wood1995,Pratt2004} in recent
  years. The system parameters for UX~UMa are usually taken from \citet{Baptista95} and are based on
  the derived parameters of a WD eclipse, which however has been brought into question by \citet{Froning2003}.

  Despite such a rich history of investigations, UX~UMa has been almost neglected with optical
  high or medium resolution spectroscopy
  for the past 25 years. To our knowledge, no detailed red spectra of UX~UMa have been published at all,
  with the exception of several profiles of the \Halpha\ emission line \citep{Kjurkchieva2006}, and
  the far-red spectra in the wavelength range \lam\lam7661--8308 \AA\ \citep{VandePutte2003}.
  The latest extensive optical spectroscopic studies of this system in the blue wavelength range
  have been made by \citet{Schlegel1983}. In particular, they found that the optical lines exhibited
  a wide range of radial velocity semi-amplitudes, varying from line to line. Soon after \citet{Shafter1984}
  found $K_1$ for \Halpha\ equal $157\pm6$ \kms. This value of the
  semi-amplitude implies a mass ratio of $q=1$, making UX~UMa one of the few systems where the masses of
  the primary and secondary stars are the same \citep{Baptista95}. Nevertheless,
  \citet{VandePutte2003} estimated $K_2$ to be $262 \pm 14$, whereas \citet{Froning2003}
  found the extremely low $K_1$ of 70 \kms\ for far ultraviolet absorption lines.
  \citet{Suleimanov2004} gave a preliminary analysis on part of the data presented in this paper, and
  estimated the semi-amplitude of the radial-velocity variations of the \Hbeta\ emission line to be
  about 100 \kms. They also found that UX~UMa showed some features which are commonly used for the identification
  of the SW~Sex subclass stars.

  This motivated us to perform a new time-resolved spectroscopy of UX~UMa in order to study its properties in
  more detail. In this paper we present and discuss the medium-resolution spectroscopic observations of
  the system in 1999 and 2008.

\begin{table*}
\label{ObsTab}
\begin{center}
\caption{Log of observations of UX~UMa}
\begin{tabular}{cccccccll}
\hline\hline
Set     &  Date        & HJD Start  & Instrument &$\Delta\lambda$$^a$& $\lambda$~range & Exp.Time & Number   & Duration\\
        &              &  2450000+  &            &  (\AA)            & (\AA)           &  (sec)   & of exps. &  (hours)\\
\hline
SAO     & 1999-Apr-08  &  1277.384  & UAGS    &  2.5          & 3920--5220      &  300     & 9         & 0.73   \\
        & 1999-Apr-08  &  1277.504  & UAGS    &  2.5          & 3920--5220      &  300     & 19        & 1.66    \\
        & 1999-Apr-10  &  1279.324  & UAGS    &  2.5          & 3920--5220      &  300     & 34        & 3.25   \\
\hline
Loiano  & 2008-Mar-20  &  4546.521  & BFOSC   & 3.3           & 6100--8100      &  600     & 15       & 2.56   \\
        & 2008-Mar-22  &  4548.453  & BFOSC   & 3.3           & 6100--8100      &  600     & 7        & 1.16   \\
        & 2008-Mar-24  &  4550.331  & BFOSC   & 3.3           & 6100--8100      &  600     & 15       & 2.57   \\
\hline
\end{tabular}
\begin{tabular}{l}
$^a$ -- $\Delta\lambda$ is the FWHM spectral resolution \\
\end{tabular}
\end{center}
\end{table*}

\begin{figure*}
\centering
\includegraphics[width=17cm]{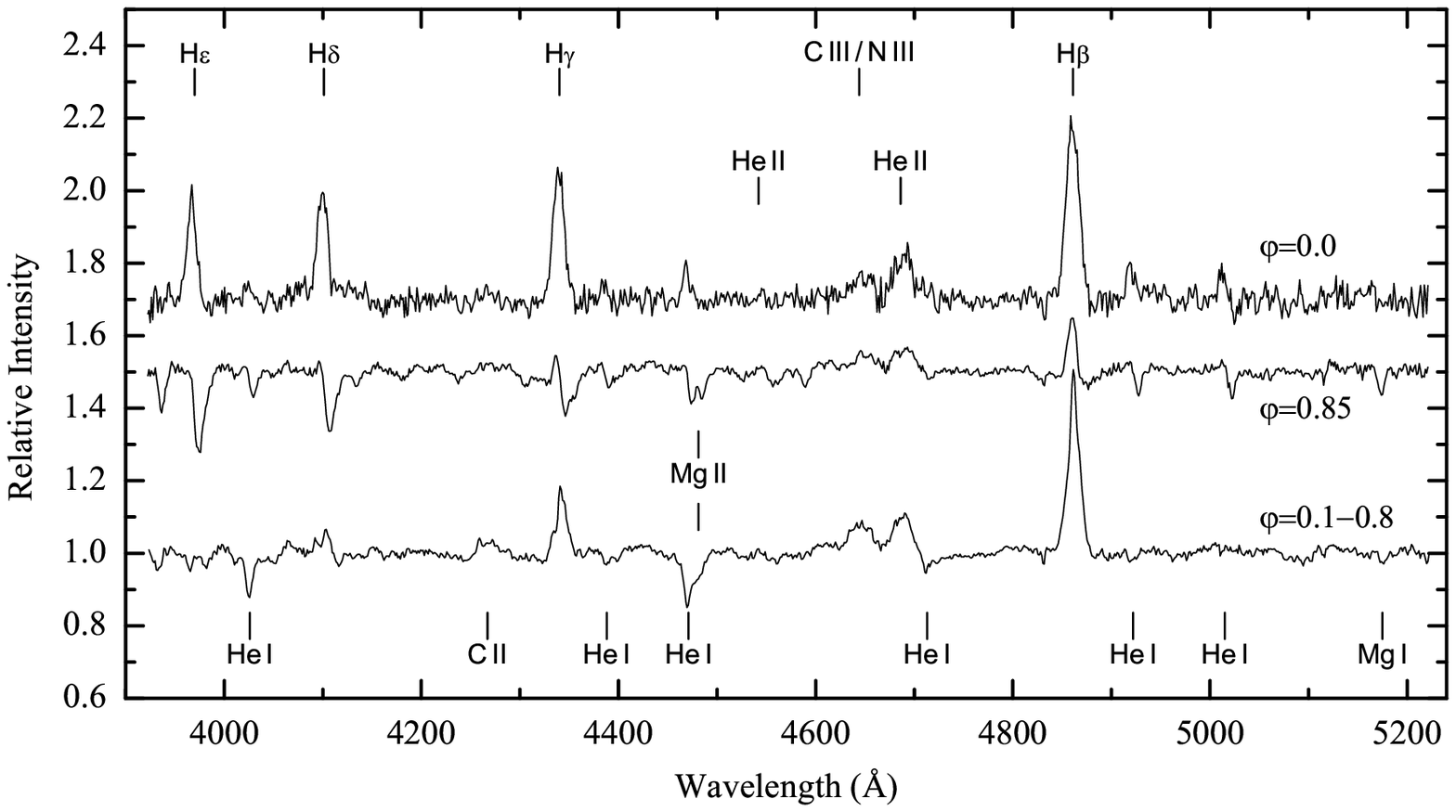}\\
\bigskip
\includegraphics[width=17cm]{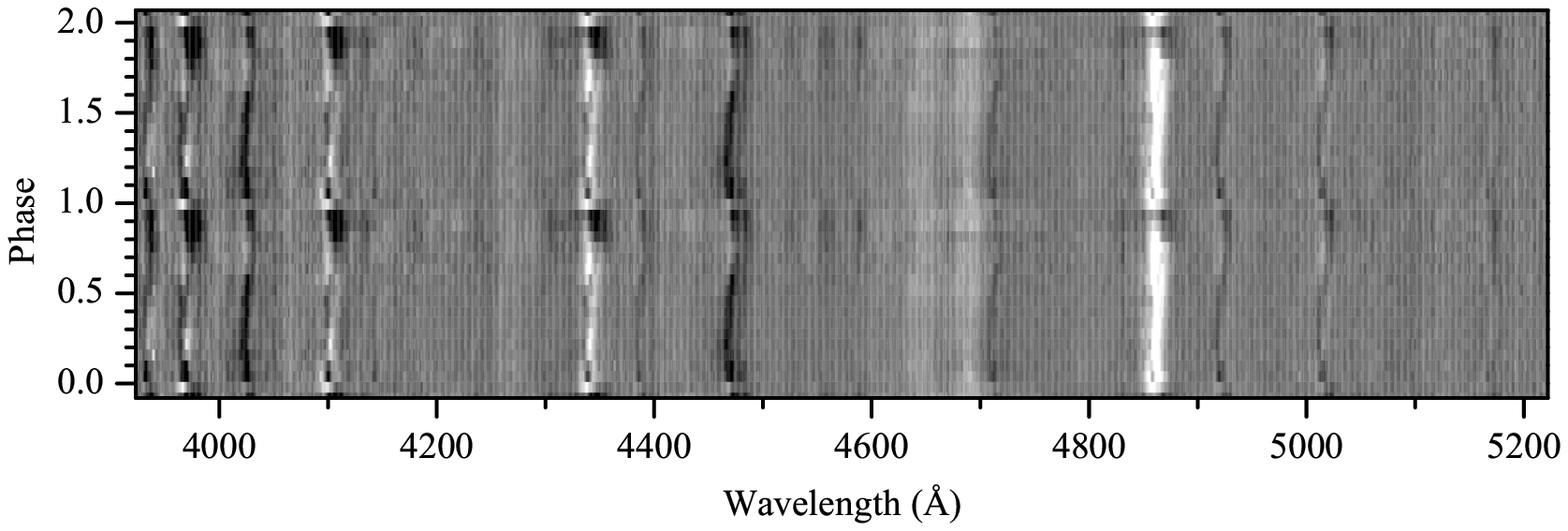}
\caption{Average (top panel) and trailed (bottom panel) spectra of UX~UMa from the SAO set.
The average spectrum is divided in three phase intervals. The top curve shows the spectrum in mid-eclipse
($0.025<\varphi<0.975$), the bottom curve shows the spectrum outside eclipse (0.08 $<\varphi <$ 0.78), and
the middle curve shows the spectrum outside but just prior to eclipse (0.78 $<\varphi <$ 0.92).
In the trailed spectrum, displayed twice for clarity, numerous emission and absorption features are
visible. Note, that all the lines consist of a mixture of absorption and emission components.
White indicates emission.}
\label{aver_spec_blue}
\end{figure*}

\begin{figure*}
\centering
\includegraphics[width=17cm]{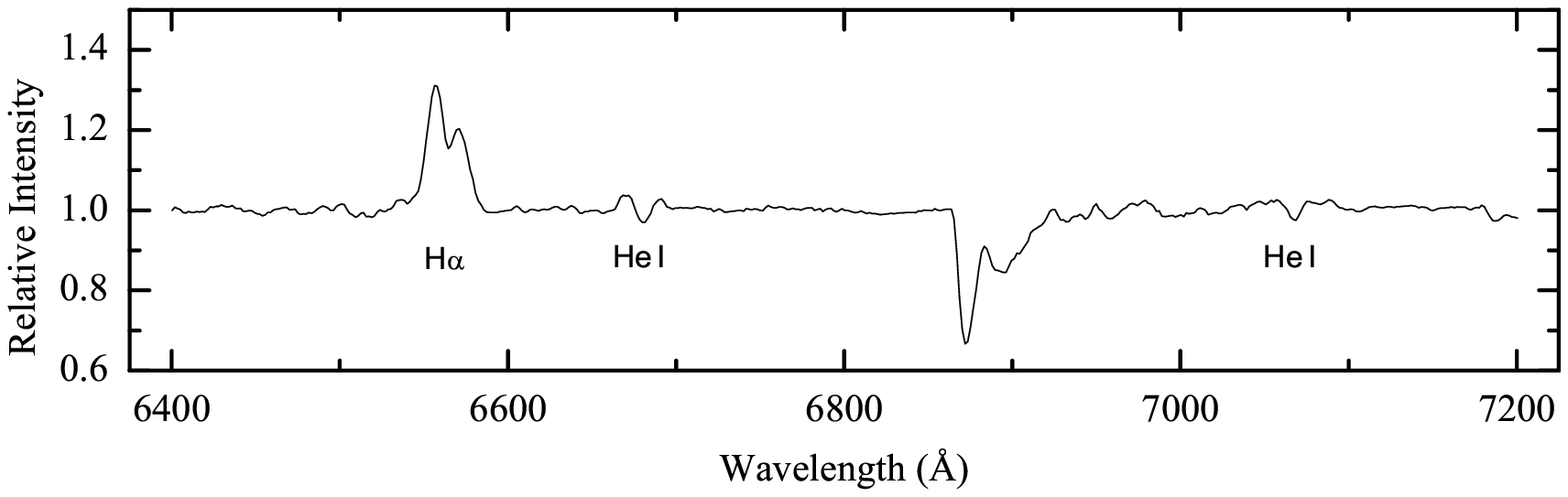}
\bigskip
\includegraphics[width=17cm]{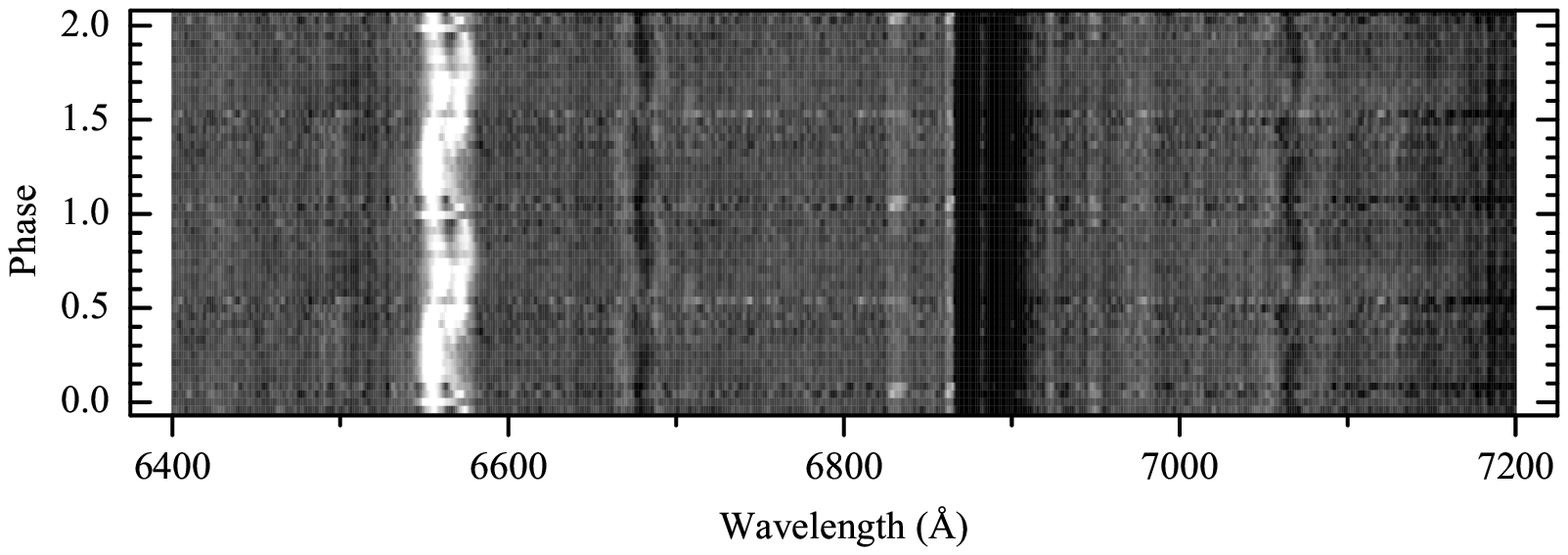}
\caption{The red average (top panel) and trailed (bottom panel) spectra of UX~UMa from the Loiano set.}
\label{aver_spec_red}
\end{figure*}

\section{Observations and Data reduction}
\label{ObsSec}

  The spectra presented here were obtained during two observing runs. The first observations
  were conducted in 1999  during two nights of April 8 and 10 at the Special Astrophysical Observatory
  of the Russian Academy of Sciences (SAO RAS) on the 6~m telescope,
  using the UAGS spectrograph, equipped with a Photometrics CCD. A total of 62 spectra in the wavelength
  range of 3920--5250~{\AA} and a dispersion of $\sim1.3$~{\AA/pix} were obtained with 300~s
  individual exposures. Corresponding spectral resolution was about 2.5~\AA.

  Further observations were obtained in 2008 during 3 nights of March 20, 22 and 24, using the
  imaging spectrograph BFOSC with a $1300\times1340$ pixels EEV CCD attached to the Cassini 1.5~m telescope
  at Loiano (Italy). Grism \#8 was used, providing nominal spectral coverage of 6100--8190 \AA\ and
  a dispersion of $\sim1.6$~{\AA/pix}. A total of 37 spectra were obtained with 600~s individual exposures.
  Corresponding spectral resolution was about 3.3 \AA.

  The reduction procedure was performed using the \midas\ (the SAO observations) and \iraf\ (Loiano)
  environments. Comparison spectra of Ar-Ne-He (SAO) and He-Ar (Loiano) lamps were used for the wavelength
  calibration. The absolute flux calibration of the spectra was also achieved by taking nightly
  spectra of the standard stars BD+33 2642 (SAO), Feige 92 and Hiltner 600 (Loiano).
  However, as the weather conditions during both observing sets were not optimal due to cirrus and poor
  seeing (down to 3\arcsec), we consider the flux calibration to be only approximate. In view of this
  fact, from now on we use spectra normalized to the continuum.

  To improve the confidence of the results presented in this paper we also acquired the phase-binned spectra.
  For this the spectra were phase-folded according to the linear ephemeris of \citet{Baptista95}:

  \begin{equation}  \label{ephemeris}
  T_{min}(HJD) = 2443904.87872(\pm3) + 0.196671278(\pm2) \times E,
  \end{equation}

  \noindent where $T_{min}$ is the moment of minimum, and then co-added the spectra
  into 17 (SAO) and 24 (Loiano) separate phase bins.

  The uncertainties of this ephemeris in orbital phase at the time of our observations are
  very small. For the Loiano observations, for which the propagated error is larger, it is less than
  $0.001$. This is much smaller than our phase resolution ($\sim0.035$) and is thus negligible.

\begin{figure*}
\centering
\includegraphics[width=16cm]{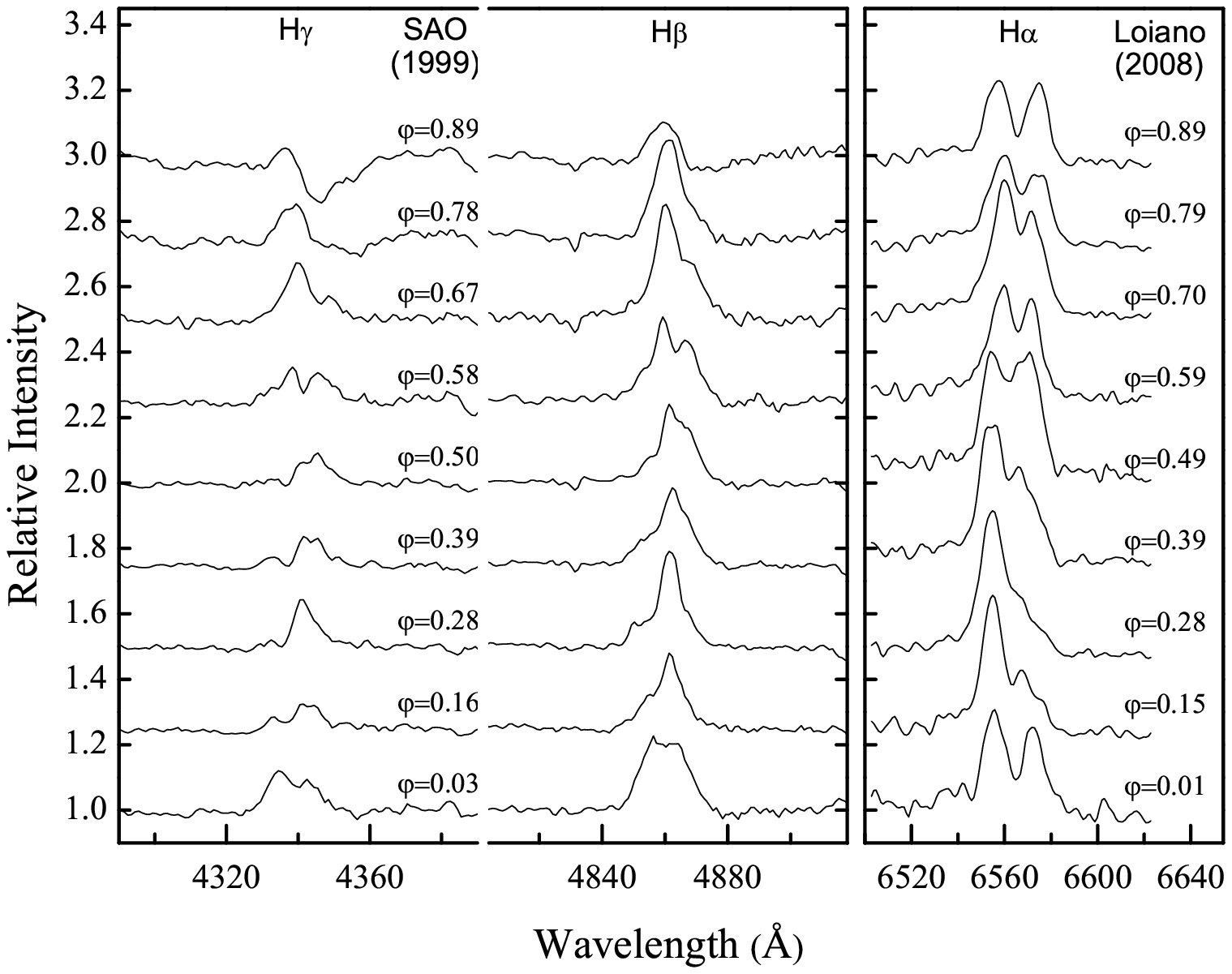}
\caption{The line profile variations of UX~UMa as a function of orbital phase. The left and right panels show
the spectra from the SAO (1999) and Loiano (2008) observations respectively. The data have been averaged
into 9 binary phase bins, and each resulting spectrum is normalized to the continuum and displaced along
the intensity axis by 0.25.}
\label{fig_prof}
\end{figure*}

\section{Data analysis and results}
\label{DatAnSec}

\subsection{Averaged and trailed spectra}
\label{AveSpecSec}

  Optical spectra of NLs show a wide range of appearances.
  According to \citet{Warner}, the UX~UMa
  stars have persistently broad Balmer absorption-line spectra, whereas the RW~Tri stars, for example,
  have pure emission-line spectra (even though occasionally with absorption cores). It is interesting to note,
  however, that such segregation is quite vague, as even UX~UMa itself does not exactly follow this definition.

  The mean spectra of UX~UMa in the blue wavelength range, from the SAO dataset, are shown in
  Fig.~\ref{aver_spec_blue} (top panel). These spectra are an average and a combination of spectra in three
  phase intervals, uncorrected for orbital motion.

  The spectra are dominated by strong and broad emission lines of
  hydrogen. The high-excitation lines of \HeII\ $\lambda4686$ and the \CIII/\NIII\ $\lambda4640$--$4650$
  blend are quite strong as well. In addition, \HeII\ $\lambda4542$ and \CII\ $\lambda4267$  are
  observed. All these emission lines exhibit, in the averaged spectra, symmetric single-peaked
  profiles. However, in the phase-folded spectra the emission line-profile variations are quite complex
  and resemble those observed in SW~Sex stars. For example, the Balmer lines \Hbeta\ and \Hgamma\
  are single-peaked but show transient absorption features at phases 0.4--0.7 turning the line profiles
  into the double-peak appearance (Fig.~\ref{fig_prof}, left panel). The strength of this absorption increases
  with increasing line excitation level.

  The absorption spectrum is also rich in features. In particular, we note that outside eclipse all the lines
  of neutral helium are in absorption, the strongest such features are \HeI\ $\lambda$4472 and
  \HeI\ $\lambda$4026. Also note the presence of the \MgI\ $\lambda5175$ and \MgII\ $\lambda4481$
  absorption lines (Fig.~\ref{aver_spec_blue}, top panel, the bottom spectrum).

  At phase $\varphi\sim$0.78, a new absorption component appeared in the red wings of the Balmer emission
  lines and rapidly became so strong that it almost flooded the higher members of the Balmer lines
  (Fig.~\ref{aver_spec_blue}, top panel, the middle spectrum). However, this absorption apparently
  did not affect the high-excitation emission lines and
  very little distorted the \HeI\ absorption lines which in fact became less deep.
  There was little qualitative change in the eclipse spectrum of UX~UMa before mid-eclipse
  ($0.025<\varphi<0.975$) when all absorptions were instantly reversed into emissions, and
  also the Balmer decrement became almost flat (Fig.~\ref{aver_spec_blue}, top panel, the upper spectrum).

\begin{figure*}
    \includegraphics[width=5.6cm]{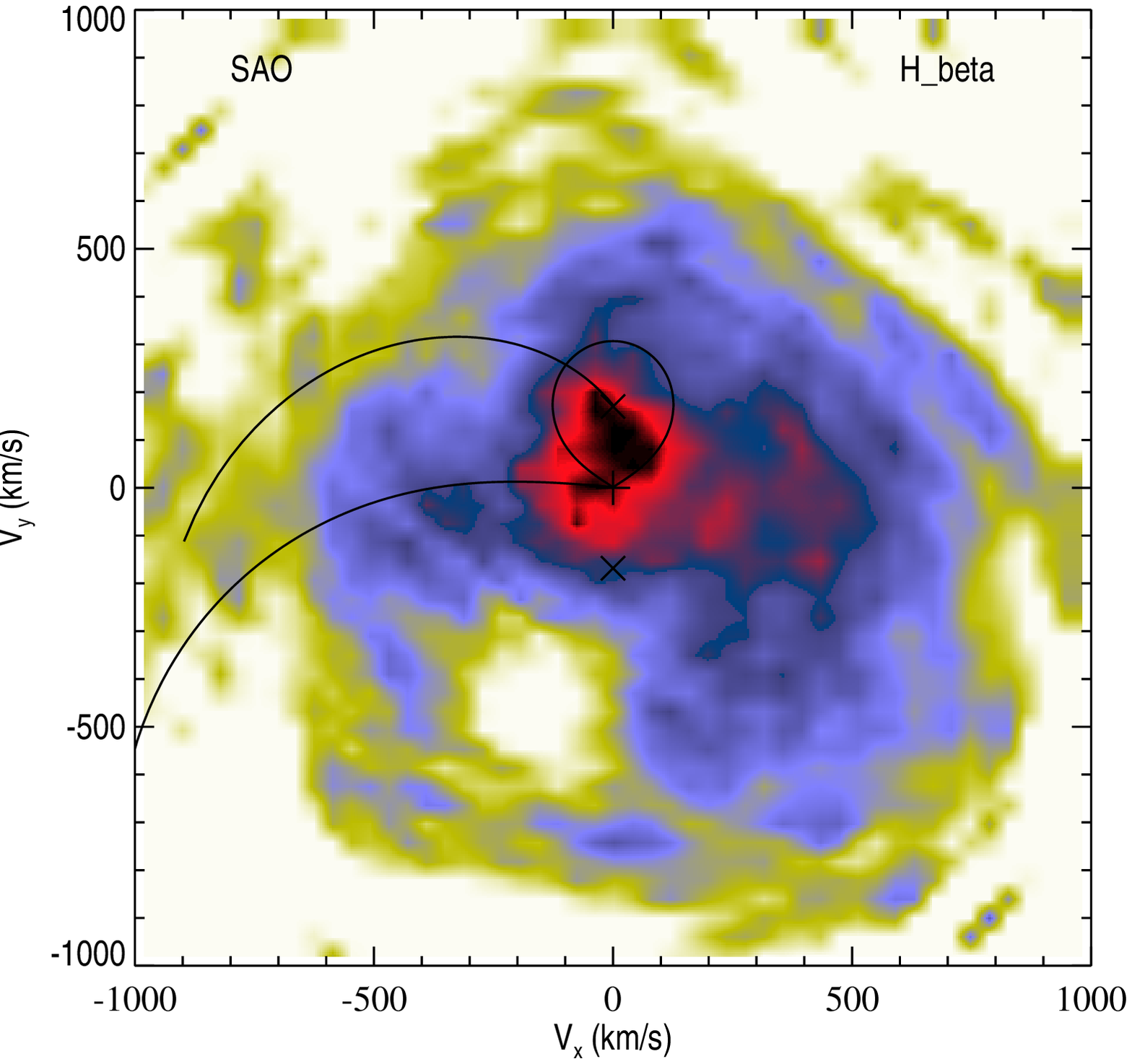}
    \includegraphics[width=5.6cm]{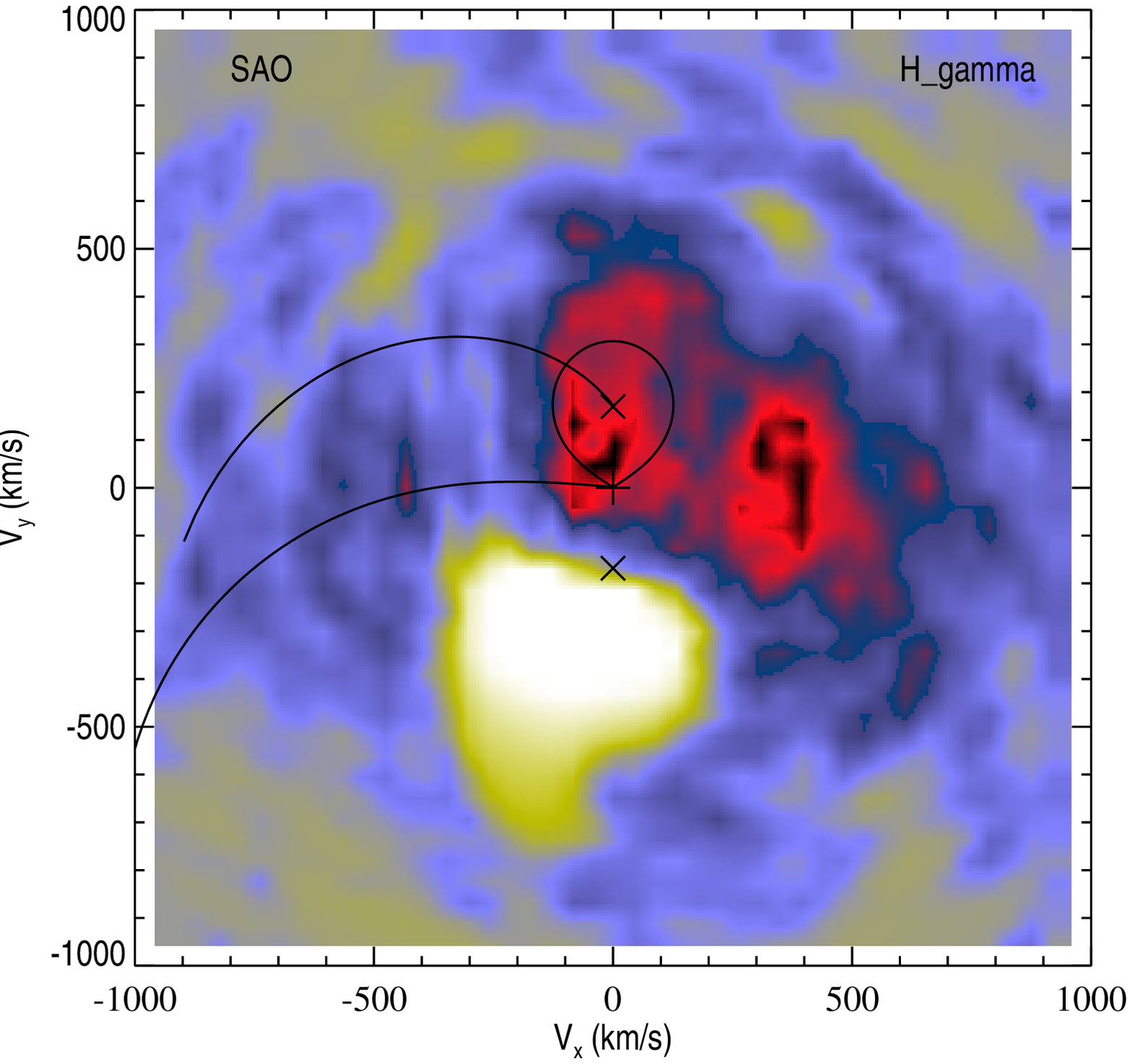}
    \includegraphics[width=5.6cm]{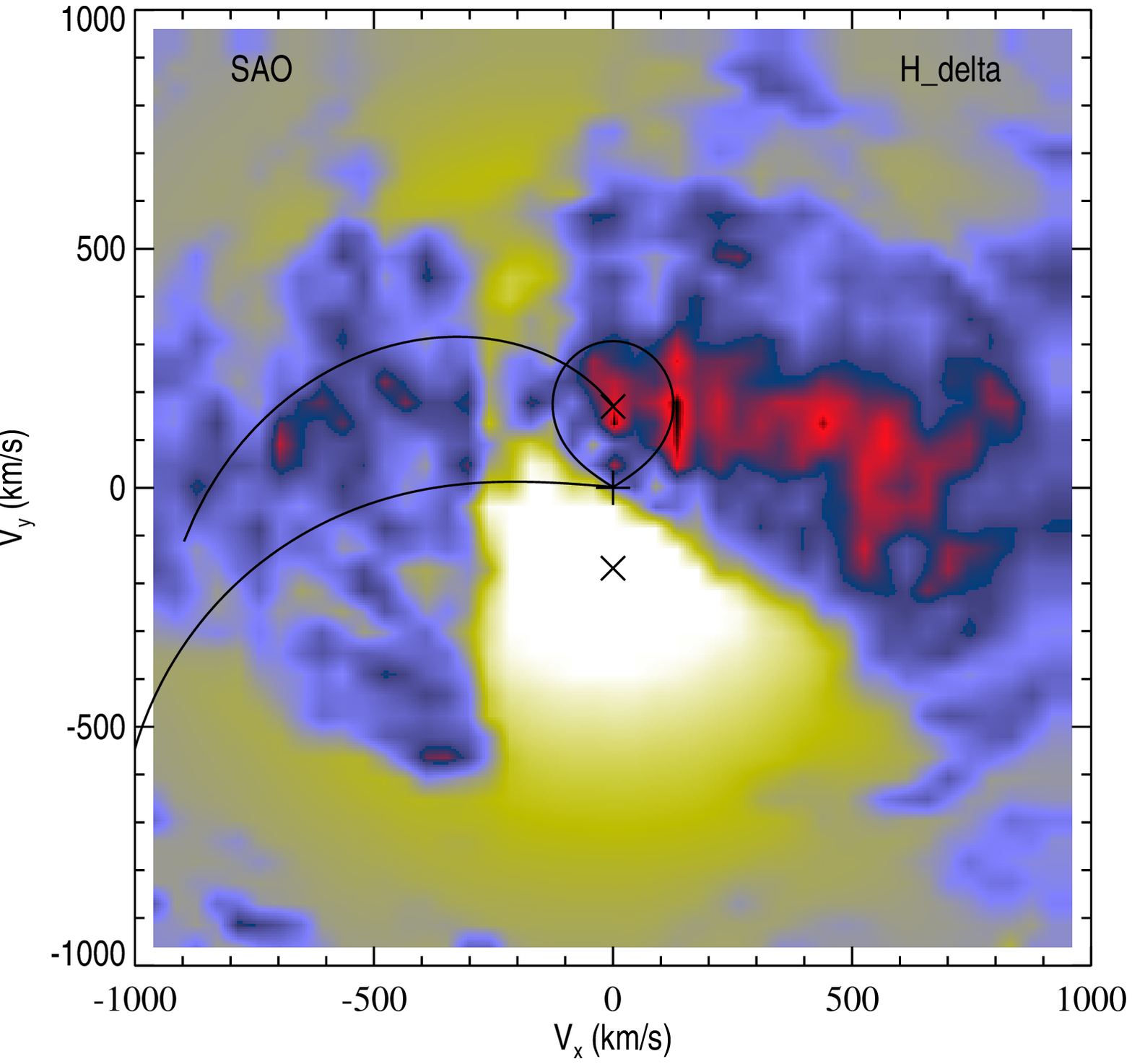}\\
    \includegraphics[width=2.8cm]{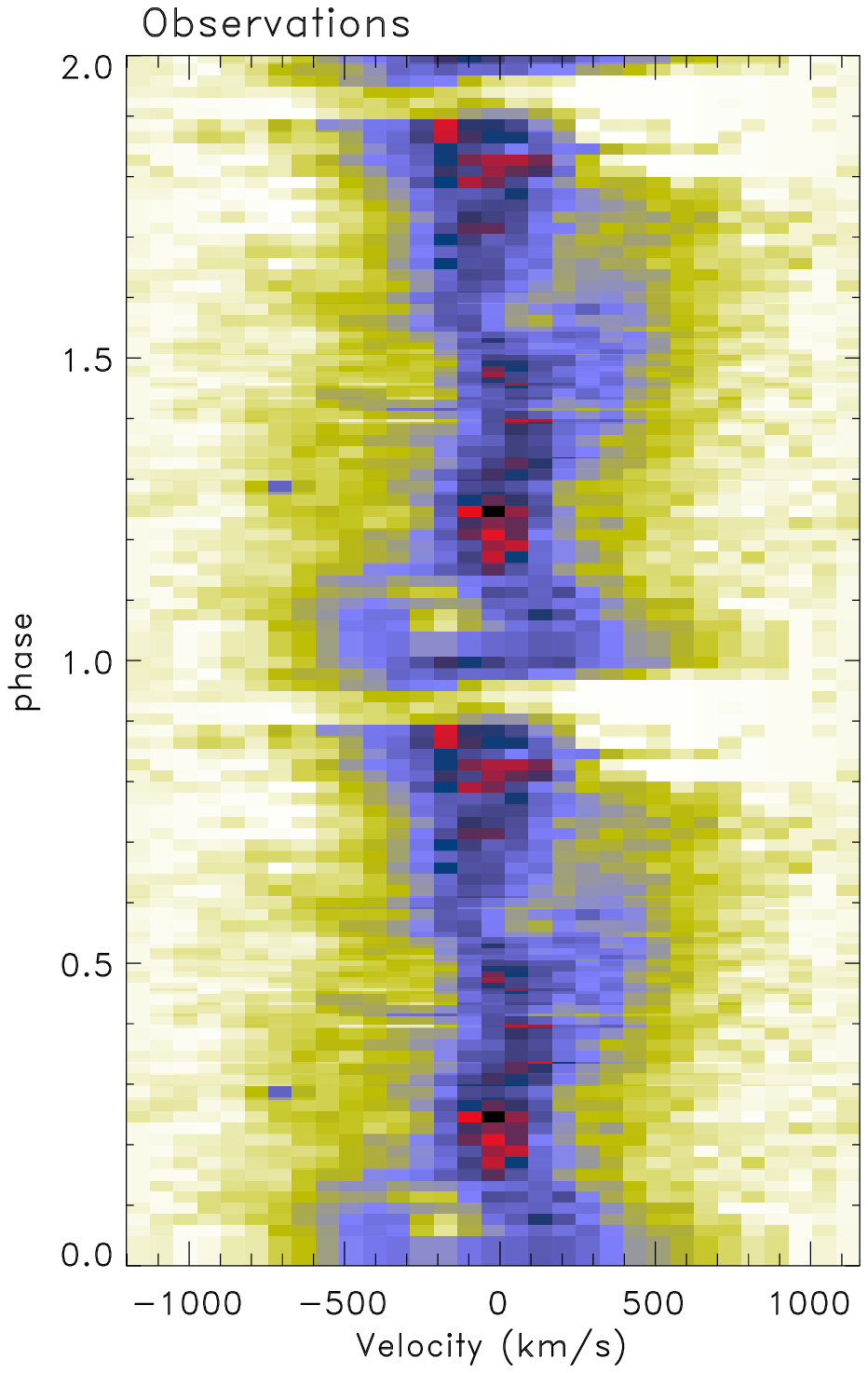}
    \includegraphics[width=2.8cm]{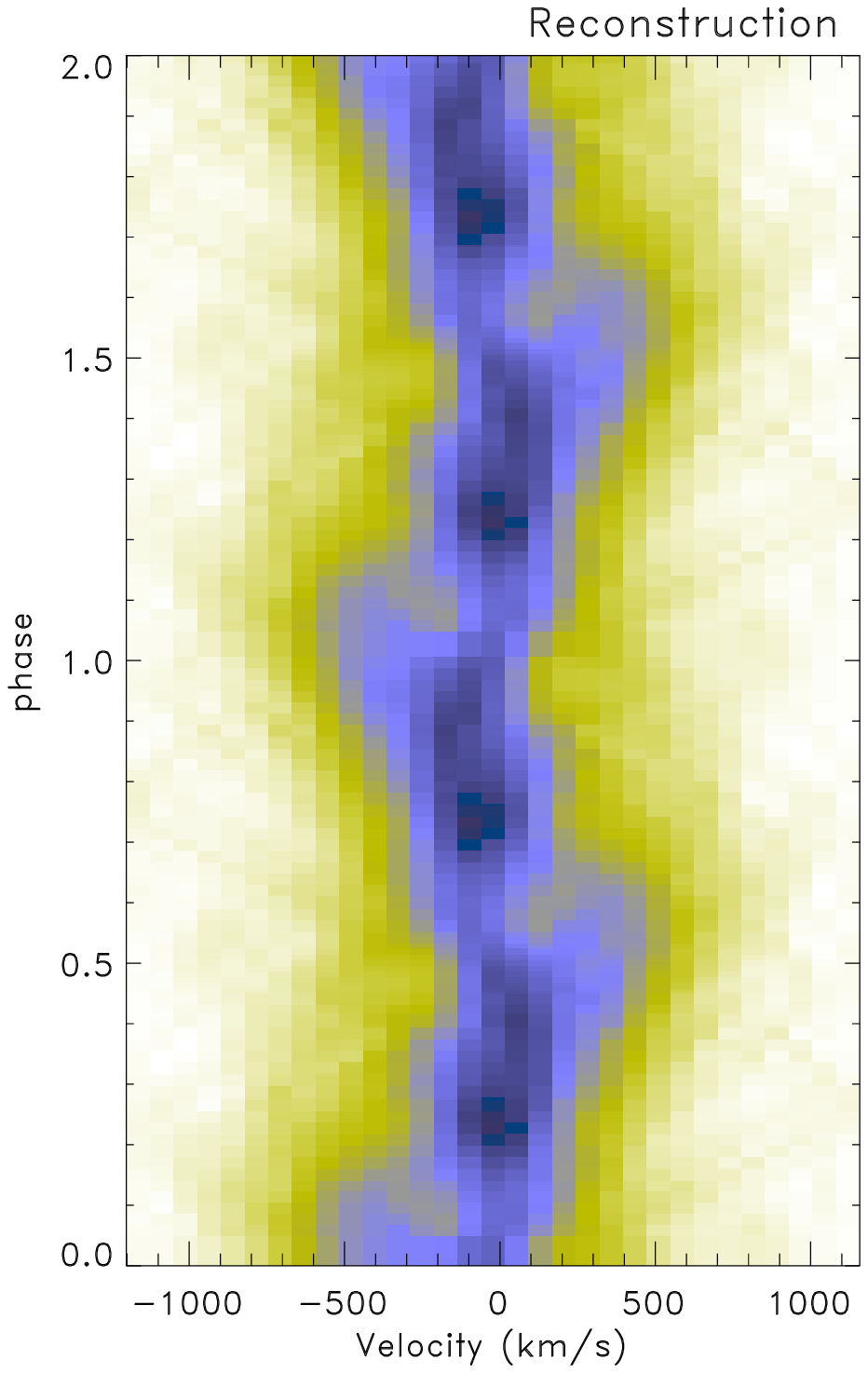}
    \includegraphics[width=2.8cm]{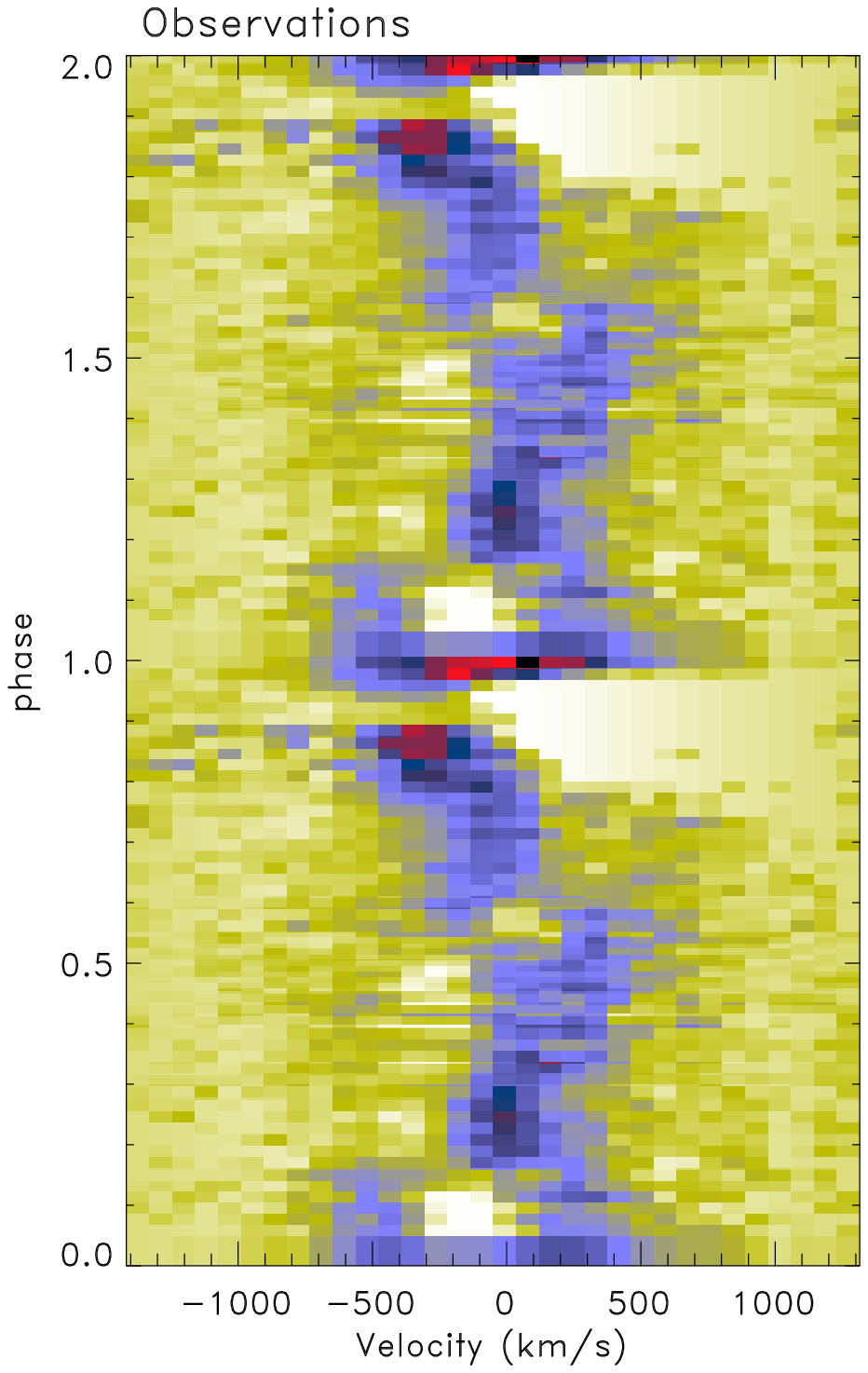}
    \includegraphics[width=2.8cm]{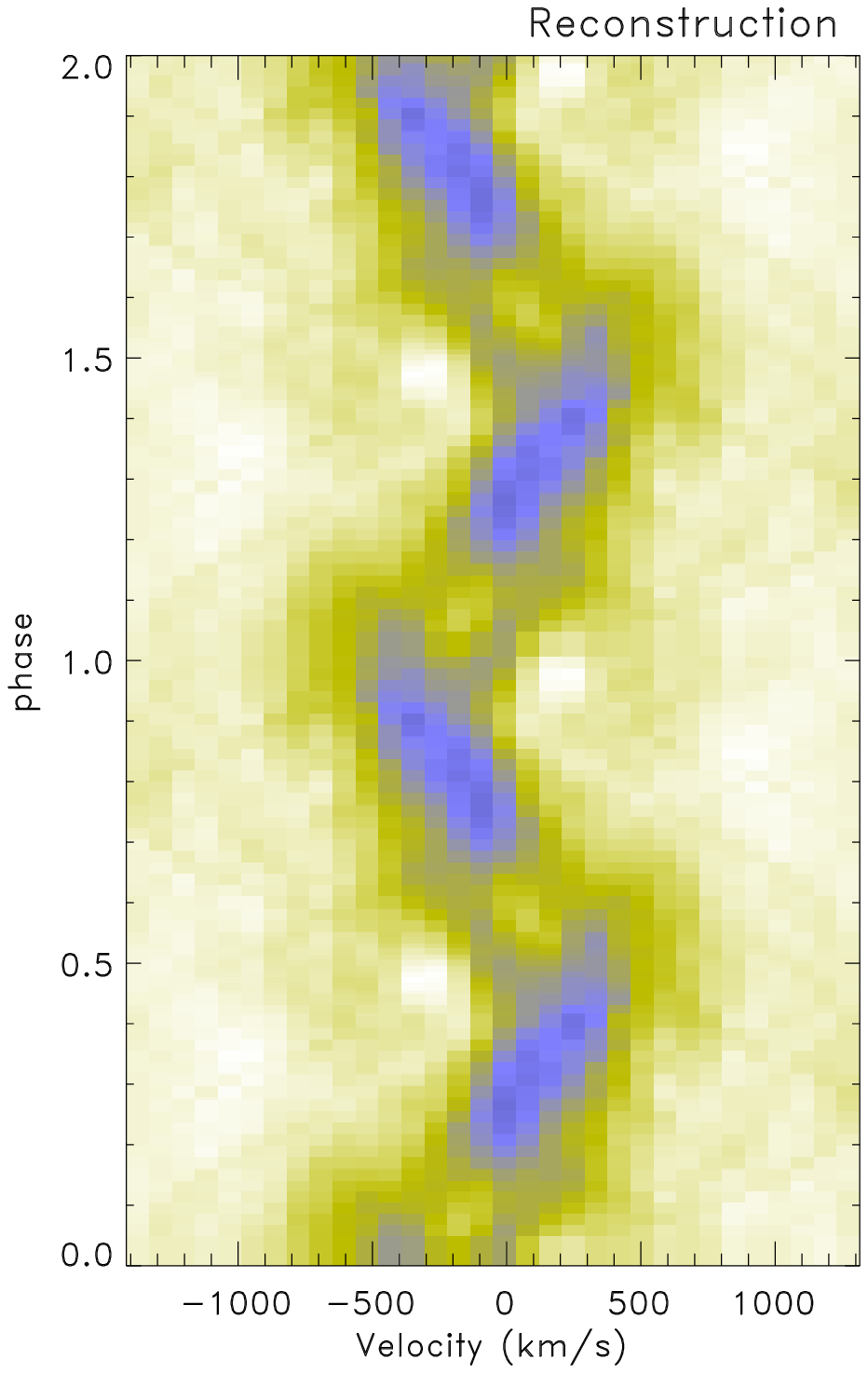}
    \includegraphics[width=2.8cm]{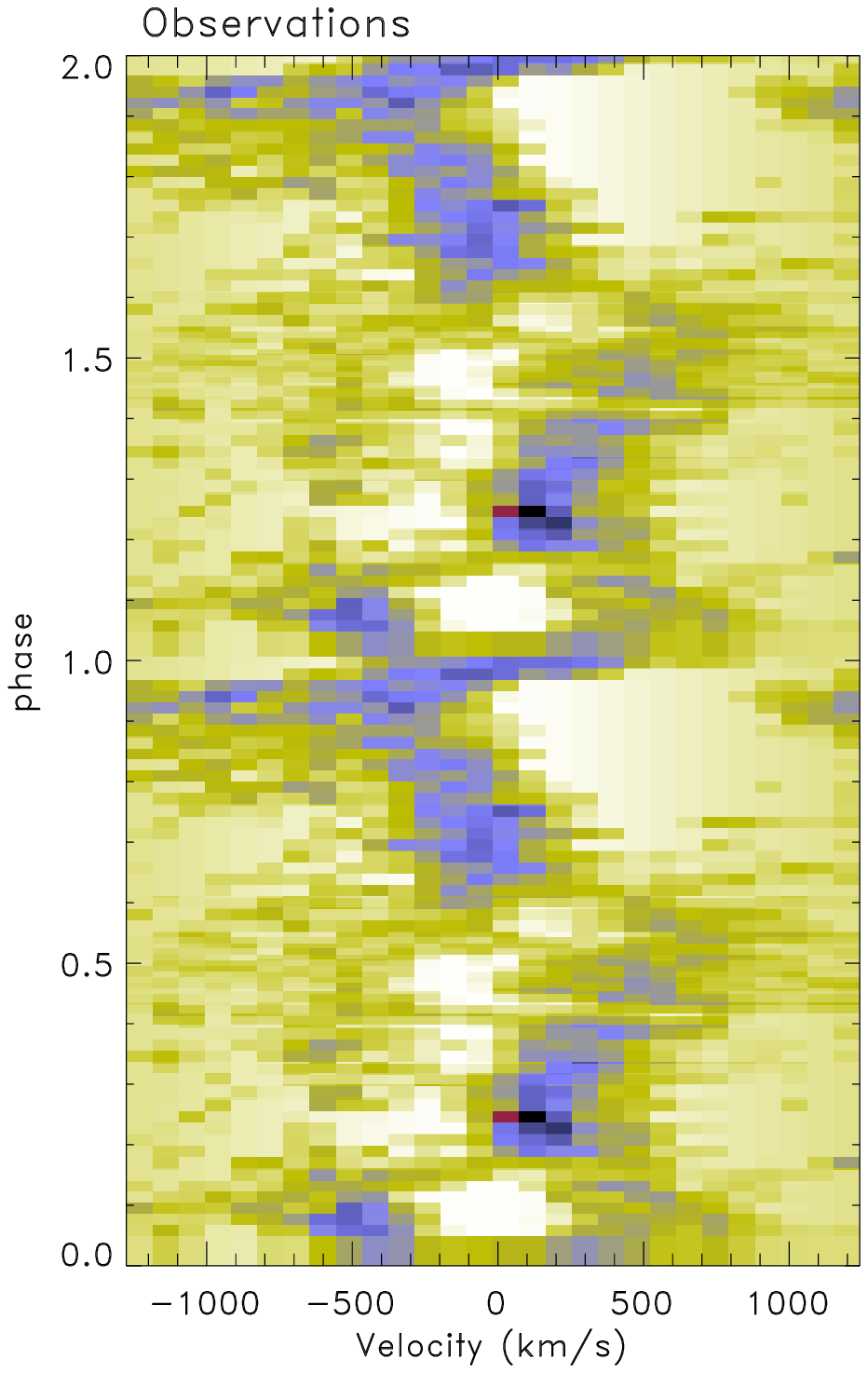}
    \includegraphics[width=2.8cm]{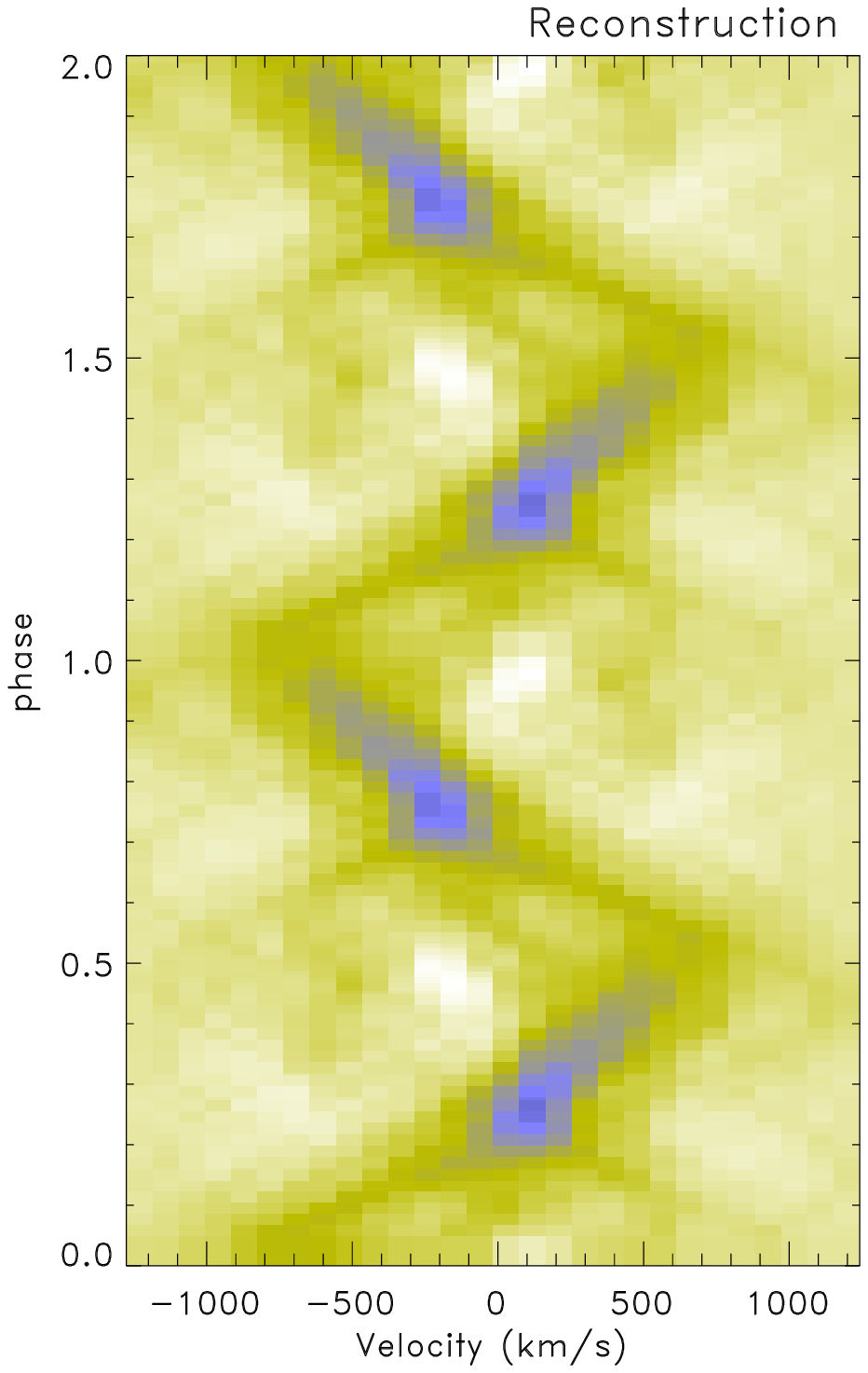} \\
    \bigskip \bigskip
    \includegraphics[width=5.6cm]{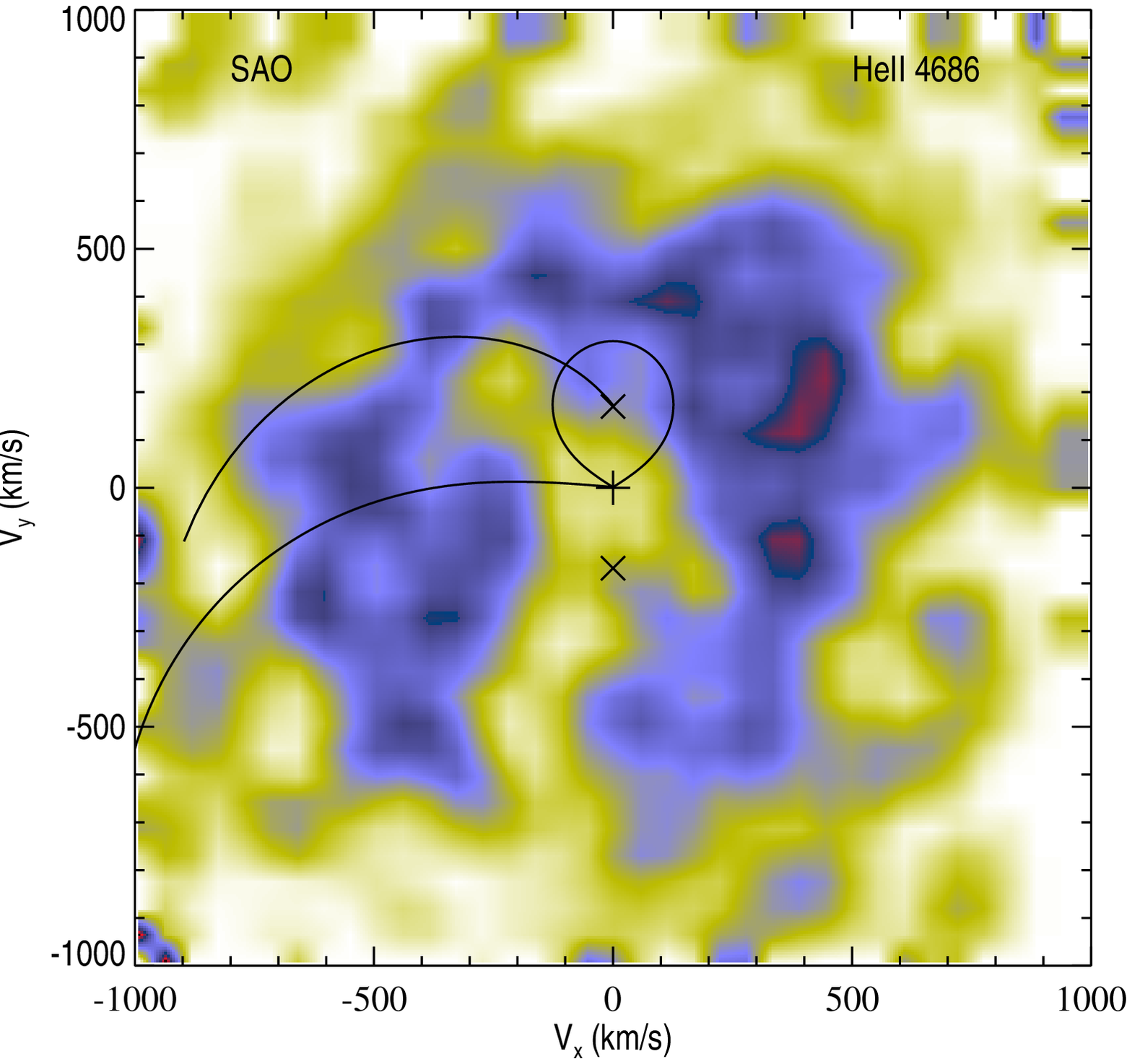}
    \includegraphics[width=5.6cm]{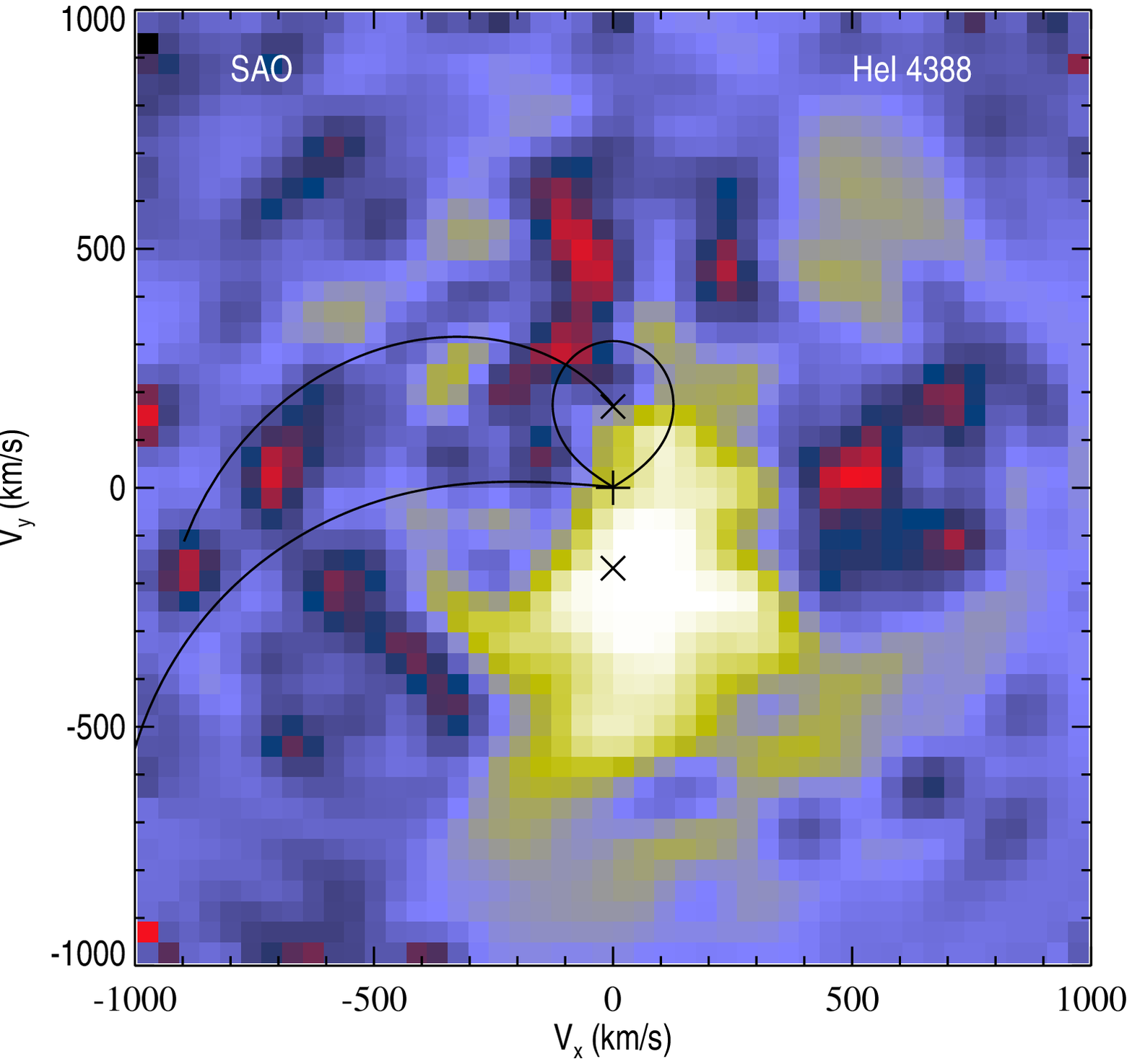}
    \includegraphics[width=5.6cm]{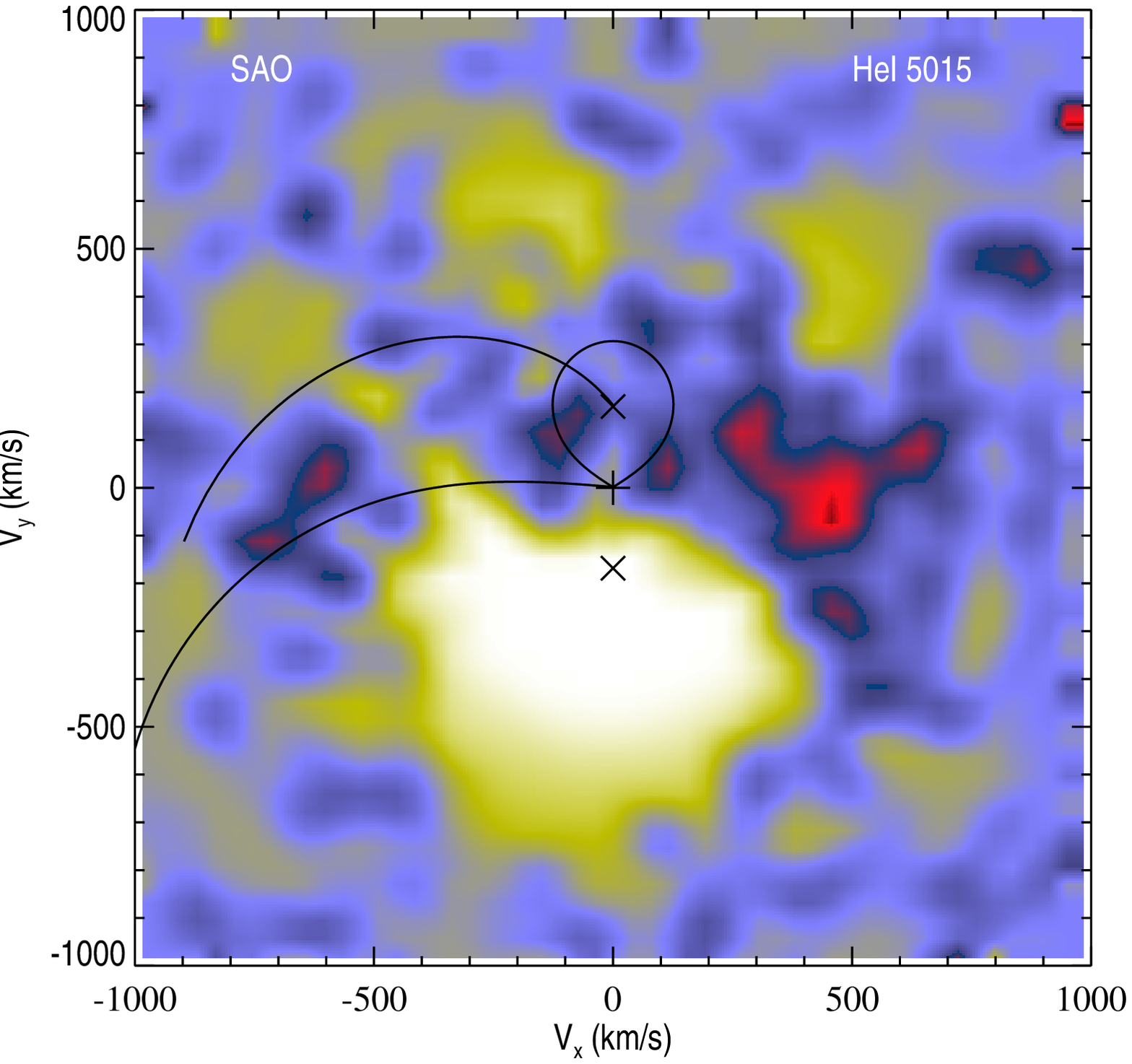}\\
    \includegraphics[width=2.8cm]{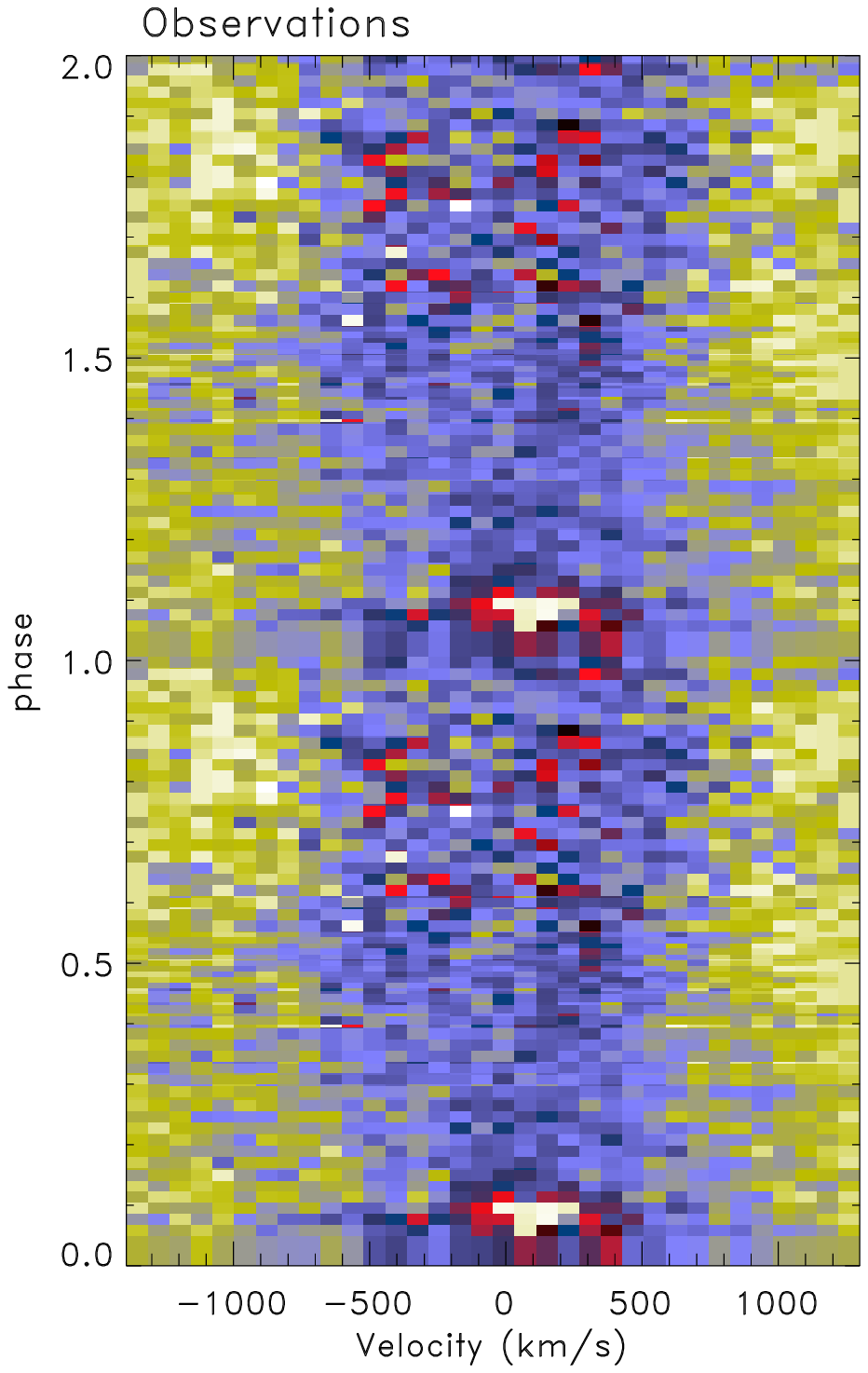}
    \includegraphics[width=2.8cm]{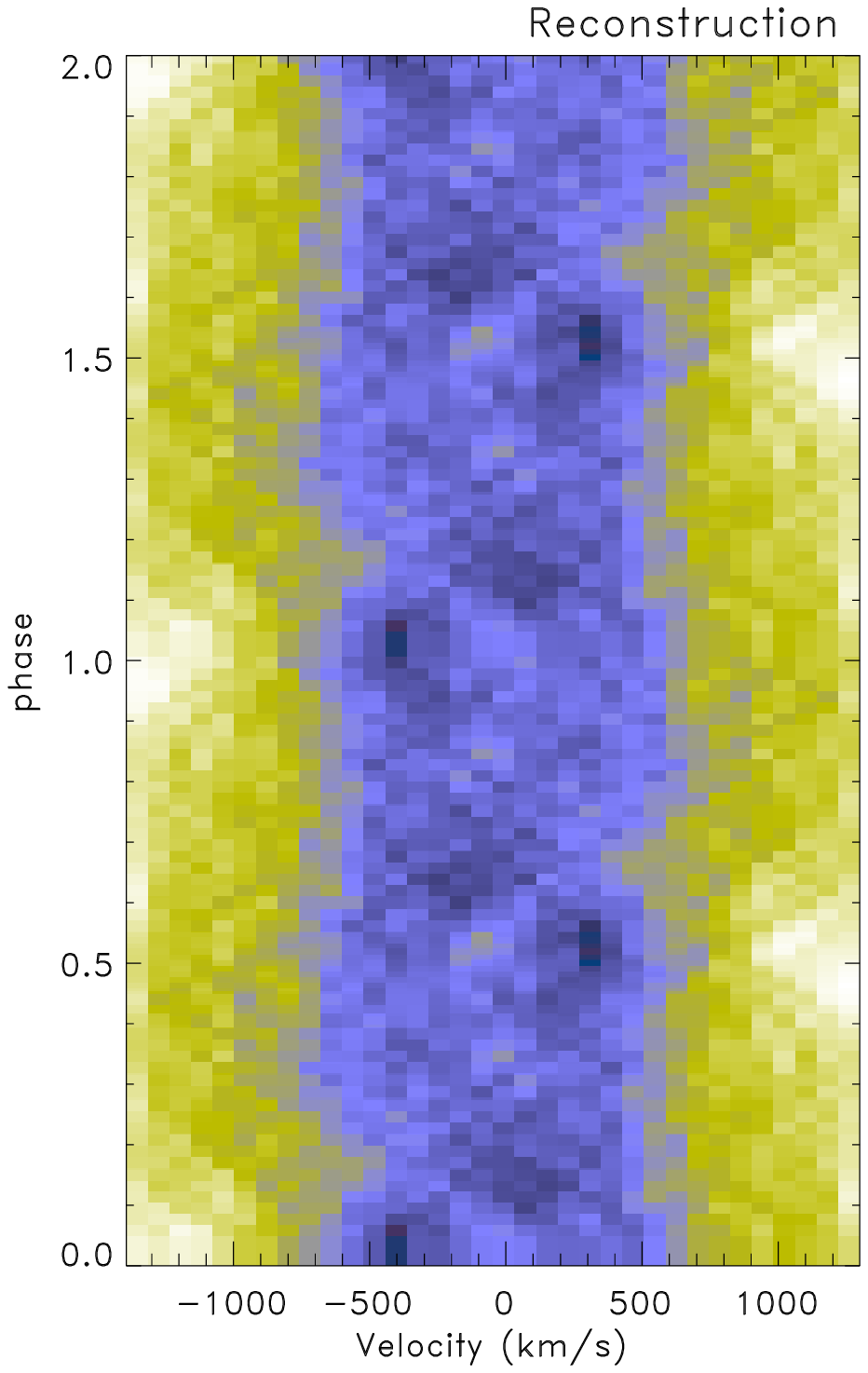}
    \includegraphics[width=2.8cm]{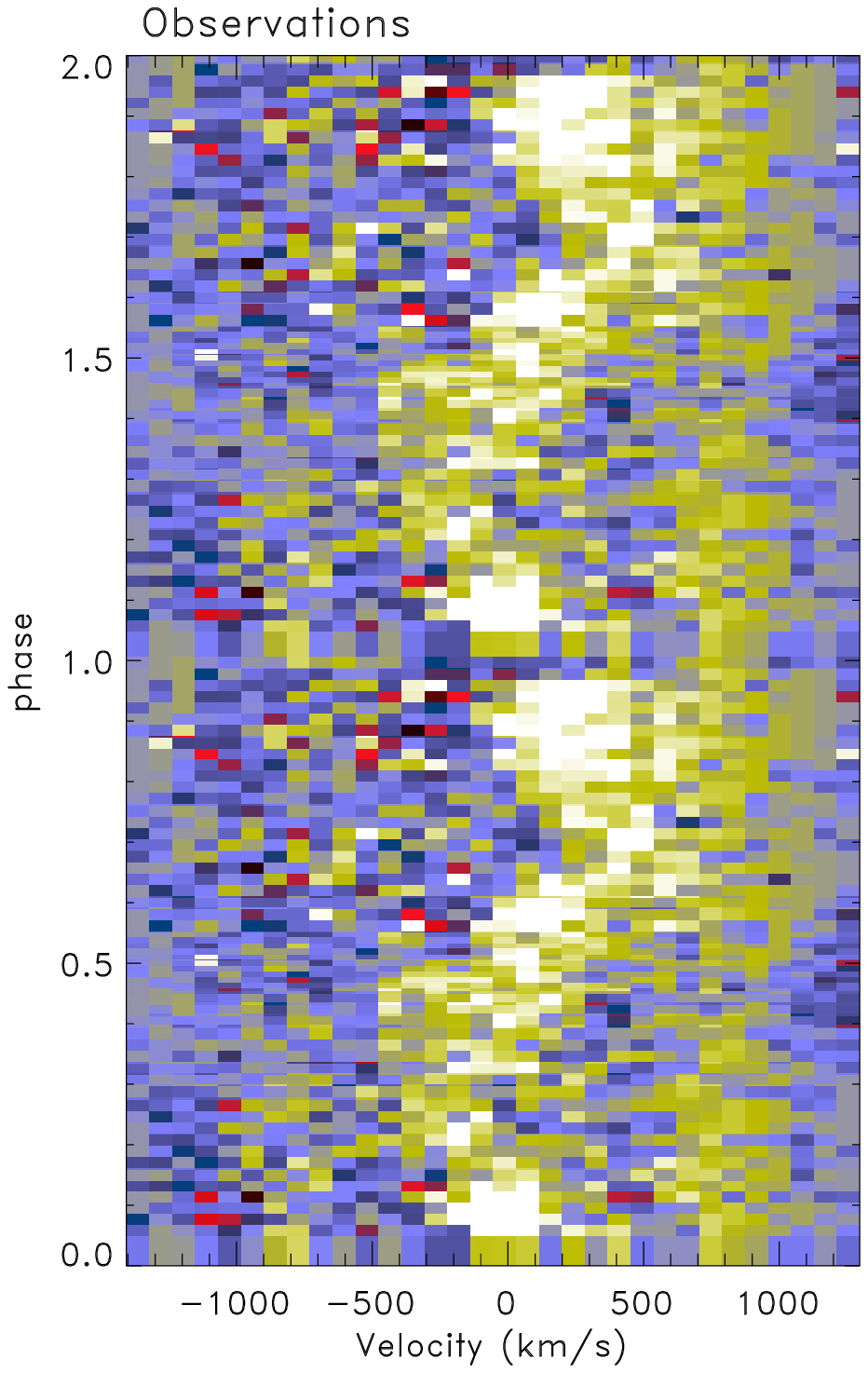}
    \includegraphics[width=2.8cm]{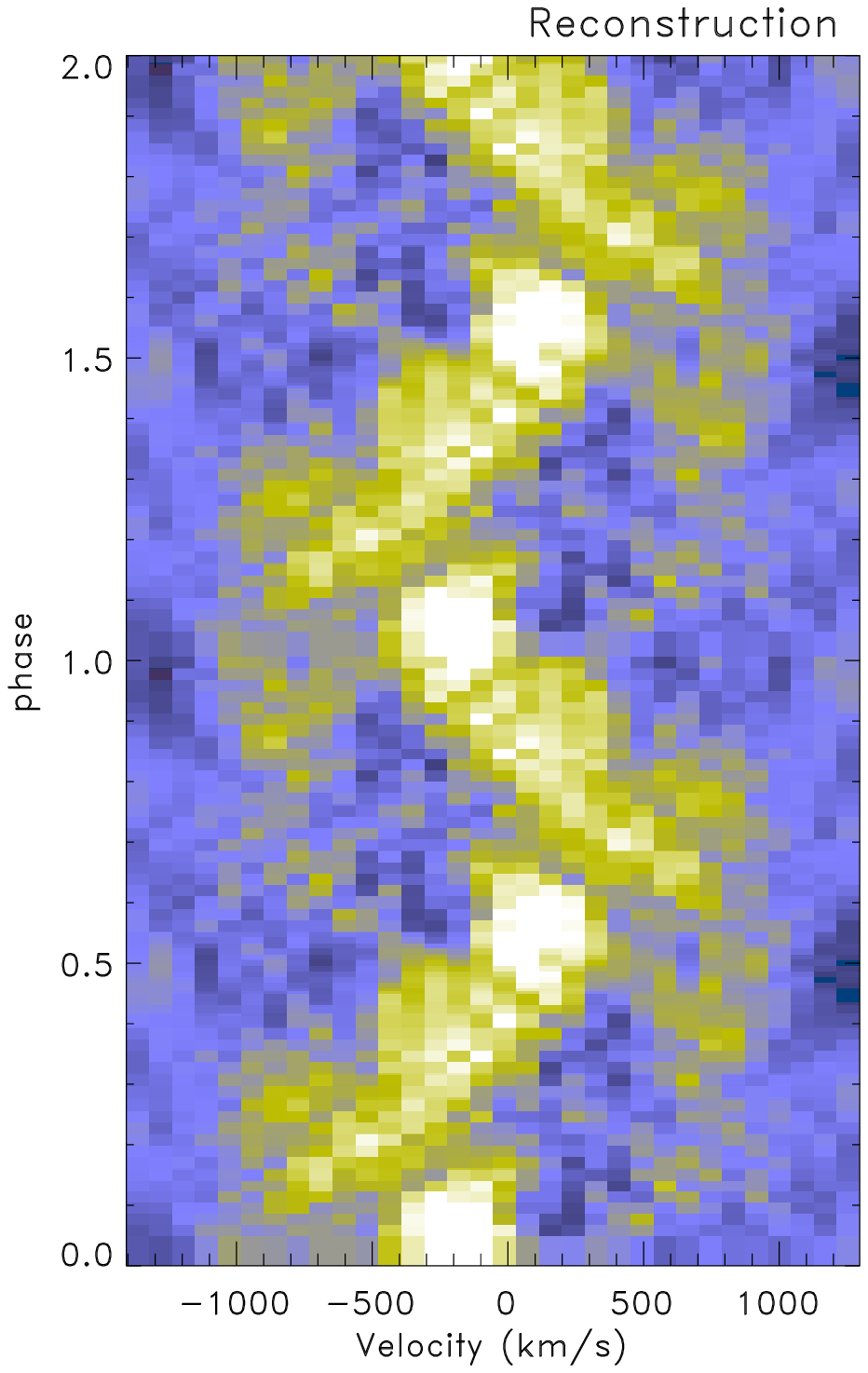}
    \includegraphics[width=2.8cm]{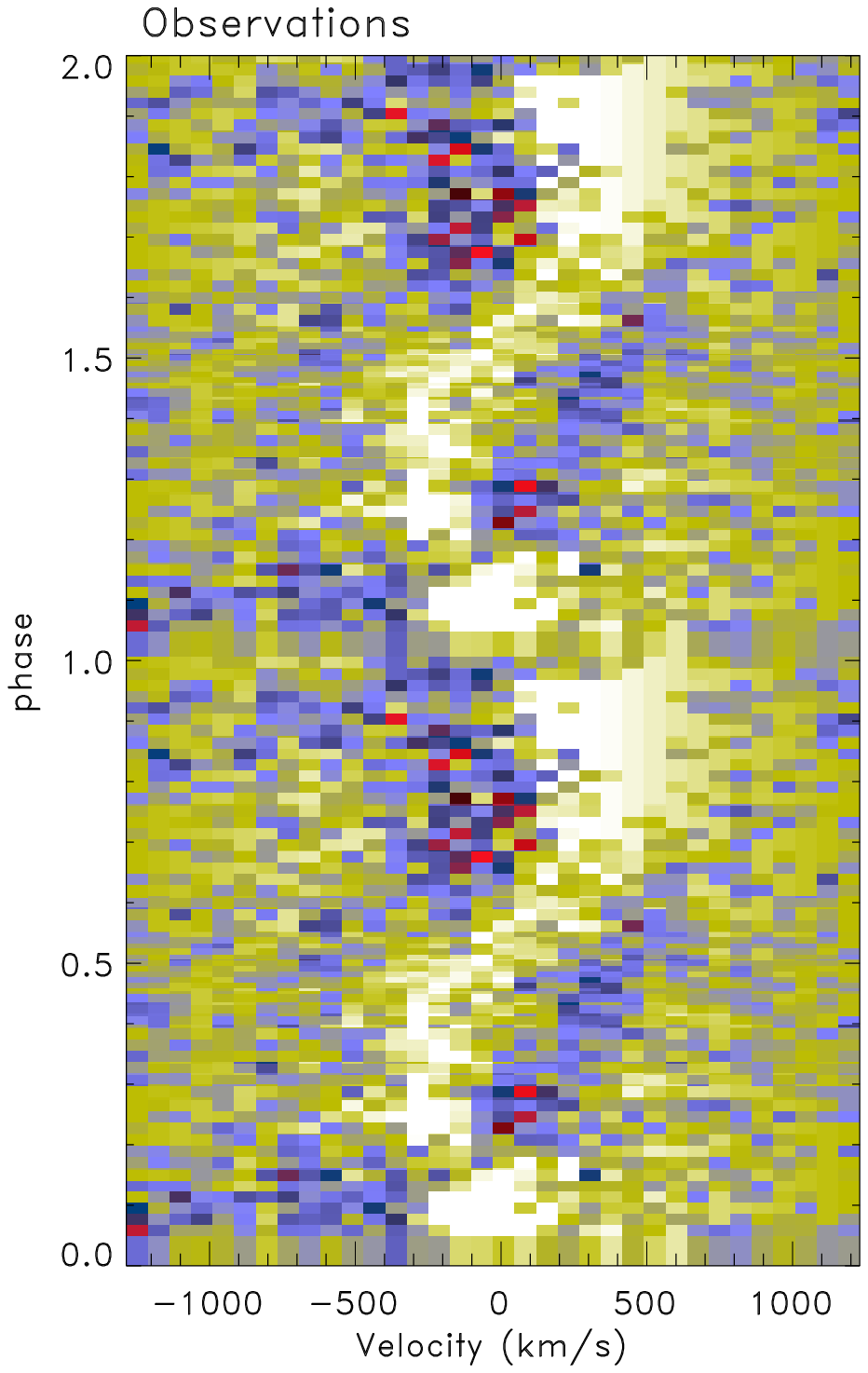}
    \includegraphics[width=2.8cm]{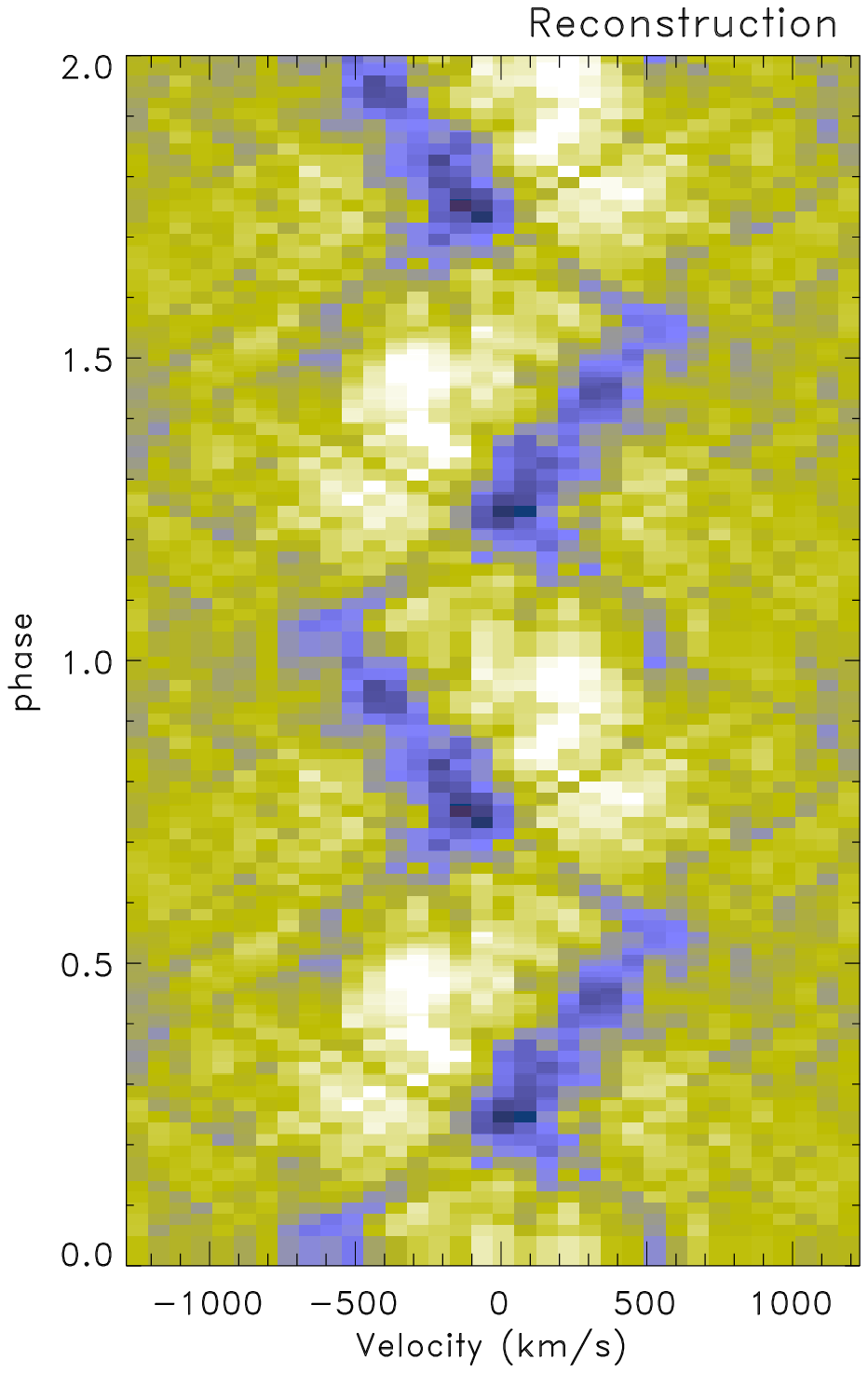}
    \caption{Doppler tomography for the \Hbeta, \Hgamma\ and \Hdelta\ emission lines (in the upper half of Figure),
    and for \HeII\ $\lambda4686$, \HeI\ $\lambda4388$ and \HeI\ $\lambda5015$ (in the bottom half of Figure)
    from the SAO set of observations. For each line the observed and reconstructed trailed spectra
    (bottom) and corresponding Doppler maps (top) are shown.   Marked on  the maps are the positions
    of the WD (lower cross), the center of mass of the binary (middle cross) and the Roche lobe of the
    secondary star (upper bubble with the cross). The predicted trajectory of the gas stream and the
    Keplerian velocity of the disc along the gas stream have also been shown in the form of the lower
    and upper curves, respectively. The Roche lobe of the secondary and the trajectories have been plotted
    using the system parameters, derived by \citet{Baptista95} -- an inclination $i=71^{\circ}$ and $M_1=M_2=0.47M_\odot$. The flux scale is from white to black.}
    \label{dopmaps1}
   \end{figure*}

  Even more details can be found in the phase-resolved trailed spectra (Fig.~\ref{aver_spec_blue}, bottom panel).
  One can clearly recognize that the phase 0.6 absorption mentioned before is in fact a manifestation of
  the sinusoidal absorption component of the emission lines seen continuously over the entire orbital
  period, which crosses from blue-shifted to red-shifted around phase $\sim$0.6.
  The semi-amplitude of its radial velocity variations can be roughly estimated to be $\sim350$ \kms, much more
  than published estimates of the radial velocity semi-amplitude of UX~UMa's WD. Thus, this absorption component
  with a nearly constant intensity originates somewhere on the far side of the accretion disc relative to
  the secondary star.
  Note also that all the lines including the deepest absorption lines \HeI\ $\lambda$4472 and
  \HeI\ $\lambda$4026, consist of a mixture of absorption and emission components.  Thus, the difference
  in the appearance of the ``emission'' and ``absorption'' lines is determined mostly by the relative
  contributions of the corresponding components. Note that the absorption component of different lines
  seemingly varies in phase with each other.
  It is also interesting to mention that the broad absorption depression observed around the red wings of
  the lines just prior to eclipse, is certainly linked with those absorption components of the lines.
  From the trailed spectra of \Hbeta\ one can distinguish two sinusoidal emission components
  with different radial velocity amplitudes and phases. One of these components, with the smaller radial velocity
  amplitude, crosses from red-shifted to blue-shifted around phase $\sim$0.5 and can probably be attributed to the
  secondary star (see also the trailed spectra in Fig.~\ref{dopmaps1}).

  Surprisingly, the appearance and behaviour of the red spectrum was completely different from the spectrum
  in the blue wavelength range. Both the hydrogen (\Halpha) and neutral helium (\HeI\ $\lambda6678$ and
  \HeI\ $\lambda7065$) lines were observed in emission and both exhibited double-peaked profiles though
  the depth of the central dip of the helium lines was below the continuum (Fig.~\ref{aver_spec_red}).
  There was no visible change in the spectrum during the eclipse and neither high-velocity absorption
  nor emission appeared.

  These clear changes between blue and red spectra may not be so surprising as these data sets are 9 years apart.
  On the other hand, care must be taken since we are not comparing the same wavelength ranges. Such differences
  between spectra can be due to different line transfer effects in different lines. We discuss this below.

\begin{figure*}
\centering
\includegraphics[width=8.0cm]{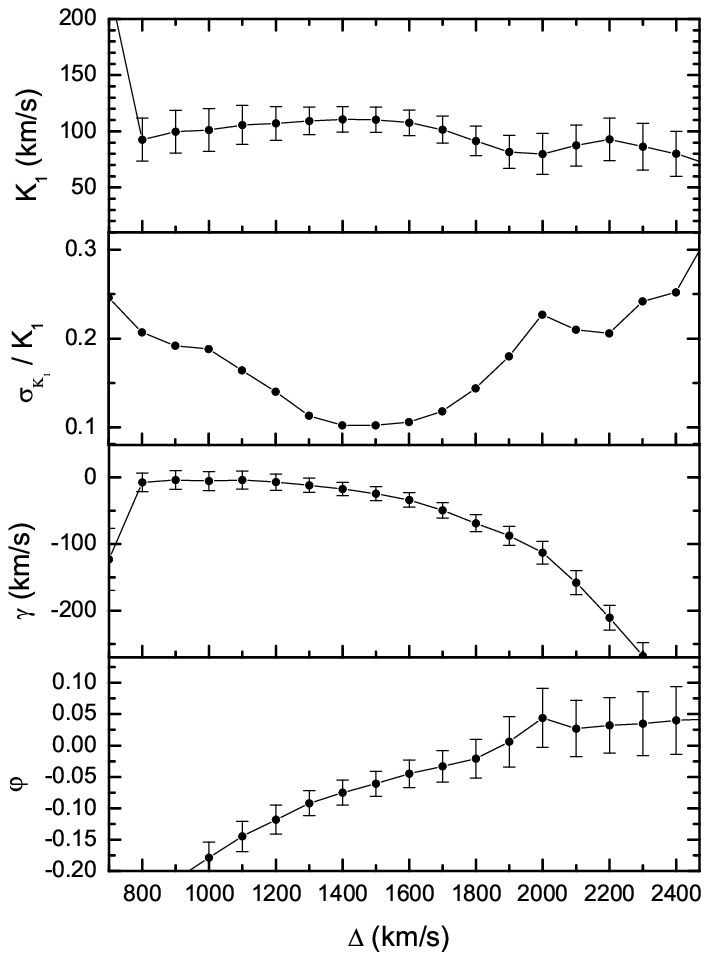} \hspace{5mm} \includegraphics[width=8.0cm]{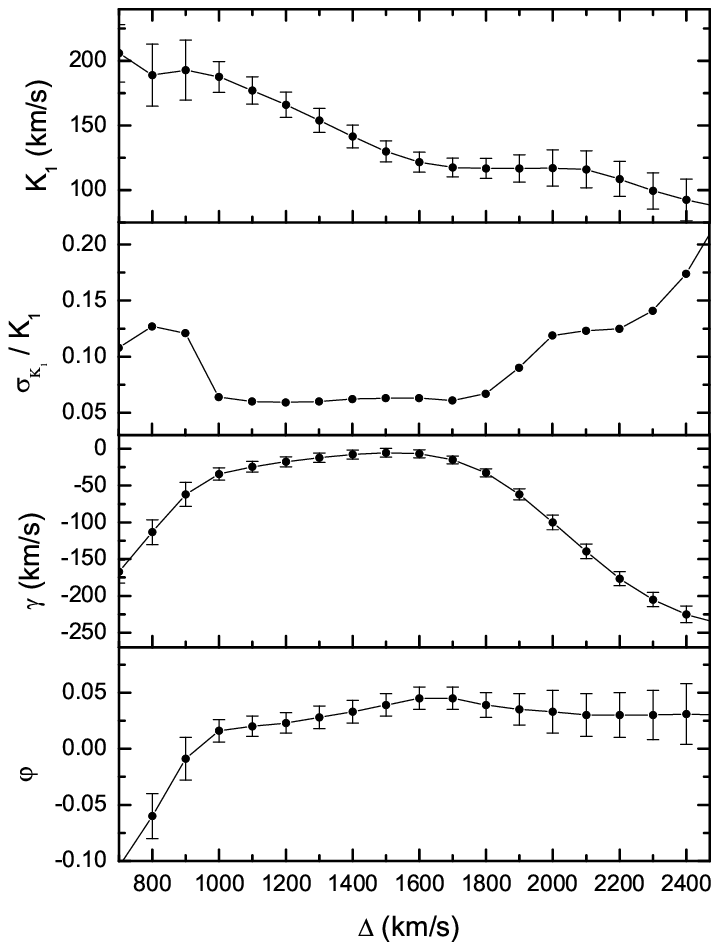}
\caption{The diagnostic diagram for SAO's H$\beta$ (left panel) and Loiano's H$\alpha$ (right panel) data,
showing the response of the fitted orbital elements to the choice of the double-gaussian separation.
The best fit is reached with the gaussian separation of $1600{-}1700$ \kms\ for \Hbeta\ and 1800 \kms\
for \Halpha.}
\label{diagram}
\end{figure*}

\subsection{Radial velocity study}
\label{SecRadVelWD}

    A long standing problem in UX~UMa is the measurement uncertainty of the radial velocity
    semi-amplitude of the WD. Obtaining an accurate value of $K_1$ has originally been one of the main aims
    of our time-resolved spectroscopy of UX~UMa.

    In CVs the most reliable parts of the emission line profile
    for deriving the radial velocity curve are the extreme wings. They
    are presumably formed in the inner parts of the accretion disc and therefore
    should represent the motion of the WD with the highest reliability.
    We measured the radial velocities using the double-Gaussian method
    described by \citet{sch:young} and later refined by \citet{Shafter1983}.
    This method consists of convolving each spectrum with a pair of Gaussians of
    width $\sigma $ whose centres have a separation of $\Delta $.
    The position at which the intensities through the two Gaussians become equal
    is a measure of the wavelength of the emission line. The measured velocities
    will depend on the choice of $\sigma $ and $\Delta $, and by varying $\Delta $
    different parts of the lines can be sampled. The width of the Gaussians $\sigma $
    is typically set by the resolution of the data.

    The analysis of the trailed spectra (Section~\ref{AveSpecSec}) and Doppler tomography
    (Section~\ref{DopMapSec}) shows that only few spectral lines can be used for
    radial velocity measurements using this method. Indeed, all the lines, with the exception of
    \HeII\ $\lambda4686$ and \CIII\ / \NIII, consist of a mixture of absorption and emission components.
    The absorption component is very strong in most of the lines, especially in \HeI\ and the metal lines.
    It forms in a compact area of the accretion disc and its radial velocity curve does not reflect
    the orbital motion of the WD. This can be the reason for unrealistic values of the parameters of
    the radial velocity variations,  obtained by \citet{Schlegel1983} using ``absorption'' lines.
    However, this absorption component is much weaker in the Balmer lines and apparently does not
    affect their wings, except for the phases just prior to eclipse.

    In order to test for consistency in the derived velocities and the zero phase, we separately
    used the emission lines \Halpha\ and \Hbeta\ in the Loiano and SAO spectra
    respectively.

    All the measurements were made using a Gaussian FWHM of 200 ${\rm km~s^{-1}}$ and
    different values of the Gaussian separation $\Delta$ ranging from 300 \kms\ to 2500 \kms\ in steps
    of 100 \kms, following the technique of ``diagnostic diagrams'' \citep{Shafter1986}.

    For each value of $\Delta$ we made a non-linear least-squares fit of the derived
    velocities to sinusoids of the form
        \begin{equation}  \label{radvelfit}
          V(\varphi,\Delta )=\gamma (\Delta )-K_1(\Delta )\sin \left[ 2\pi \left(
          \varphi-\varphi_0\left( \Delta \right) \right)\right]
        \end{equation}
    \indent
    where $\gamma$ is the systemic velocity, $K_1$ is the semi-amplitude, $\varphi_0$ is
    the phase of inferior conjunction of the secondary star and $\varphi$ is the phase
    calculated according to the ephemeris~(\ref{ephemeris}).
    During this fitting procedure we omitted spectra covering the phase ranges $\varphi = \pm0.08$, owing
    to measurement uncertainties during primary eclipse, and $\varphi = 0.78-0.92$, during which the red
    wing of \Hbeta\ was affected by the absorption.

    The resulting ``diagnostic diagrams'' are shown in Fig.~\ref{diagram}. The diagrams show the variations
    of $K_1$, $\sigma(K_1)/K_1$
    (the fractional error in $K_1$), $\gamma$ and $\varphi_0$ with $\Delta $ \citep{Shafter1986}.
    To derive the orbital elements of the line wings we took the values that
    correspond to the largest separation just before $\sigma(K_1)/K_1$ shows a
    sharp increase \citep{ShafterSzkody}. We find the maximum useful
    separation to be $\Delta_{max} \simeq 1800$ \kms\ for \Halpha, and $\Delta_{max} \simeq 1600$ \kms\ for \Hbeta.
    Note that $K_1$ is quite stable over a range of Gaussian separations around $\Delta_{max}$,
    supporting their choice. One can see that $K_1$ for \Halpha\ is practically constant for $\Delta>1700$ \kms.
    For \Hbeta\, $K_1$ is also very stable over a reasonable range in $\Delta$ of 1300--1600 \kms until the noise
    in the line wings begins to dominate.

\begin{figure}
    \centering
     \includegraphics[width=8.0cm]{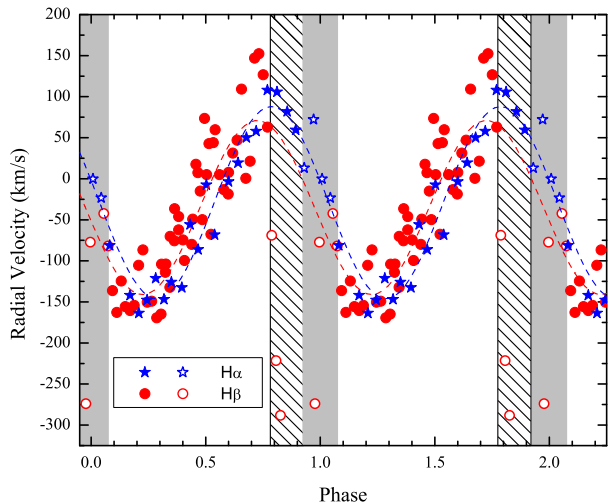}
     \caption{The \Halpha\ (blue stars) and \Hbeta\ (red circles) radial velocities folded on the orbital period.
     Two cycles are shown for clarity.
     The vertical area, shown in gray, marks the eclipsing phase range. The hatched area marks the phase range
     prior to eclipse, during which the \Hbeta\ emission line was affected by the absorption component.
     Open symbols indicate data omitted during the fitting procedure.}
     \label{FigRadVel}
   \end{figure}

    We have obtained consistent results for both the radial velocity semi-amplitudes and
    the $\gamma$-velocities for both lines. The measured parameters of the best fitting radial velocity curves
    are summarized in Table~\ref{TabRadVelEmission}, whereas the radial velocity curves are shown in Fig.~\ref{FigRadVel}.
    The formal errors are the standard deviations determined during the least-square fits for the Gaussian separation
    $\Delta_{max}$. They most likely underestimate the true errors, as they do not include a priori unknown systematic effects.
    A slightly non-sinusoidal shape of the curves can also alter the parameters.
    In the discussion to follow we adopt the value for $K_1$ to be $113\pm11$ \kms\ which is the weighted
    mean for the \Halpha\ and \Hbeta\ lines (using $\sigma(K_1)$ as a weight factor).
    In order to be on the conservative side, the error bars for all the parameters have been chosen
    by taking the largest of these two lines.

    The derived value of the radial velocity semi-amplitude $K_1$ is highly inconsistent with that one
    of \citet{Shafter1984}. It is not easy to explain why his result is so different, as he did not
    present his analysis in detail. A possible reason might be poor spectral resolution for Shafter's
    data ($\approx$ 10\AA) and as a consequence the wider Gaussians used in the double-Gaussian method.

    It is well known that the application of the double Gaussian method to CVs often gives quite uncertain or
    incorrect results, as the emission lines arising from the disc may suffer several asymmetric distortions.
    Taking these potential sources of errors into consideration, we believe we could avoid them.
    In our study of UX~UMa, both the obtained values of $K_1$ are consistent with each other. Though the
    detected asymmetric emission structure of UX~UMa (particularly, the dark spot in the lower-left of
    the \Hbeta\ Doppler map, and the spiral pattern in the \Halpha\ tomogram) may potentially influence
    the velocity determination, we hope we could avoid this as those strong structures are situated well
    inside of the chosen Gaussian separation. The correct phasing of the radial velocity curve also
    strengthens our confidence. On the other hand, we cannot exclude the possibility that some other
    yet-unknown effects may distort our results.

\begin{table}
\caption[]{Elements of the radial velocity curves of UX~UMa derived from emission lines \Halpha\ and \Hbeta.}
\begin{center}
\begin{tabular}{lccc}
\hline\hline
\noalign{\smallskip}
Emission line \& & K$_{1}$ & $\gamma$-velocity $^a$ & $\varphi_{0}$$^b$ \\
dataset & (km s$^{-1}$) & (km s$^{-1}$) &  \\
\noalign{\smallskip}
\hline
\noalign{\smallskip}
\Halpha\ - Loiano & 117$\pm$8  &  -59$\pm$6  & 0.037$\pm$0.012 \\
\Hbeta\ - SAO     & 106$\pm$11 &  -54$\pm$9  & -0.025$\pm$0.020 \\
\noalign{\smallskip}
\hline
\noalign{\smallskip}
\textbf{Adopted Value} & \textbf{113$\pm$11} & \textbf{-57$\pm$9} & \textbf{0.020$\pm$0.020} \\
\noalign{\smallskip}
\hline
\noalign{\smallskip}
\end{tabular}
\end{center}
$^a$ The measured $\gamma$-velocities are heliocentric. The mean value was obtained after correction for the solar motion.\\
$^b$ Phases were calculated according to ephemeris~(\ref{ephemeris}).
\label{TabRadVelEmission}
\end{table}

   \begin{figure*}
    \includegraphics[width=5.6cm]{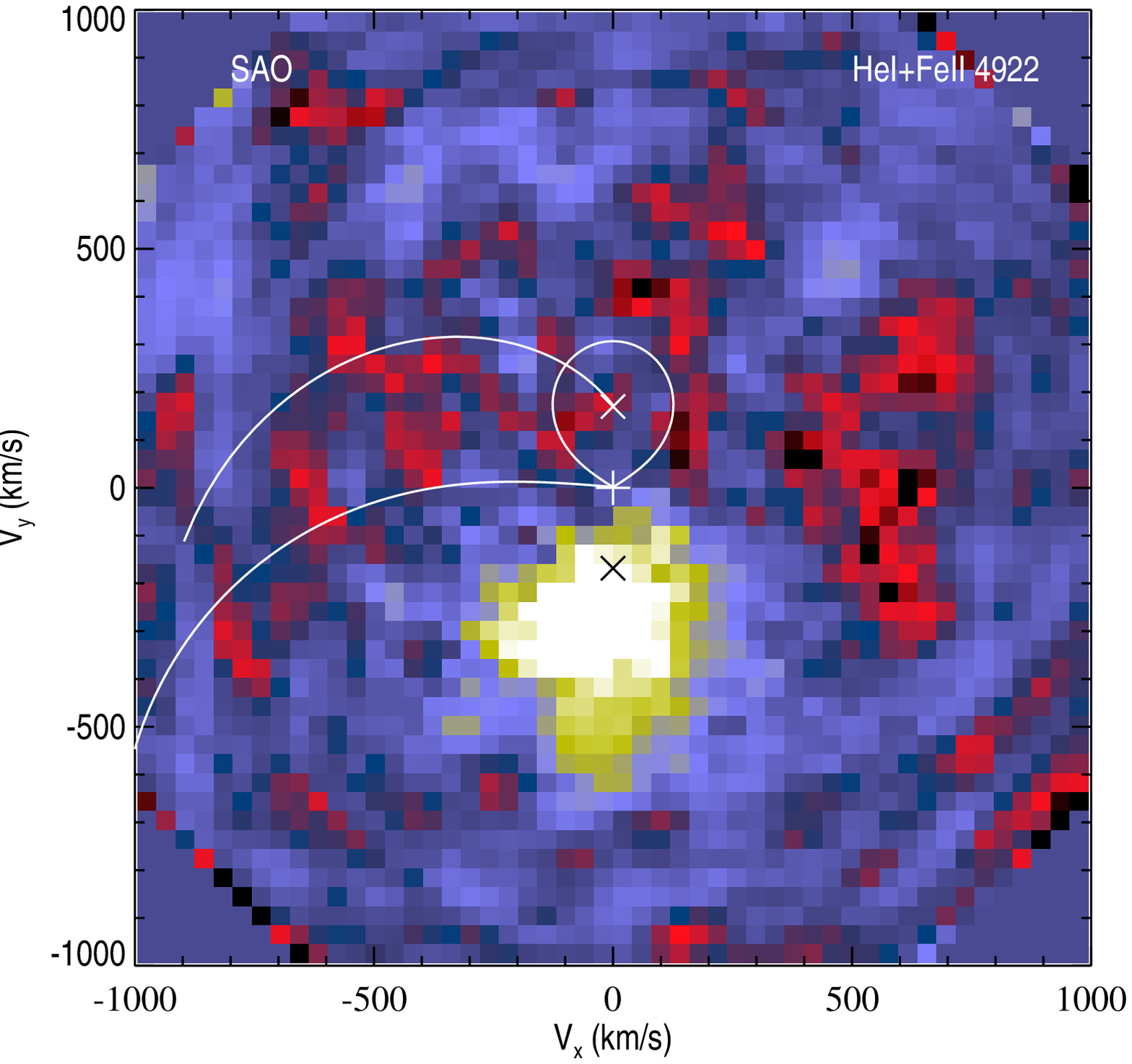}
    \includegraphics[width=5.6cm]{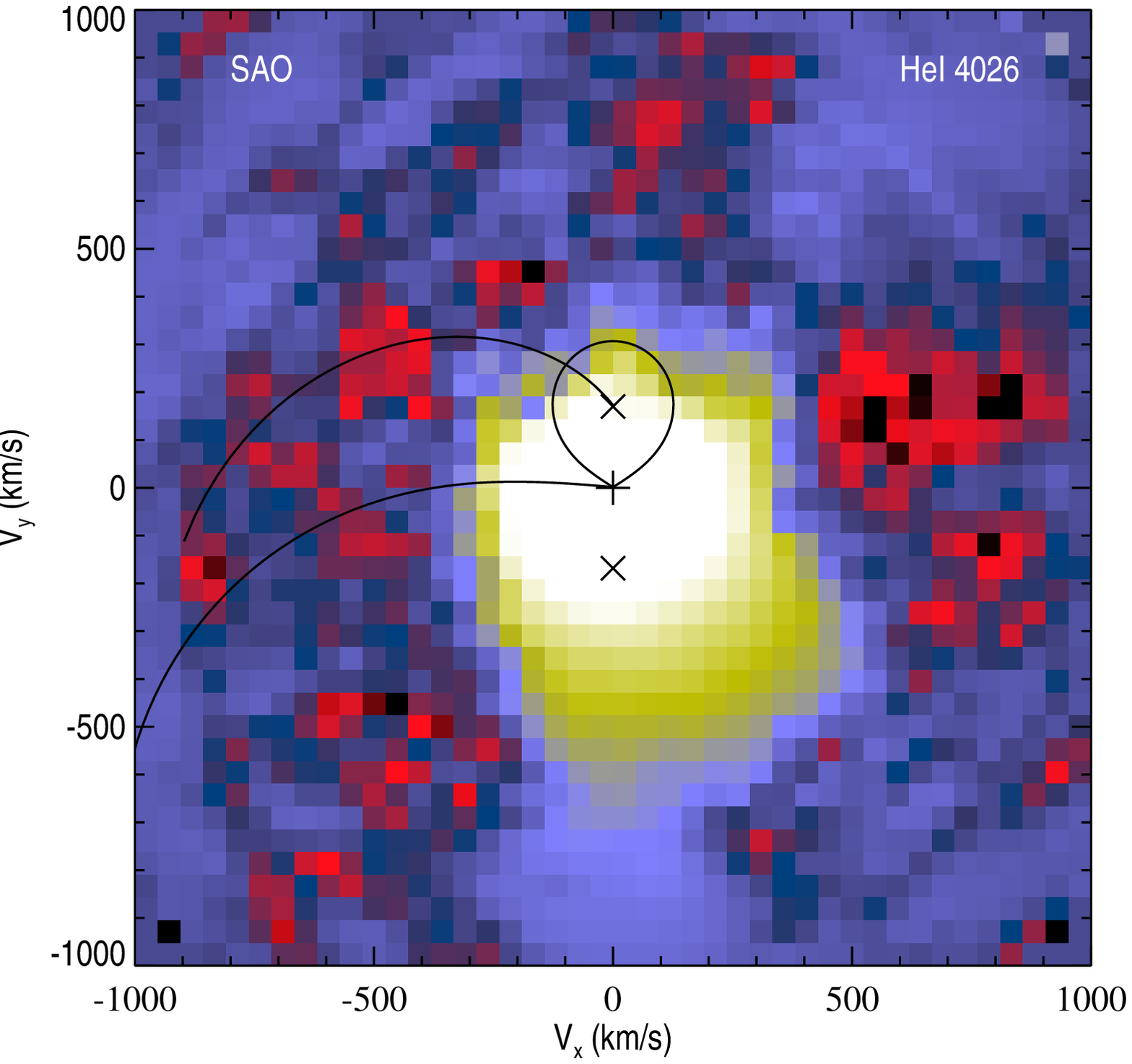}
    \includegraphics[width=5.6cm]{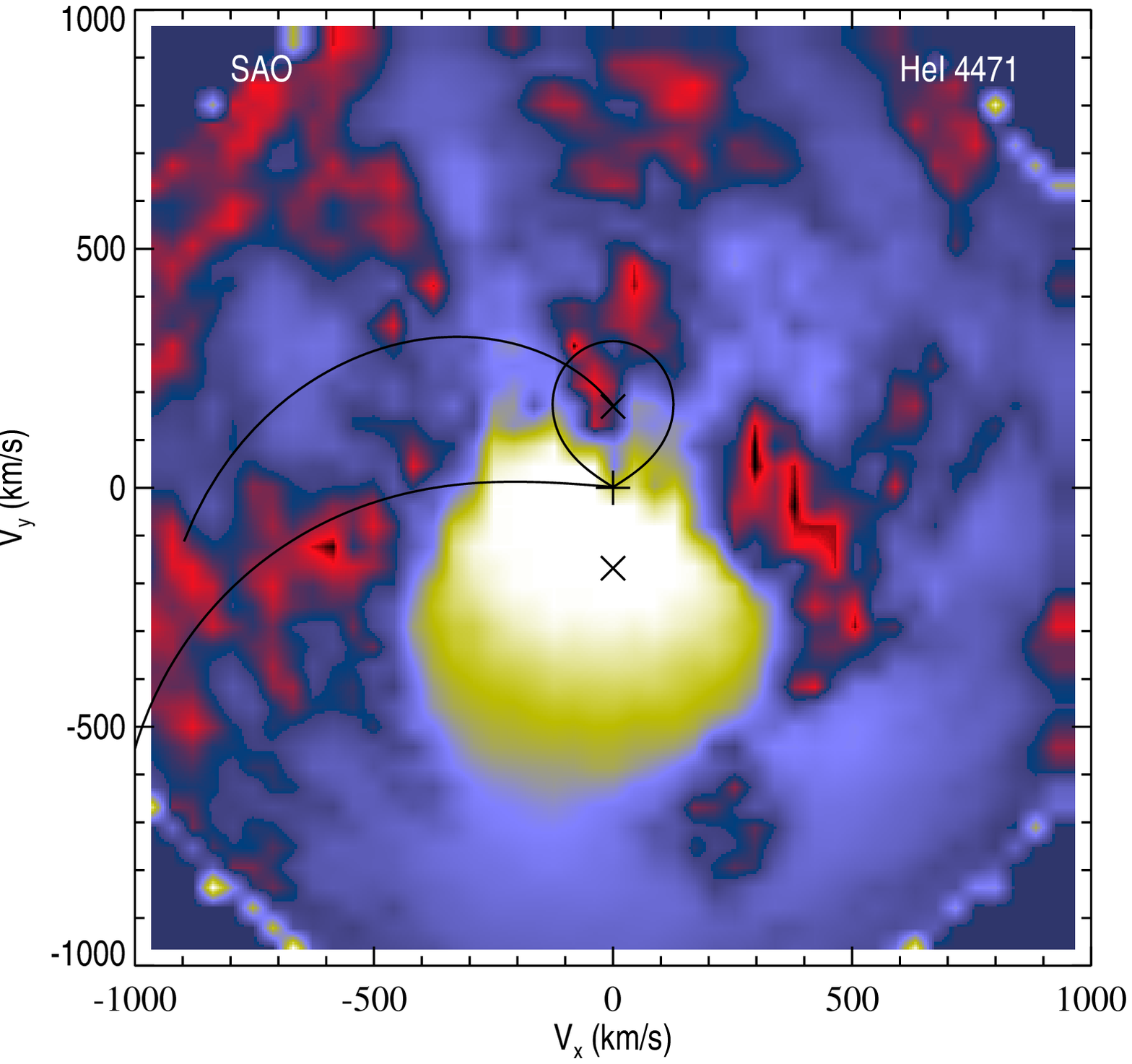}\\
    \includegraphics[width=2.8cm]{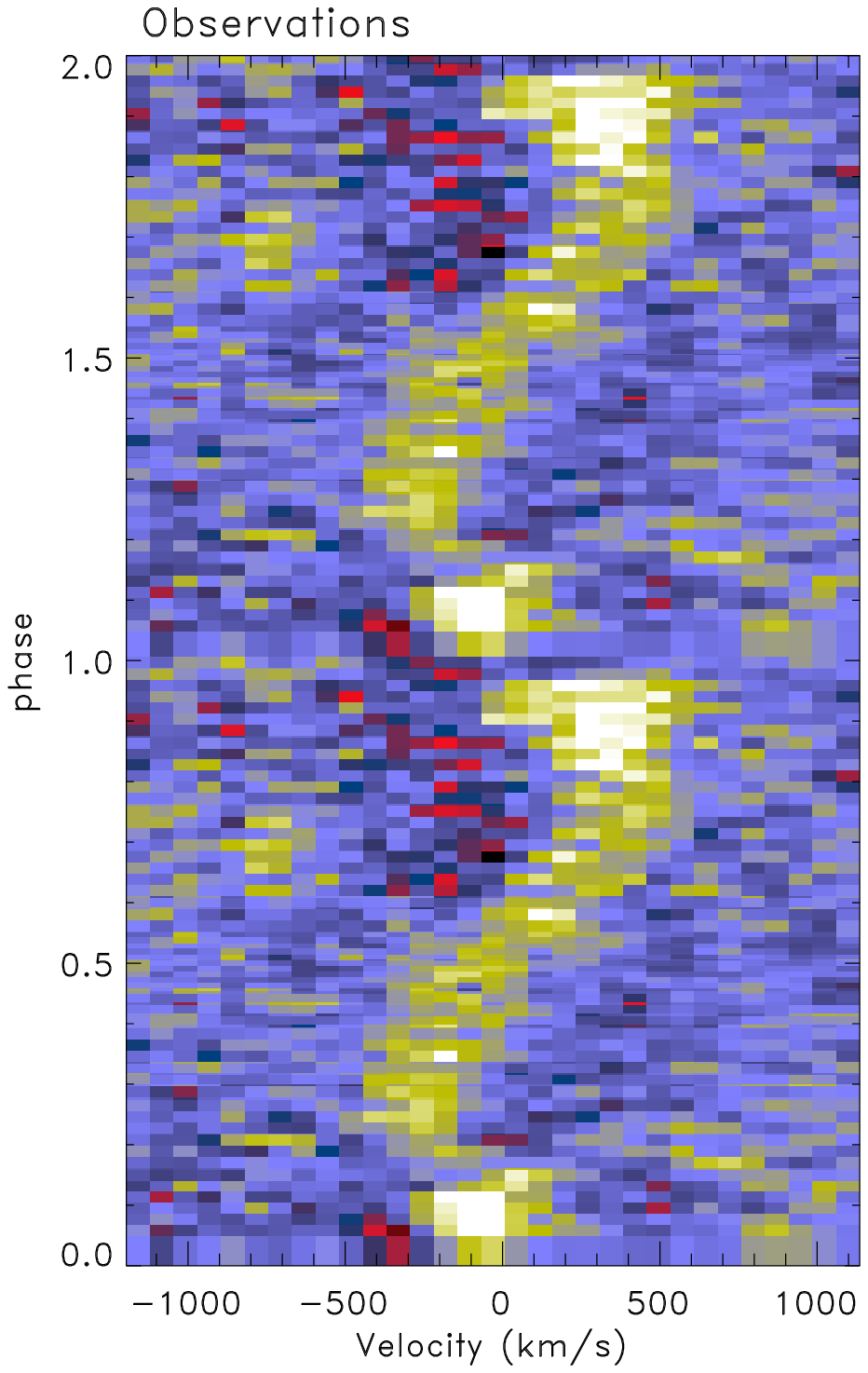}
    \includegraphics[width=2.8cm]{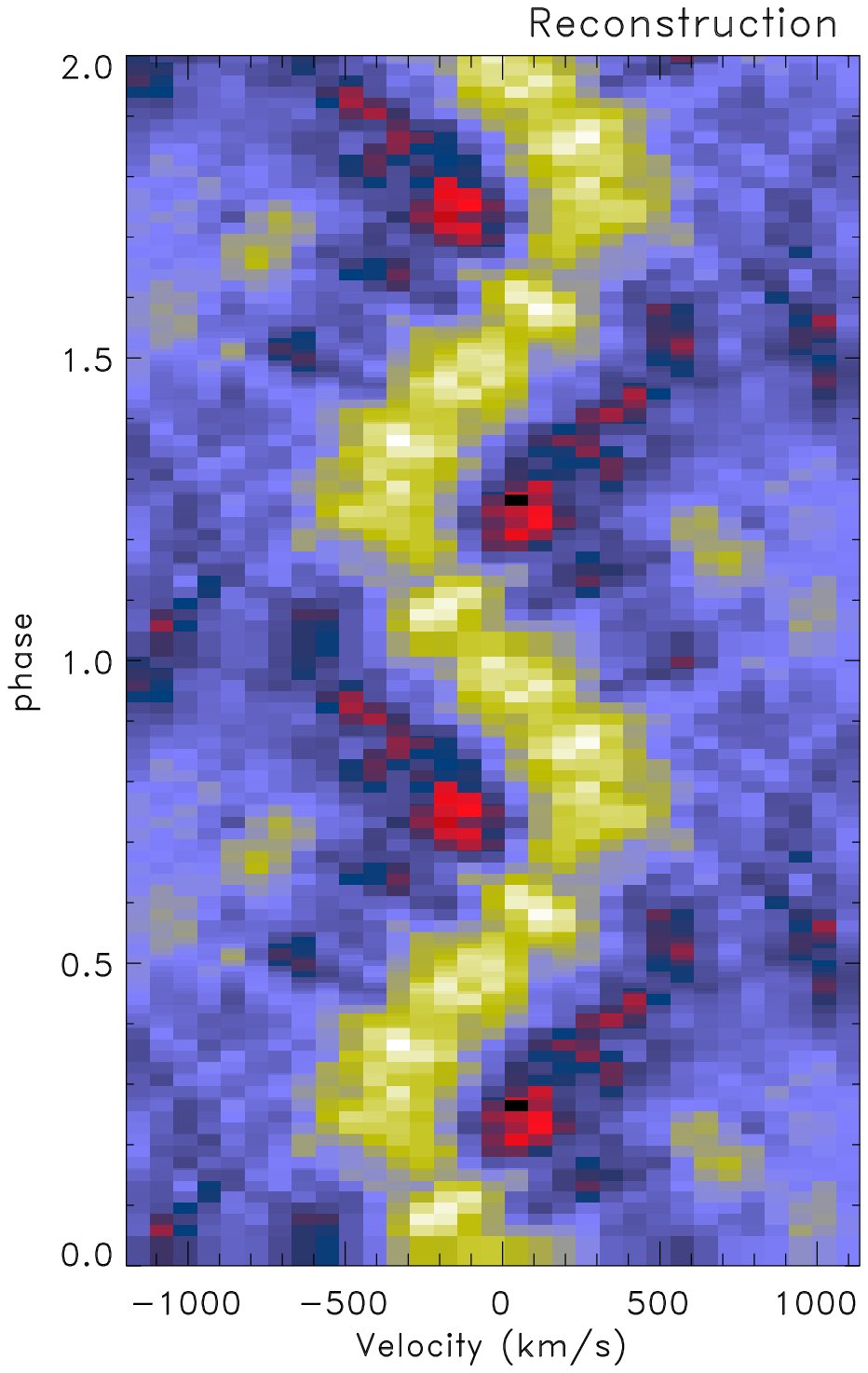}
    \includegraphics[width=2.8cm]{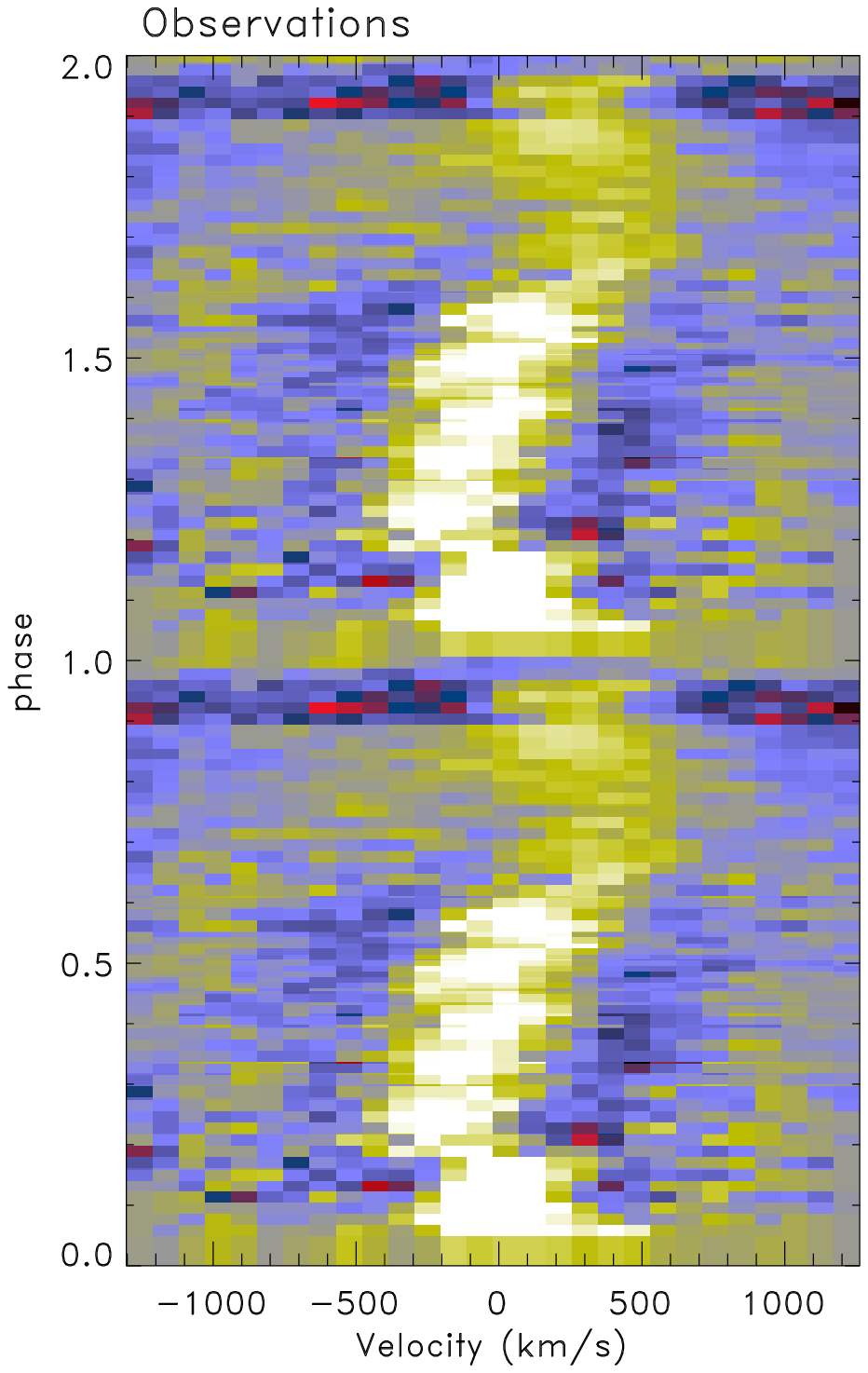}
    \includegraphics[width=2.8cm]{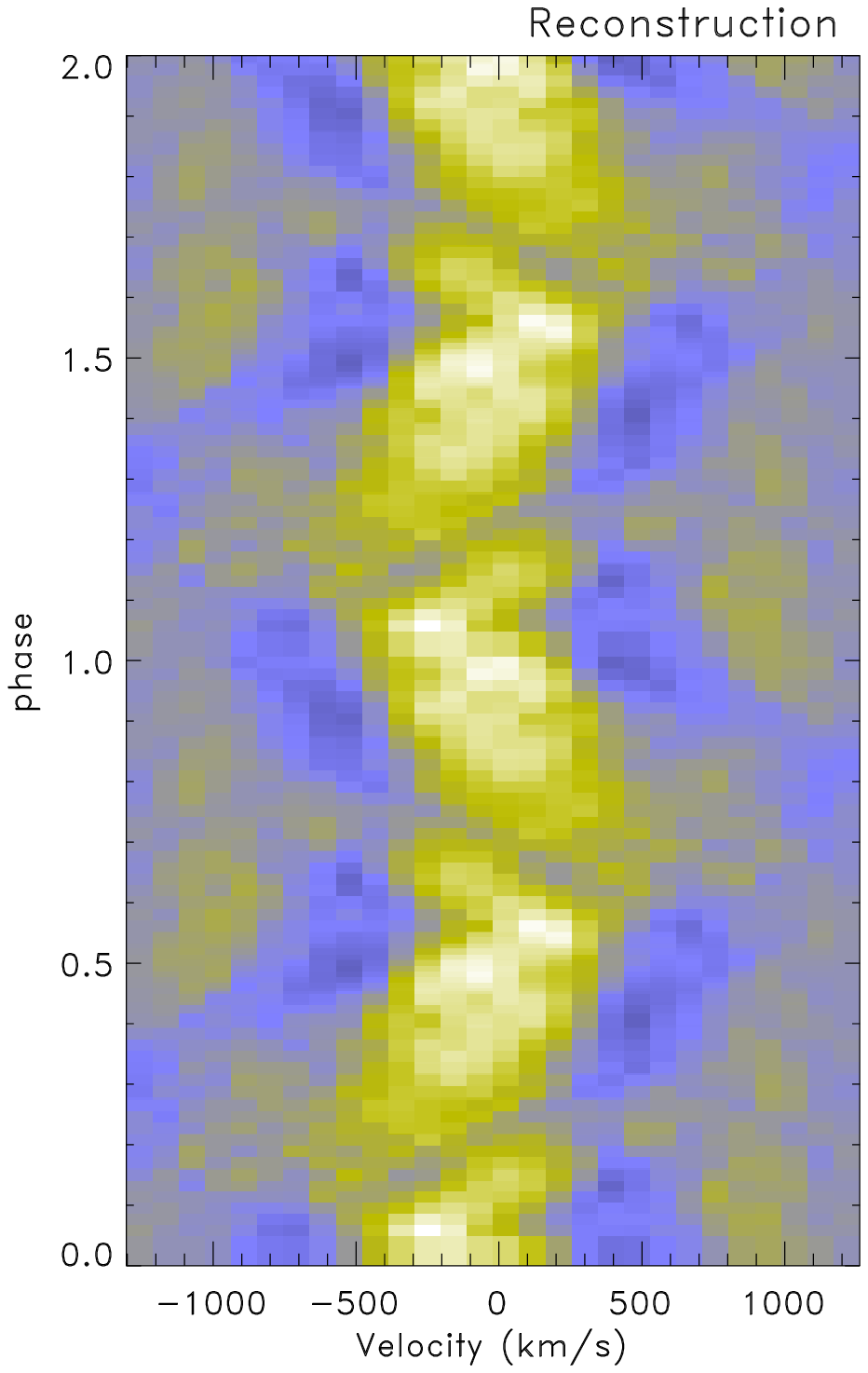}
    \includegraphics[width=2.8cm]{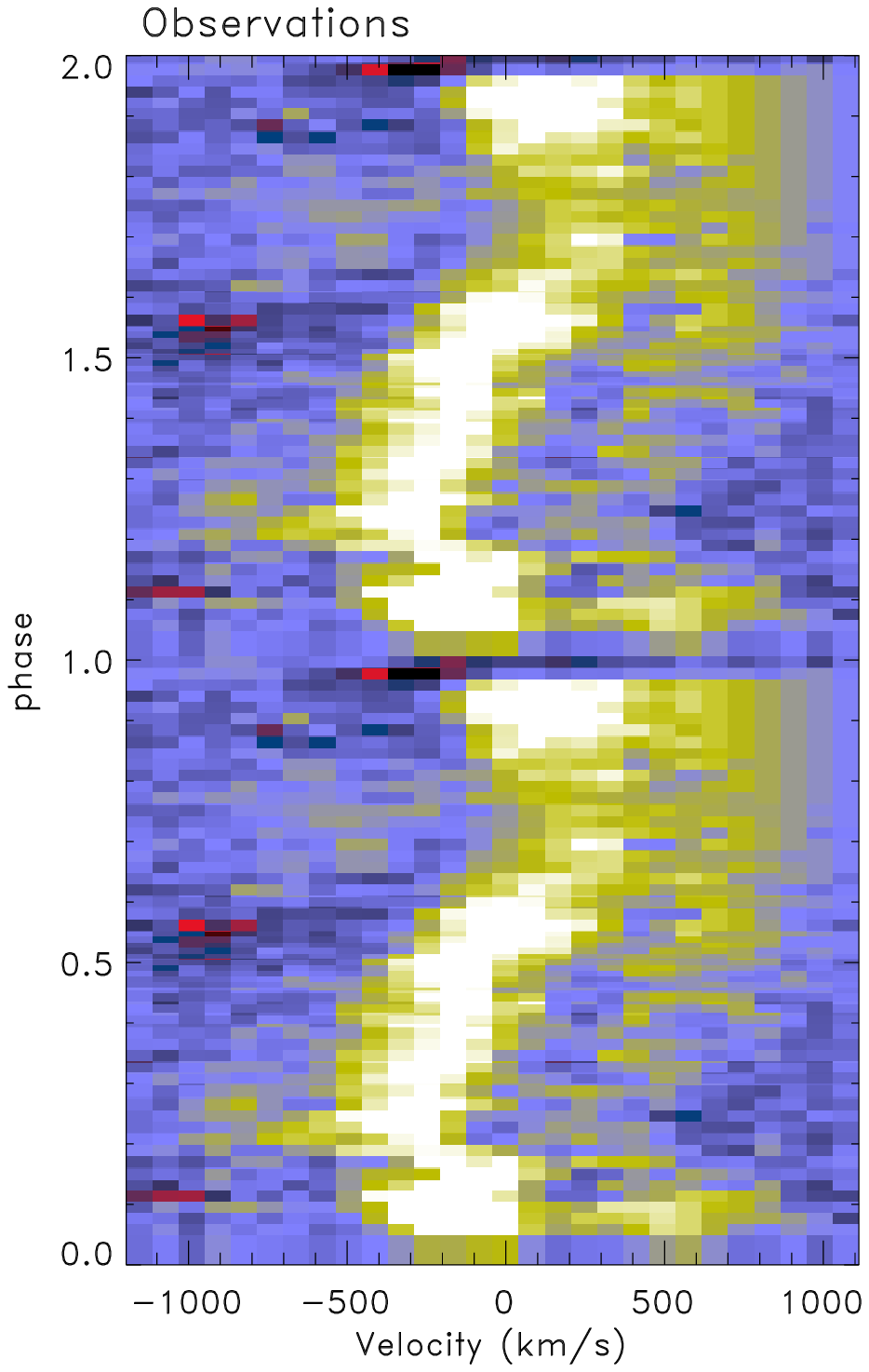}
    \includegraphics[width=2.8cm]{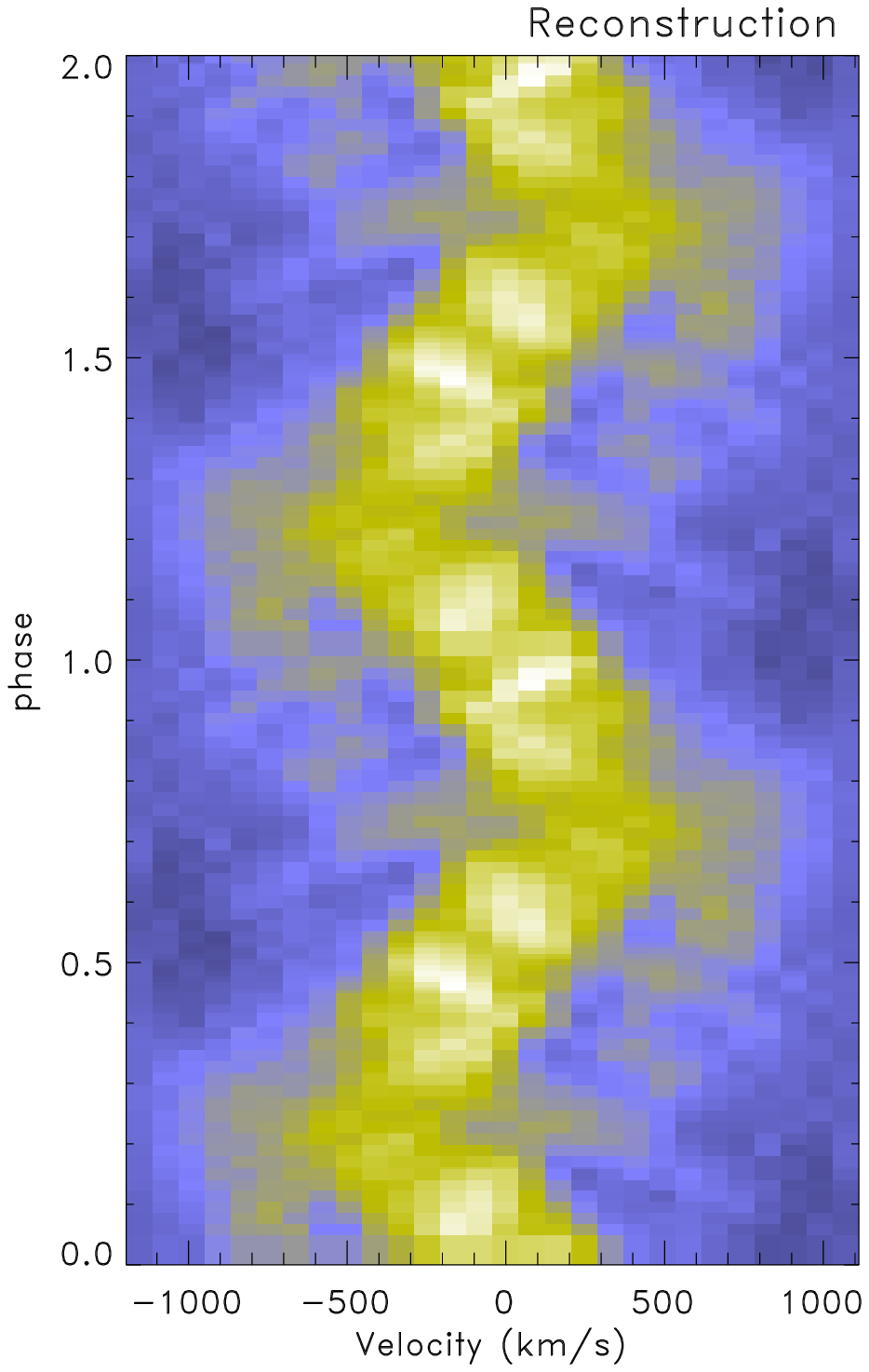}

    \bigskip \bigskip

    \includegraphics[width=5.6cm]{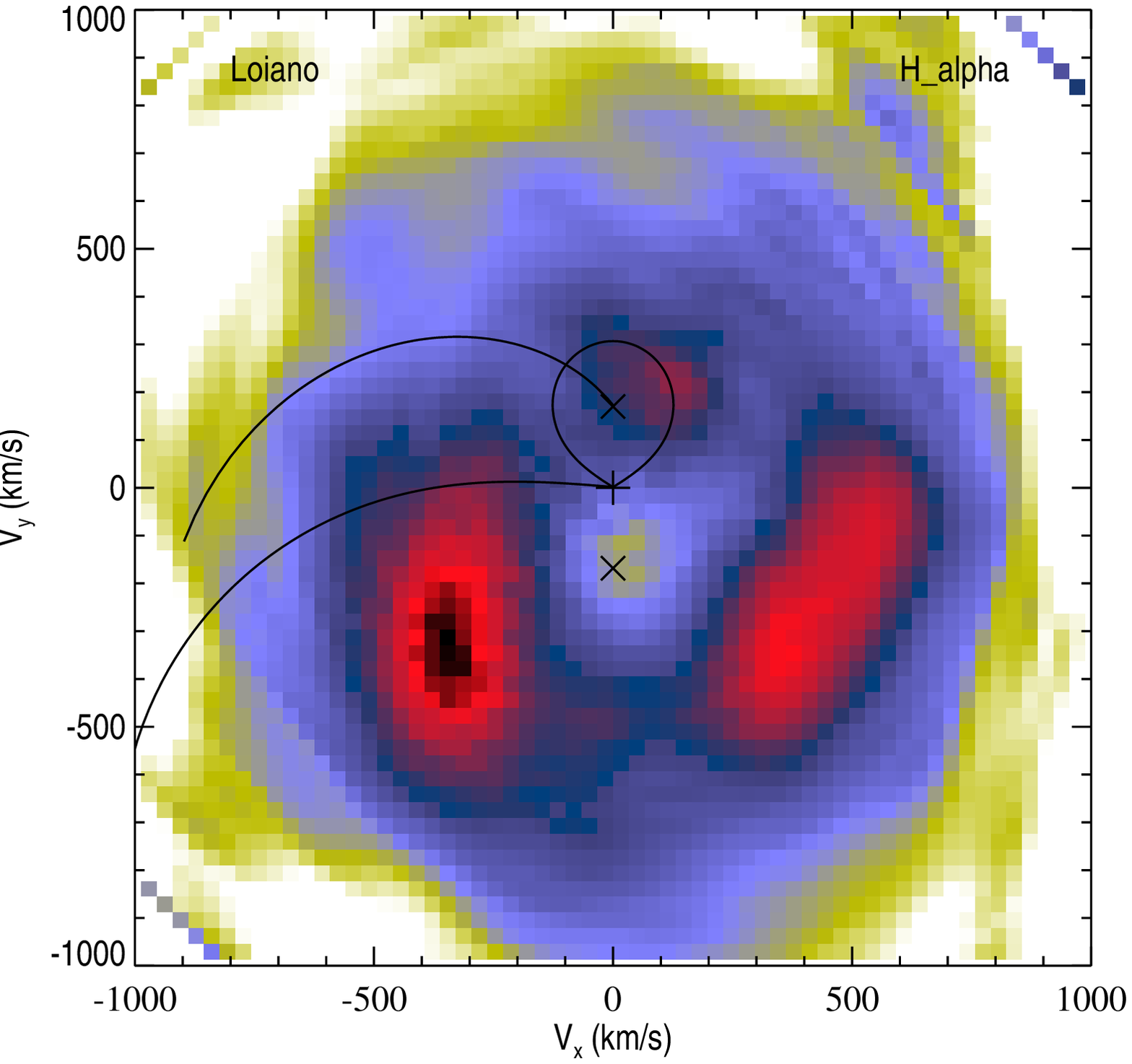}
    \includegraphics[width=5.6cm]{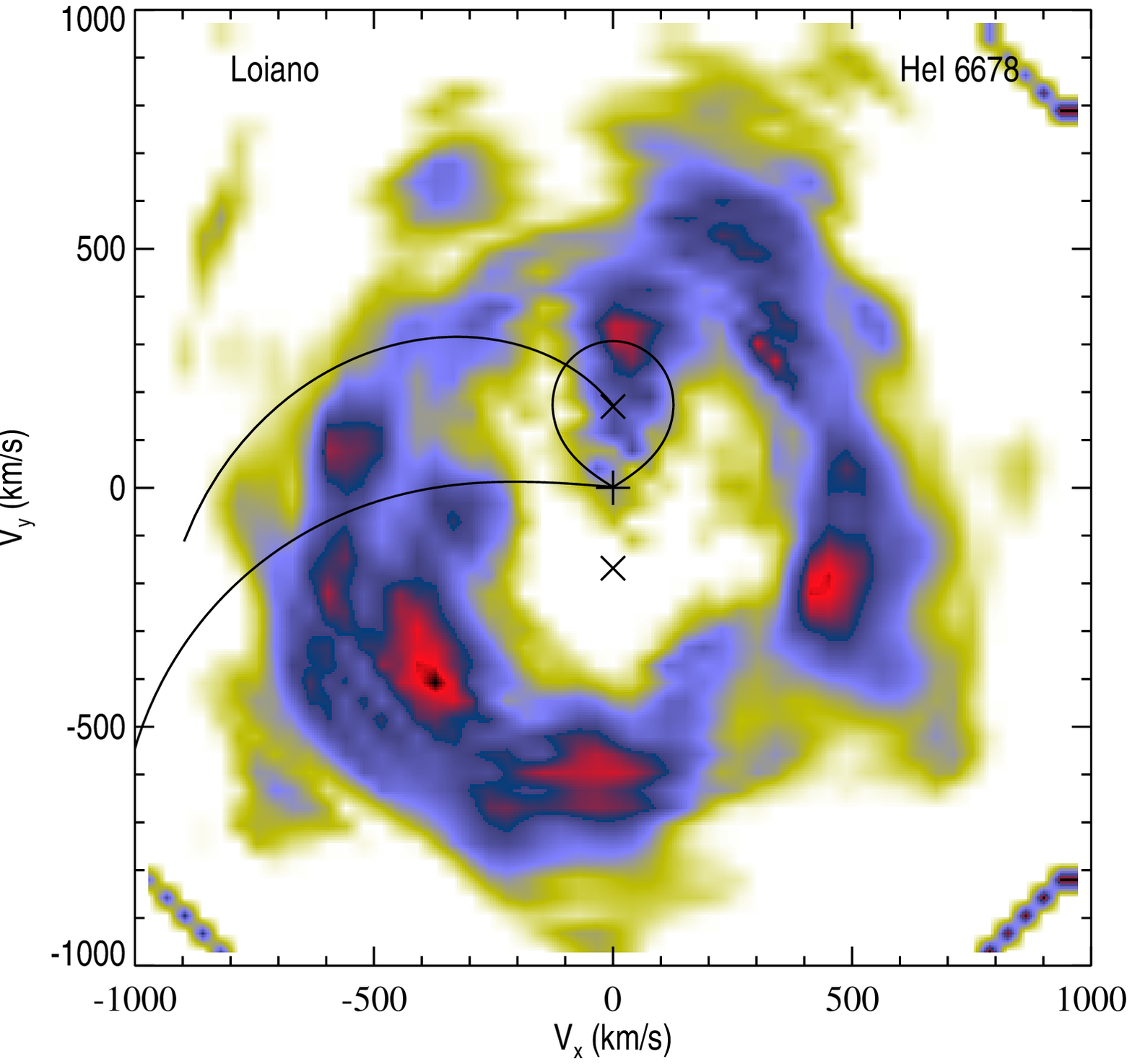}
    \includegraphics[width=5.6cm]{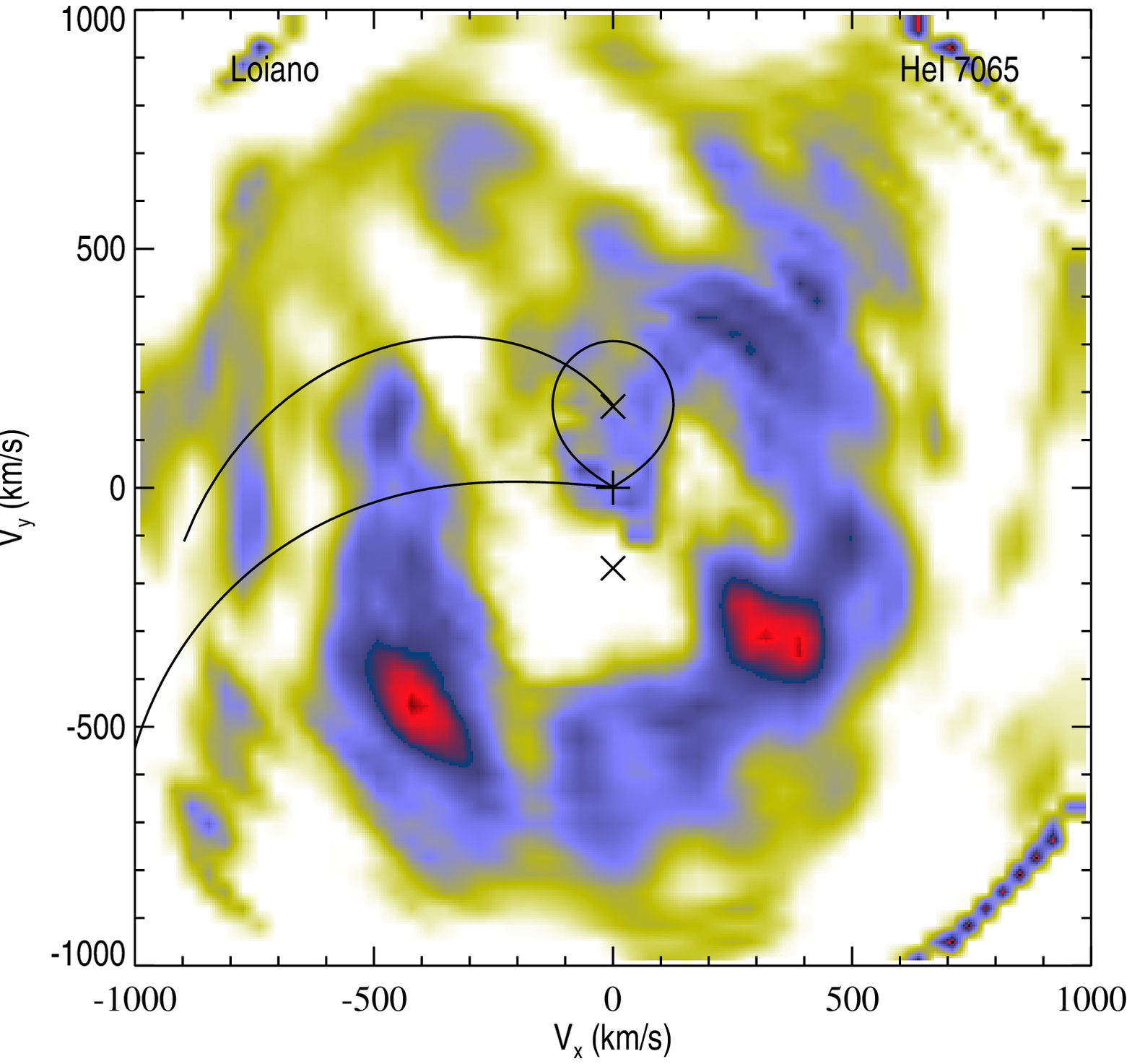}\\
    \includegraphics[width=2.8cm]{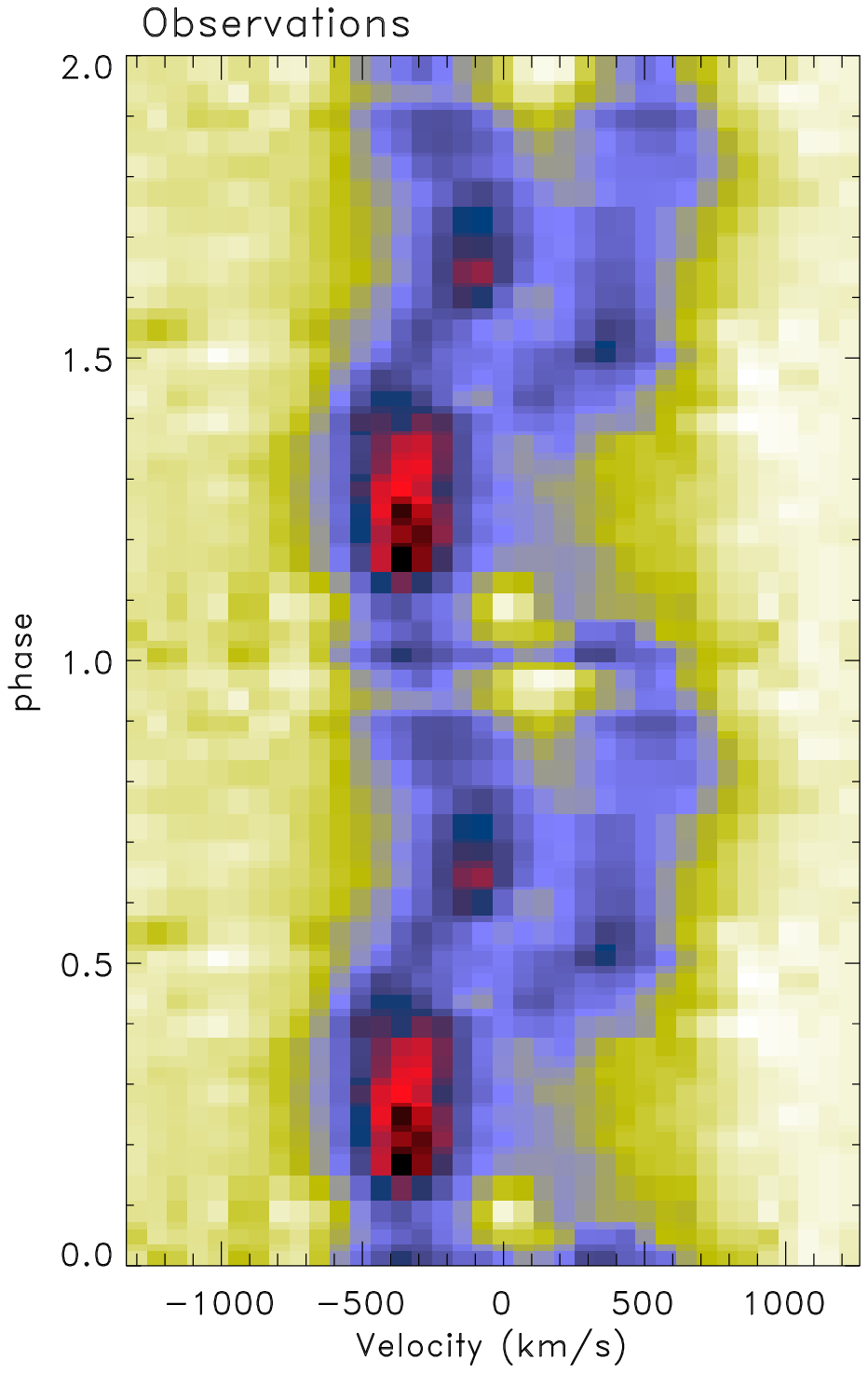}
    \includegraphics[width=2.8cm]{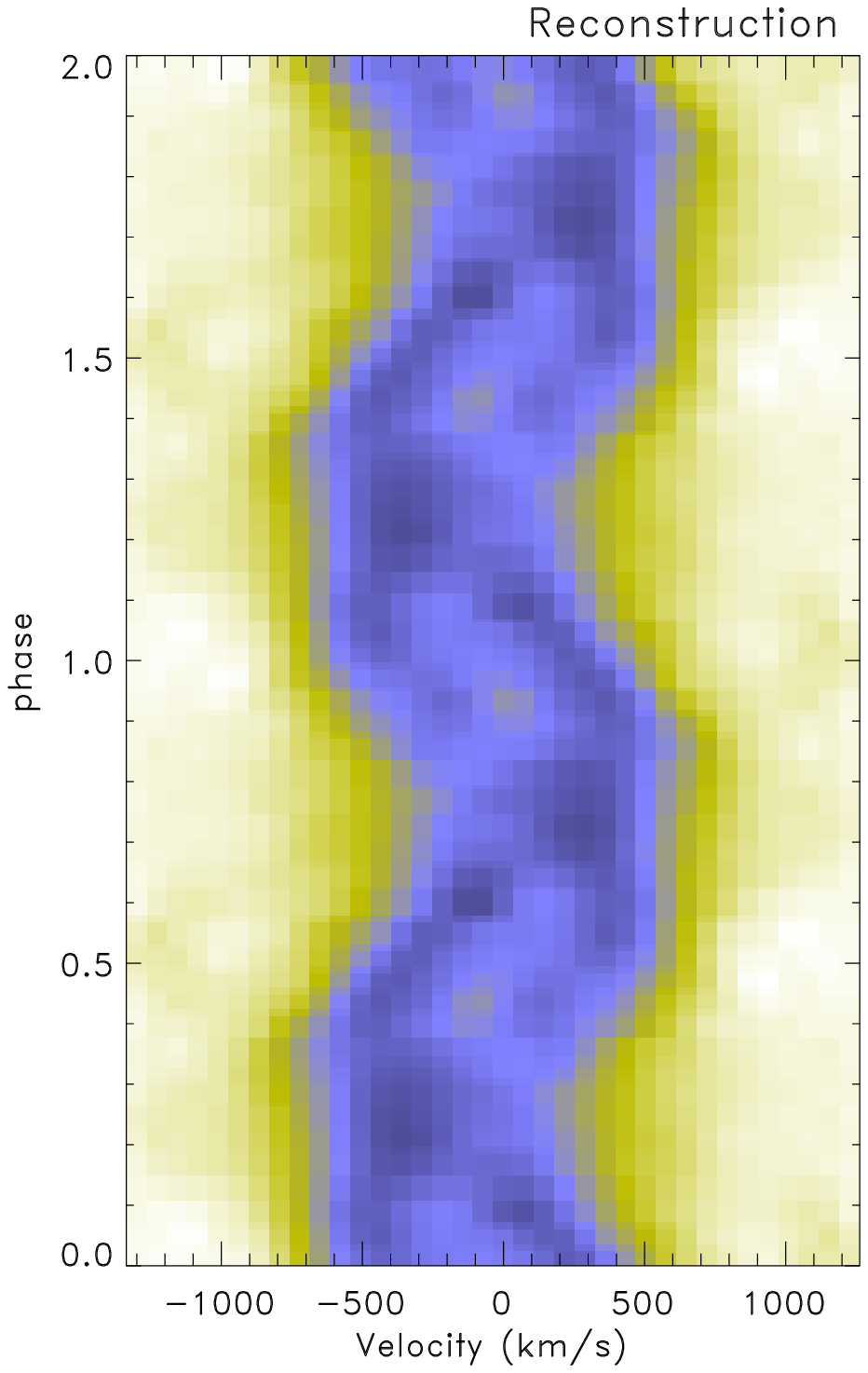}
    \includegraphics[width=2.8cm]{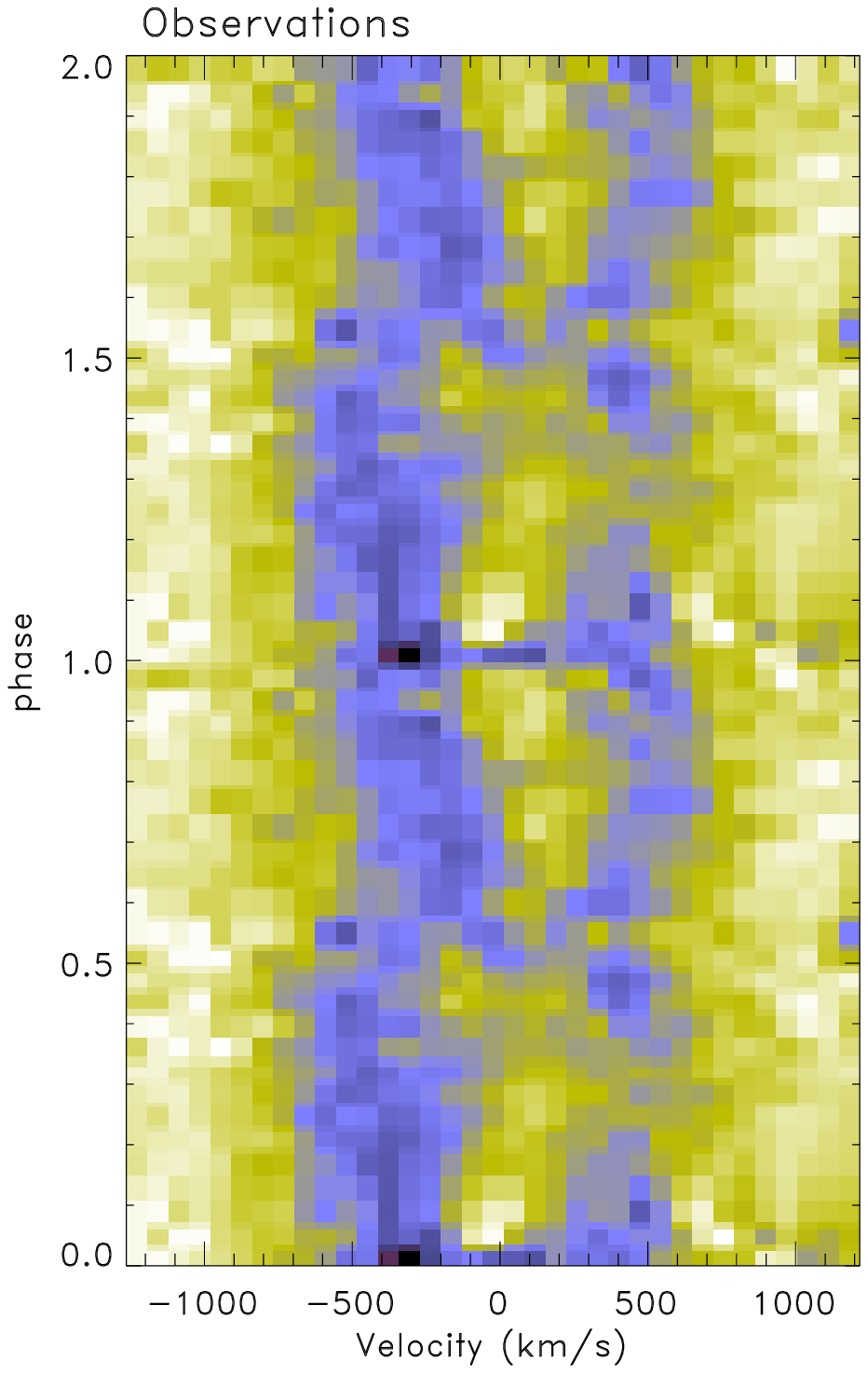}
    \includegraphics[width=2.8cm]{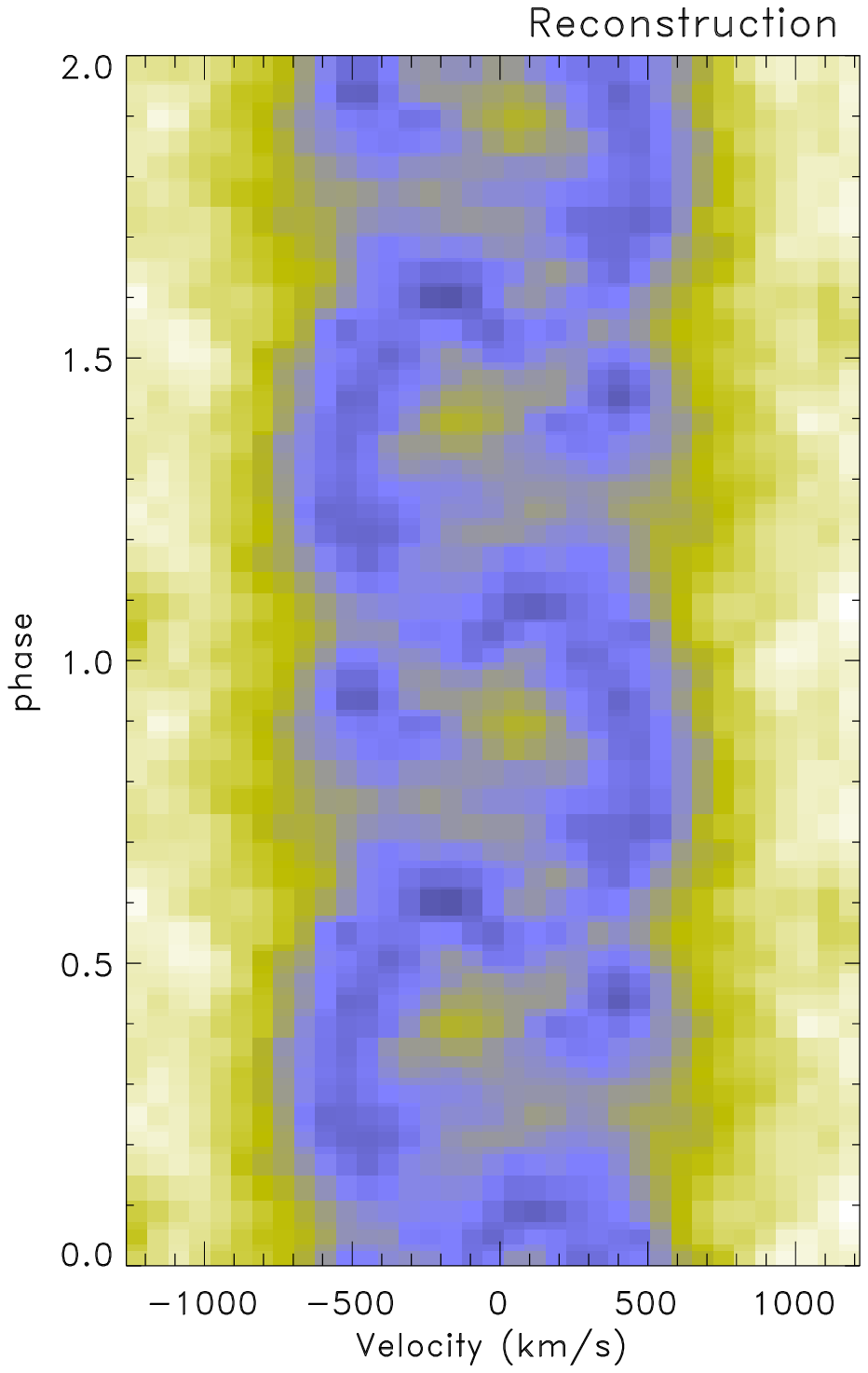}
    \includegraphics[width=2.8cm]{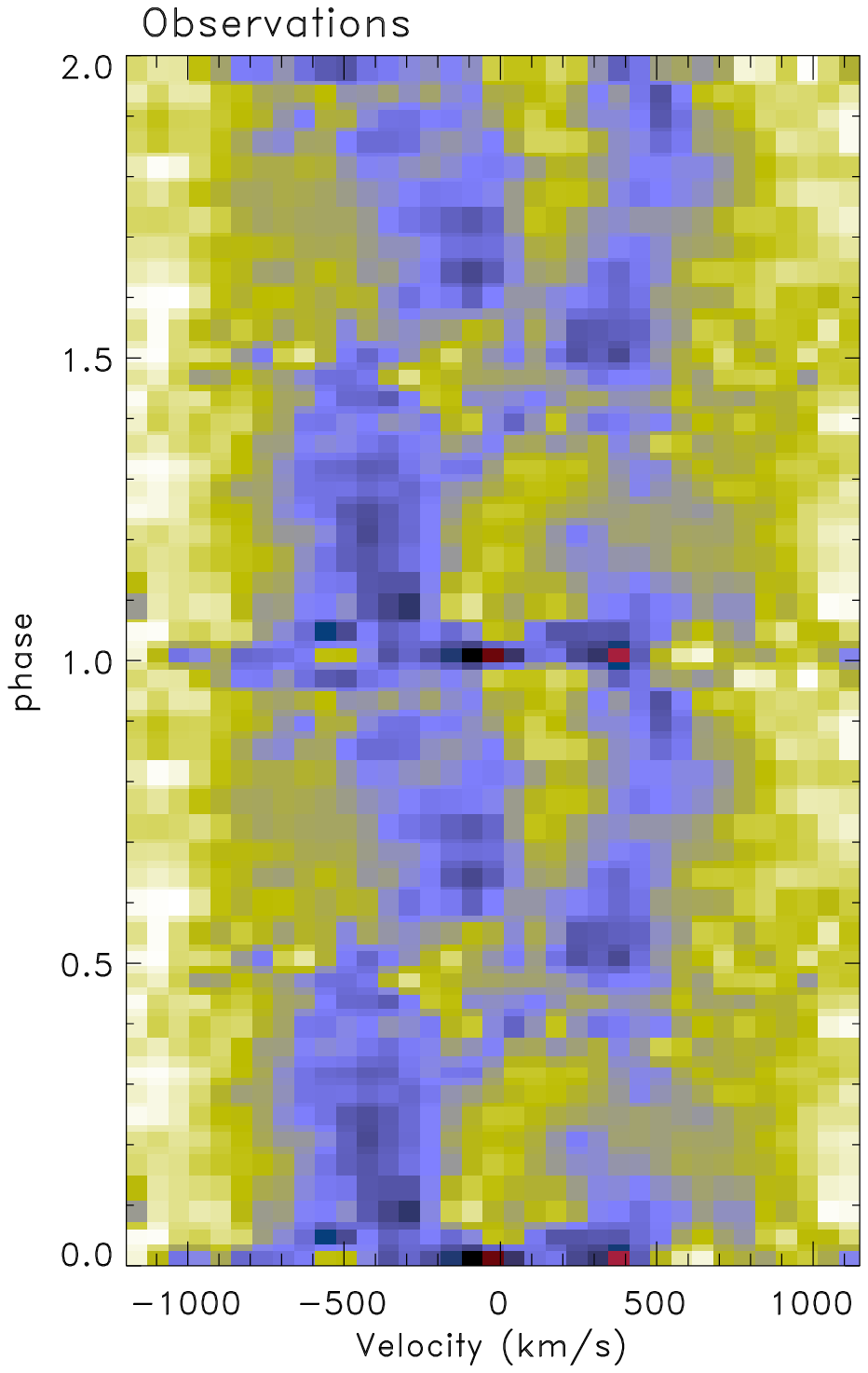}
    \includegraphics[width=2.8cm]{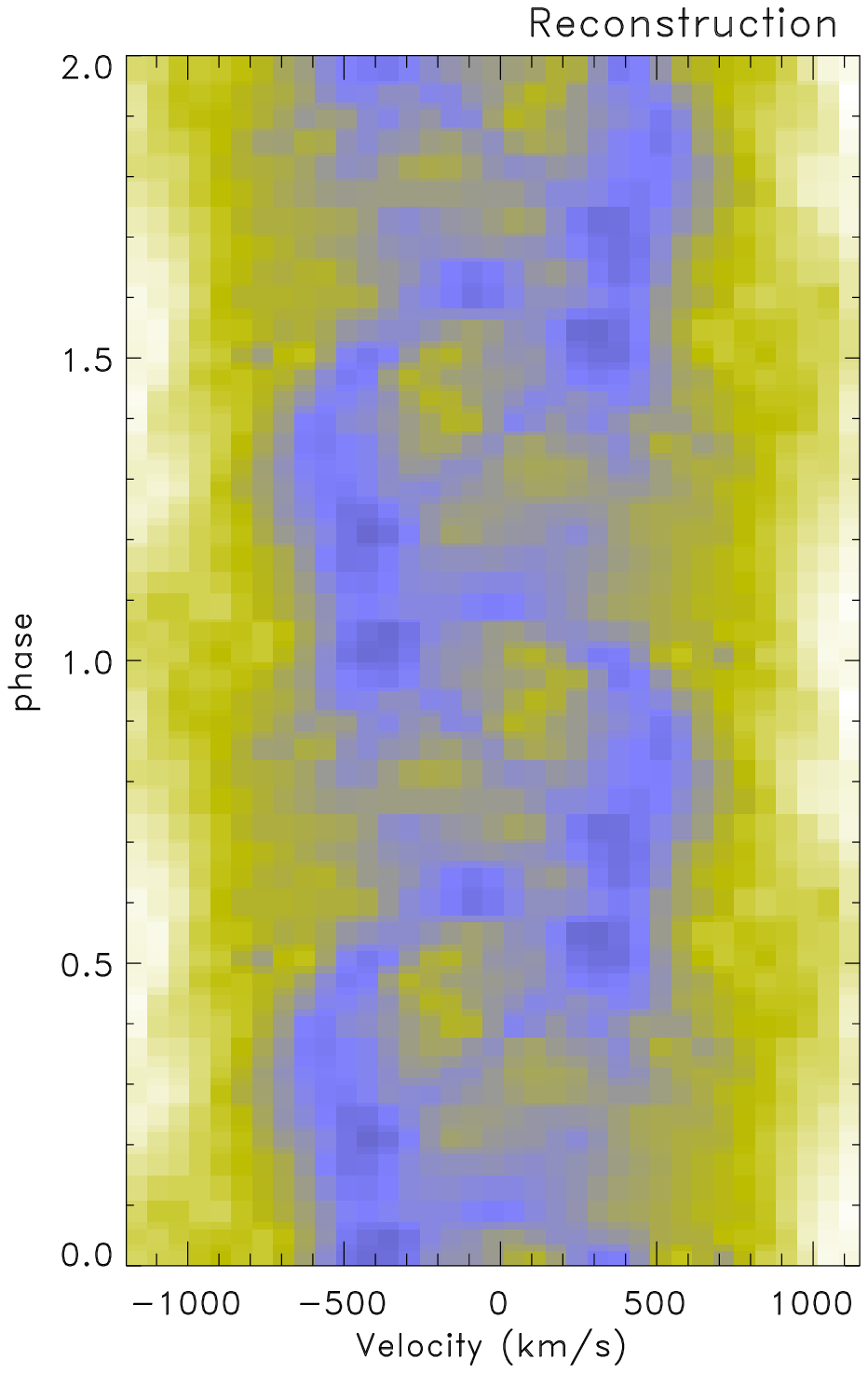}
    \caption{Doppler tomography for the \HeI~$\lambda$4922, $\lambda$4026 and $\lambda$4471 lines from
    the SAO dataset (in the upper half of Figure),
    and for the \Halpha, \HeI~$\lambda$6678 and \HeI~$\lambda$7065 emission lines from the Loiano
    set of observations (in the bottom half of Figure). For each line the observed and reconstructed trailed
    spectra (bottom) and corresponding Doppler maps (top) are shown. Marked on  the maps are the positions
    of the WD (lower cross), the center of mass of the binary (middle cross) and the Roche lobe of the
    secondary star (upper bubble with the cross). The predicted trajectory of the gas stream and the
    Keplerian velocity of the disc along the gas stream have also been shown in the form of the lower
    and upper curves, respectively. The Roche lobe of the secondary and the trajectories have been plotted
    using the system parameters, derived by \citet{Baptista95} -- an inclination $i=71^{\circ}$ and $M_1=M_2=0.47M_\odot$. The flux scale is from white to black.}

    \label{dopmaps2}
   \end{figure*}

\section{Doppler Tomography}
\label{DopMapSec}

  The orbital variation of the spectral lines profiles indicates a non-uniform
  structure for the accretion disc.
  In order to study the emission structure of UX~UMa we have used Doppler tomography.
  Full technical details of the method are given by \citet{Marsh-Horne} and \citet{Marsh2001}.
  Examples of the application of Doppler tomography to real data are given by \citet{Marsh2001}.

  Figures~\ref{dopmaps1}--\ref{dopmaps2} show the tomograms of the different spectral lines
  from the SAO and Loiano sets of observations, computed using the code developed by
  \citet{Spruit}.
  The figures also show trailed spectra in phase space, from which the maps were computed, and their
  corresponding reconstructed counterparts.
  As the gradual occultation of the emitting regions during eclipse is not taken into account,
  we constrained our data sets by removing eclipse spectra covering the phase
  ranges $\phi = 0.92{-}1.08$.

  To assist the interpretation of the Doppler maps, the positions of the WD (lower cross),
  the center of mass of the binary (middle cross) and the Roche lobe of the secondary star
  (upper bubble with the cross) are marked. The predicted trajectory of the gas stream and the
  Keplerian velocity of the disc along the gas stream have also been shown in the form of the lower
  and upper curves, respectively. The Roche lobe of the secondary and the trajectories have been plotted
  using the system parameters, derived by \citet{Baptista95} -- an inclination $i=71^{\circ}$
  and $M_1=M_2=0.47M_\odot$.

  The spectral lines in UX~UMa (especially \HeI) require negative values in some parts of the tomogram in order to fit
  the absorption component of the line. To avoid this, we have followed an approach of Marsh et al (1990).
  Prior to the reconstruction, we have added a Gaussian of FWHM = 600 \kms\ to the data in order
  to avoid the need for negative pixel values. We then removed the Gaussian equivalent from the calculated
  tomogram to produce the final result.
  This procedure has no effect on the goodness of the fit or entropy when the latter is determined over small scales.

  The appearance of all the tomograms from the SAO dataset is extremely unusual. We start to
  discuss them from the \Hbeta\ emission line as this is the strongest line and it produces a map of the
  highest quality.
  Due to the non-double-peaked line profile of \Hbeta\, we did not expect a Doppler map to have
  an annulus of emission centered on the velocity of the WD, and the observed tomogram does not show it.
  Instead of this the map displays a very non-uniform distribution of the emission. It appears to be
  dominated by emission consistent with an origin on the secondary star,
  from the relatively weak bright-spot located in
  the region of interaction between the gas stream and the outer edge of the accretion disc,
  and also from the leading side of the accretion disc (the right side of the tomogram),
  whose interpretation is ambiguous and will be discussed later.

  But the most intriguing and undoubtedly unique feature of UX~UMa's tomogram is a ``dark spot''
  -- a compact \textit{absorption} area with FWHM $\la$ 250-300 \kms,
  centered on ($V_x\sim-160$\kms, $V_y\sim-440$\kms) --
  which is associated with the absorption S-wave in the trailed spectra.
  This region is situated far from the region of direct interaction between the stream and the disc
  particles, and from this point of view the location of any structures here is unexpected. However,
  many NL systems and especially the SW~Sex stars show the emission in the lower-left quadrant of the
  tomogram but, to the best of our knowledge, UX~UMa is the only CV that exhibits the absorption here.


  The appearance of the \Hgamma\ and \Hdelta\ maps is similar to the \Hbeta\ map, though ``the dark spot''
  here is even deeper and more extended in velocity space.
  The \HeI\ Doppler maps also have a similar morphology to the Balmer lines. The main difference
  between them is a different contribution of the above-mentioned components. So, emission from
  the secondary star is, at best, very weak whereas the absorption is strong and
  extended across a wide range of velocities. This reason makes the tomograms look noisy, as the
  emission component in most \HeI\ lines
  is rather weak and we have had to apply an aggressive intensity scaling for these lines.
  The dark spot in the different lines is closely spaced on the tomograms, though it is difficult
  to determine its exact coordinates when it is as extended as in \HeI\ $\lambda4026$ and $\lambda4471$.
  However, it seems that the dark spot in \HeI\ slightly shifted along azimuth respectively
  to the Balmer lines.

\begin{figure}
\centering
\includegraphics[width=8.0cm]{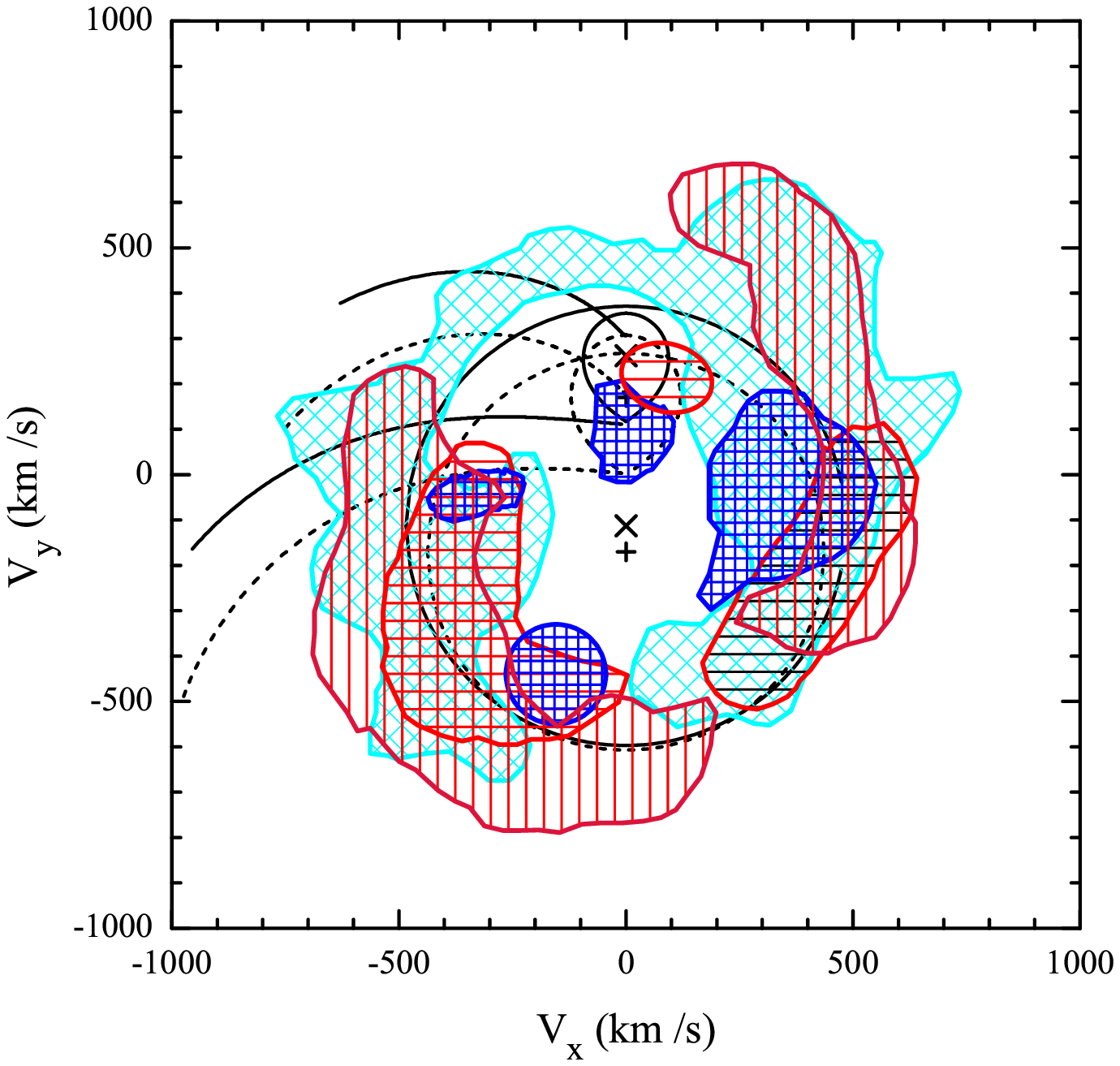}
\caption{Schematic representation of the essential components of the Doppler maps of UX~UMa
(Figures~\ref{dopmaps1}--\ref{dopmaps2}). Different patterns correspond to different emission lines:
\Hbeta\ (the blue vertical/horizontal cross-hatched area), \HeII\ \lam4686 (the cyan oblique
crosshatched area), \Halpha\ (the red horizontal hatched area) and the \HeI\ emission lines from the
red wavelength range (the crimson vertical hatched area).
Also plotted the positions of the WD, the Roche lobe of the secondary star,
and the predicted trajectory of the gas stream and the Keplerian velocity of the disc along the gas stream,
using the dynamical solution from this paper (solid lines) and
the system parameters, derived by \citet{Baptista95} (dashed lines).
The circles represent the largest radius of the accretion disc determined by tidal limitations.}
\label{alldopmaps}
\end{figure}

  The high-excitation emission lines show little variation with time and produce quite noisy tomograms with
  little details. However, a most prominent contribution of the emission here is from the quasi-circular
  structure with a radius of $\sim500$ \kms centered on the velocity of the WD (see the \HeII\ $\lambda4686$ map
  in the bottom half of Fig.~\ref{dopmaps1}).
  Note also that \HeII\ $\lambda4686$ does not show any emission on the hemisphere of the donor star that faces
  the WD and boundary layer, and that the circular structure, most likely linked with the accretion disc, has
  an interuption in the region of the dark spot. There is also enhanced emission from the leading side of
  the accretion disc.

  The red spectra from the Loiano dataset have completely different appearance in comparison with the blue
  spectra, and as a result the Doppler maps in this wavelength region appear distinctly different
  (Fig.~\ref{dopmaps2}, bottom half).
  The double-peaked profiles of the \Halpha\ and \HeI\ emission lines, attributed to an accretion disc,
  produce an azimuthally asymmetric annulus of emission centered on the velocity of the WD, revealing
  a prominent two armed pattern. This pattern resembles the signature generated by
  spiral structure in the disc, predicted numerically by a number of researchers \citep{Sawada1986,Steeghs1999}.
  Comparing the blue and red tomograms, one can also speculate on the same origin of the emission
  from the leading side of the accretion disc (in the blue spectra) and of one of the arms in the
  red spectra (Fig.~\ref{alldopmaps}). In conclusion, we also note that
  in contrast to the SAO observations, there is no evidence for both the dark spot
  and the gas stream/disc impact region emission in the Loiano dataset, while the emission from
  the area where the secondary star is expected,
   is weaker and with slightly different velocity coordinates.

  Thus, combining all the results above, we conclude that most of the spectral line components in UX~UMa
  appears to originate from four distinct components: the secondary star, the accretion disc with two
  arch-like structures, the bright spot from the gas stream/disc impact region, and the dark spot in the lower-left
  quadrant of the tomograms (Fig.~\ref{alldopmaps}). All these components exhibit a different
  appearance in different spectral lines and in different epochs. A more detailed discussion of
  the emission structure of UX~UMa follows in the next section.

\section{The accretion disc and emission/absorption structure of UX~UMa}
\label{AccDiscStrSect}

It is widely accepted that the NL stars experience high mass transfer rates \citep{Puebla2007,Ballouz2009}
and their accretion disc is optically thick. As follows from the theoretical calculations, an intrinsic
spectrum of such discs should be close to stellar, with absorption Balmer and \HeI\ lines
\citep{Suleimanov1992,Diaz1996}. The irradiation of the standard optically thick disc by the hot central
source is not enough to replace the absorption by emission \citep{Kromer2007}.
Emission lines in these systems are believed to originate in a relatively
hot extended region above the orbital plane. This extended matter is heated due to the
external irradiation by the WD, boundary layer and inner disc. In the inner disc, despite the higher
temperatures, the absorption cannot be replaced by emission because the irradiated flux here is
insignificant in comparison with the intrinsic flux. Therefore, there are two spatially
resolved emitting regions -- the inner disc, which has mostly photospheric spectrum with absorption lines,
and the outer disc, emitting spectrum with emission lines. Note also that the emission component tends to be
more pronounced in high inclined eclipsing systems where the optically thin hot plasma is seen along the disc
surface and has more significant optical depth (in comparison with face-on discs, but $\tau$ still remains $\la1$)
and therefore larger emission flux, whereas the absorption line core depth decreases with
the orbital inclination \citep{Diaz1996,Kromer2007}. This picture is confirmed by observations of "disc-on" NL
stars (TT Ari, V603 Aql) and DNe during outbursts, especially in the UV part of the spectra.

UX~UMa is a high inclined eclipsing system and it does not show broad absorptions similar to low inclined TT~Ari
\citep{Stanishev2001}. However, it has a steep Balmer decrement outside eclipse phases which suddenly becomes
almost flat at mid-eclipse (Fig.~\ref{aver_spec_blue}). This transformation could also be easily explained
within the framework of the above-described model: the inner disc with its photospheric spectrum is shielded around
mid-eclipse phases by the secondary star, whereas an optically thin line emission region above
the orbital plane is still seen and becomes the main contributor to the spectrum.

The oscillator strength for \Halpha\ is significantly larger than those for the higher members of
the Balmer series, thus detection of accretion disc structures is more sensitive in this line.
Indeed, usually \Halpha\ has larger equivalent widths of emission components than the other Balmer lines.
Therefore, we can expect that the bright spot and the secondary star in UX~UMa should be seen in \Halpha\
better than in \Hbeta\ (similar behaviour is expected for the ``blue'' and ``red'' \HeI\ lines).
As such, the changes between the SAO and Loiano spectra most likely testify of time dependence within the system,
though part of these changes can be due to different line transfer effects in different lines.

\subsection{Dark spot}
\label{DarkSpotSect}

  Whilst the emission in the lower-left quadrant of the tomogram is one of the distinguishing characteristics
  of the SW~Sex-type systems \citep{Hellier2000}, it has also been observed in many CV stars of other types
  (WZ~Sge -- \citealt{WZSge}, BF~Eri -- \citealt{BFEri}) and some low mass X-ray binaries (LMXB),
  (XTE~J2123{-}058 -- \citealt{Hynes2001}, GX9{+}9 -- \citealt{Cornelisse2007}).
  Different ideas have been proposed to explain this phenomenon and other probably related
  observational peculiarities, such as a high-velocity emission S-wave, transient absorption features in
  the Balmer and \HeI\ emission line cores and X-ray/UV dips in the phase range $\sim 0.3{-}0.6$.
  Among these mechanisms invoked are stream overflow \citep{Hellier-Robinson,Hellier1996},
  a WD-anchored magnetic field \citep{Williams1989,Dhillon1991},
  magnetic propeller anchored in the inner disc \citep{horne1999},
  magnetic accretion \citep{Rodriguez2001,Hameury2002}.

\begin{figure}
\centering
\includegraphics[width=8cm]{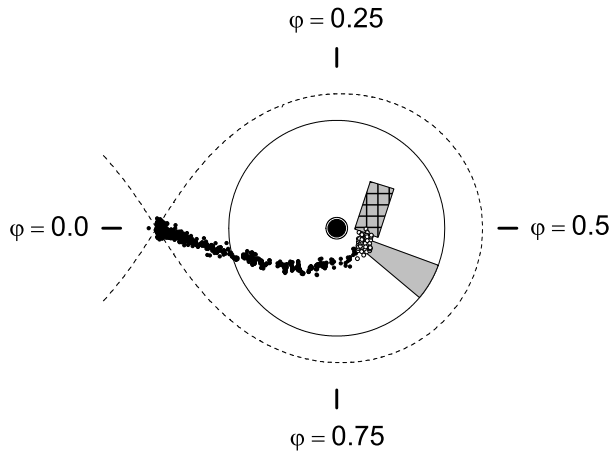}
\caption{A schematic model of UX~UMa. The overflowing stream creates the lump of matter which
prevents the outer sector of the disc, shown in gray, to be irradiated by the hot inner disc.
The lump and the falling stream also shields a part of the accretion disc from the observer.
The cross-hatched area indicates the region of the disc obscured at the orbital phase $\sim0.8$.}
\label{model}
\end{figure}

  We believe that the compact area of absorption which produced the absorption S-wave in the trailed spectra
  and the dark spot in the lower-left part of UX~UMa's tomograms, can be explained
  in the context of the stream-disk overflow model. Numerous theoretical and numerical studies
  indicate that free flowing gas could pass over the WD and hit the back side of the accretion disc
  \citep{LubowShu,Lubow,ArmitageLivio1996,ArmitageLivio1998,Hessman1999}. In particular, intensive
  3-D smoothed particle hydrodynamics (SPH) simulations show \citep{Kunze2001} that a substantial
  fraction of the stream overflows the outer disc edge and impacts the disc close to the
  circularisation radius at orbital phase $\sim0.5$. \citet{Kunze2001} conclude that the presence
  of substantial quantities of gas above and below the disk plane can cause X-ray absorption in CVs
  and LMXBs around orbital phase 0.7, if the inclination is at least $65^{\circ}$. We suggest the
  same might be happening in UX~UMa, and the dark spot in the lower-left quadrant of the tomogram,
  situated around $\varphi \sim 0.6$, might be caused by the deflected stream flow. In case of a rather
  small disc and the high mass transfer rate or its sudden increase, the overflowing stream might
  produce a dense lump of matter in the place where it re-enters the disc. This lump and/or
  the overflowing stream itself can prevent the outer sector of the disc to be irradiated by the hot
  inner disc that will cause a predominance of absorption over emission here (Fig.~\ref{model}).
  As a result we will see a narrow absorption in the spectrum and a dark spot in the tomogram.
  Considering that the optical depth of the matter in the \HeI\ lines is smaller in comparison with
  the Balmer lines due to the lower helium abundance and the oscillator strengths of the \HeI\ lines,
  it may explain the displacement of the dark spot in the \HeI\ lines relative to the Balmer lines.

  As the lump will not only occult the outer disc from irradiation but will also shield some part of the
  disc from the observer, it can also very naturally explain the absorption in the red wings of emission
  lines observed during the phase range $\varphi \sim 0.78{-}0.95$. The opacity of the falling stream is
  most likely anisotropic, increasing along the stream. Thus, around these phases the lump can occult
  the hottest and brightest innermost part of the accretion disc with the greatest range of the radial
  velocities, leading to weakening of the emission component in the red wing of spectral lines
  (Fig.~\ref{model}).

  The supposition that UX~UMa had a rather small accretion disc during the SAO observations
  (smaller than during the Loiano set)
  is supported by Doppler tomography. All the tomograms from this data set show weak,
  but clearly seen emission from the bright spot -- the region of interaction between the gas
  stream and the outer edge of the accretion disc. In most of the lines this emission lies over
  the predicted trajectory of the gas stream, though in a few lines the emission has the Keplerian
  velocity of the disc along the gas stream. The singularity and difference of UX UMa's bright spot
  from those usually observed in CVs consist in its extent. The bright spot in UX~UMa appears
  to be extended over hundreds kilometers per second along the trajectories, which could be due
  to the matter stream deflecting vertically or penetrating deeply into the disc before finally
  being deviated.

  No such thing has been observed during the Loiano observations. There are no signs of
  the bright spot lying over either of the trajectories. The accretion disc is large and thick
  (see the following discussion), the interaction between the stream and the outer part of the disc
  is less energetic, the lesser part of the stream matter overflows the disc and -- as a consequence --
  there is no dark spot in the lower-left part of the tomograms.

\subsection{Spiral pattern}

  The accretion disc in UX~UMa manifests itself as a non-uniform circular structure, with a radius of $\sim400$ \kms,
  in the Doppler maps from the Loiano set of observations. The disc emission
  is not azimuthally symmetrical, but exhibits the two arch-like structures, superposed on the diffuse ring.
  This pattern resembles the non-axisymmetric features observed in the accretion discs of several DN systems
  when they were in outburst (IP~Peg -- \citealt{Steeghs1997}, U~Gem -- \citealt{Groot2001}), and also in
  a few NLs (V3885~Sgr -- \citealt{Hartley2005}, V1494~Aql -- \citealt{Hachisu2004}). The presence of such
  structures has been originally interpreted as an observational confirmation for the existence of
  the tidally induced spiral shocks, predicted by hydrodynamic models of accretion discs in close binaries
  \citep{Sawada1986,Spruit1987,Steeghs1999,Boffin2001}. Nevertheless, \citet{Smak2001}
  and \citet{Ogilvie2002} have questioned such interpretation, as it requires the presence of large or
  unusually hot discs, and proposed alternatives to the spiral shocks to explain the phenomenon.
  \citet{Ogilvie2002} has shown that some regions of the outer disc would be tidally thickened and may
  therefore be irradiated by the WD, boundary layer and/or inner disc. The pattern of thickening would be
  slightly spiral, but no waves or shocks would be involved in the process. \citet{Smak2001} has
  wondered if the similar effect should work in CVs with stationary accretion (i.e., novae and NL
  systems), whose outer radii of the large accretion discs are controlled by tidal effects. He reviewed
  the collection of Doppler tomograms published by \citet{Kaitchuck1994} and identified similar
  arch-like structures in tomograms of several NL-systems, including UX~UMa.
  Note, however, that this identification does not exclude the shock origin of these structures but only
  supports the existence of non-Keplerian flows in the accretion disc of UX~UMa.

  The appearance of the two armed pattern in UX~UMa is a bit different in the \Halpha\ and \HeI\ lines.
  The structures here are not well-defined spiral arms, though they are possibly connected, overlapping
  in azimuth ranges where tidal distortions such as spirals are expected.
  The \HeI\ lines are much weaker than \Halpha\ but the spiral-like structure in their tomograms closely
  resembles the Doppler maps of those DN systems where spiral structure is basically indisputable.
  In the \Halpha\ tomogram, the brightest part of the right arm is placed somewhat lower. Note, that
  both the theory and the numerical simulations tell us that non-linear effects can significantly modify
  the appearance of the spiral pattern (e.g. \citealt{Godon1998,Steeghs1999}).

  Despite the dissimilarity between the red and blue tomograms of UX~UMa from our observations,
  they have (the only) common component -- extended bright area from the leading side of the accretion
  disc. The second, lower-left arm cannot be detected in the blue Balmer and \HeI\ tomograms as it
  is hidden by the dark spot, if exists. However, the \HeII\ $\lambda4686$ map shows some extended emission
  in the lower-left area (Fig.~\ref{alldopmaps}). Furthermore, \citet{Kjurkchieva2006}
  employed a spiral arm model to explain the observed features of the light curves of UX~UMa from
  their 9-year photometry, and  the model reproduced well the observational data. Thus, the spiral-like pattern
  is apparently a regular structure in the disc of UX~UMa.

  At present there are arguments in favour and against both models, proposed to explain the spirals
  (see \citealt{MR2004}, and references therein). The NL systems with their large, hot and stable
  accretion discs can be considered as the best systems in which to analyse these structures.
  Nevertheless, despite the wealth of observations of these stars, and some twenty published Doppler
  maps, only very few NL systems have shown signs of spiral structure in their discs (V347~Pup --
  \citealt{Still1998}, V1494~Aql -- \citealt{Hachisu2004}, V3885~Sgr -- \citealt{Hartley2005}).
  From this point of view UX~UMa is an attractive candidate for high time and spectral resolution
  spectroscopic observations.

\label{SecondaryStarSect}
\subsection{Secondary star}

  The secondary star manifests itself in the trailed spectrum of \Hbeta\ as a narrow sinusoidal emission component,
  which is mapped onto the inner hemisphere of the secondary's Roche lobe. The emission from this area is clearly seen
  in all the Balmer Doppler maps (in \Hbeta\ this is the strongest emission component),
  however its distribution over the mass donor is again different in the two data sets.
  In the SAO spectra the whole donor hemisphere, facing the WD, appeared to be a source of emission, with the emission
  intensity becoming stronger near the first Lagrangian point. Such distribution almost certainly exclude the possibility
  of photospheric line emission and the most likely cause of this emission is irradiation by UV-light from the accretion disc,
  boundary layer or bright spot.

  The \Halpha\ emission from this area in the Loiano spectra is much weaker. It is slightly shifted in a clockwise direction
  and apparently concentrates towards poles of the irradiated hemisphere (Fig.~\ref{alldopmaps}). The origin of this emission
  is not so clear and its link with the companion star can be put in doubt\footnote{Note, that simulations show that
  the Maximum Entropy Method applied during the Doppler tomography reconstruction can perhaps be partly responsible for the above effects
  by blurring/smoothing the compact emission sources (see, e.g. \citealt{HarlaftisMarsh1996}}.
  If this emission indeed belongs to the secondary star
  then the appearance of the \Halpha\ tomogram suggests irradiation with shielding of the mass donor's equatorial
  regions by the the disc's outer rim \citep{Sarna1990}. From this we estimate a height-to-radius ratio
  (H/R) to be $\sim0.25$ (for the system parameters of \citealt{Baptista95}).
  This is an extremely high H/R ratio, at least from a theoretical point of view, although there are some
  other observations suggesting highish H/R's (e.g. \citealt{Mason1997} found that the vertical structure of
  the accretion disc of UX~UMa can extend to H/R$\sim0.36$ above the orbital plane). On the other side,
  the lack of emission from the secondary star during Loiano observations will mean that something completely blocked
  the irradiation sources. Otherwise the secondary should be easily seen in \Halpha\ as it was previously seen in \Hbeta.
  In this case the H/R ratio can be even greater.

  It is interesting to note that there is no evidence for shielding in the SAO tomograms.
  This means the accretion disc had the smaller H/R ratio at that time.
  This might have happened if the disc was significantly smaller\footnote
  {It cannot be excluded also the smaller vertical extension of the irradiation source at that time.
  However, this is unlikely that such extension can be comparable with the outer disc thickness unless
  the disc is not small enough.}.
  The presence of spiral waves or the tidally thickened regions in the accretion disc during the Loiano run,
  and only uncertain indications of it during the SAO observations indirectly confirms this idea, as the
  tidally induced spiral structures require the presence of a large disc.

  The detected emission from the secondary star in the Doppler maps allows us to estimate $K_2$.
  The radial velocity of the inner face of the secondary about the centre of mass of the system is
  lower than the radial velocity of the centre of mass of the secondary, thus a measurement of the velocity
  of the emission feature from the tomograms will set a lower limit to the radial velocity semi-amplitude
  $K_2$ of the secondary. From the Doppler map of \Hbeta\ we find $K_2>180$ \kms, whereas the \Halpha\ tomogram
  increases this limit to $\sim200$ \kms. Thus, this limit does not contradict
  with the value of $K_2=262$ \kms\ obtained by \citet{VandePutte2003}, though the latter appears to be
  slightly overestimated.
  Indeed, \citet{VandePutte2003} obtained $K_2$ using the far-red absorption lines, which may arise
  in the back (non-irradiated) side of the secondary. However, this hypothesis requires further study.

\section{Discussion}
\label{DiscSec}

\subsection{Long-term variability}
  The appearance and behaviour of the spectra obtained in 1999 and 2008 indicates that UX~UMa may have been
  in different states during those observations. In 1999 the accretion disc of the system appeared to be
  smaller than in 2008 and we suggest that it might have been due to the smaller mass transfer rate in 1999.
  Unfortunately, we are unable to make a conclusion regarding whether one epoch of observations was notably
  brighter than the other. The absolute flux calibration of our spectroscopy is not completely reliable,
  and also the spectra were obtained in different wavelength ranges. Bearing in mind that UX~UMa's colours can
  change significantly with time \citep{Dmitrienko1994}, it does not make sense to compare the fluxes of
  the spectra. Amateur observations also cannot help as they provide us with only sparse data.


  However, a variation of the mass-transfer rate in the NL systems is not unusual, and UX~UMa is not an exception.
  Many researchers mentioned UX~UMa's peculiar light curves which could be interpreted as being due to a much
  lower relative luminosity of the hot spot and accordingly to the lower mass-transfer rate. \cite{Smak1994}
  noted that the standard and peculiar light curves of UX~UMa appear to occur with comparable frequency.
  \citet{Knigge1998} reported on a fairly substantial increase in the accretion rate in UX~UMa by $\ga50\%$
  within three months.

 \subsection{System parameters}
 \label{SysParDiscSect}

  Most of the parameter estimates for UX~UMa were found in the past by means of analysis of WD eclipse features in
  the light curve in the different wavelengths, whereas the value of the radial velocity semi-amplitude $K_1$ was
  mostly used in order to support or confirm a photometric solution \citep{Smak1994sys,Baptista95}.
  The parameters, derived by \citet{Baptista95} and commonly adopted in the literature (an inclination $i=71^{\circ}$
  and $M_1=M_2=0.47M_\odot$) were determined using the UV light curve and the Shafter's value of $K_1=160$ \kms.

  However, the value of $K_2$ obtained by \citet{VandePutte2003}, along with our estimates of $K_1$ and $K_2$,
  immediately refute the Baptista et al.'s parameters. We can determine the parameters of UX~UMa using a traditional
  spectroscopic solution based on the derived radial velocity semi-amplitudes.
  In order to calculate the masses of the binary, we assume that the semi-amplitudes of the
  measured radial velocities represent the true orbital motion of the stars.
  Combining our $K_1=113\pm11$ \kms and taking $K_2=262\pm14$ \kms\ from \citet{VandePutte2003}
  and $P$ from equation~(\ref{ephemeris}), we find the mass ratio
  $q \equiv M_2/M_1 = K_1/K_2 = 0.43 \pm 0.07$, the masses of each component of the system
   \begin{equation}
    \label{M1sini}
      M_1 \sin^3 i = {P K_2 (K_1 + K_2)^2 \over 2 \pi G} = 0.75 \pm 0.20 M_{\odot}\,,
   \end{equation}
   \begin{equation}
    \label{M2sini}
      M_2 \sin^3 i = {P K_1 (K_1 + K_2)^2 \over 2 \pi G} = 0.32 \pm 0.10 M_{\odot}\,,
   \end{equation}
   \noindent and the projected binary separation
   \begin{equation}
      a \sin i = {P (K_1 + K_2) \over 2 \pi} = 1.46 \pm 0.10 R_{\odot}\,.
   \end{equation}

  The orbital inclination of UX~UMa was determined in various papers, ranging from $i=65^{\circ}$ to
  $75^{\circ}$. Using this, we now have the pure dynamical solution for the parameters of UX~UMa:
  $M_1$=0.83--1.01 ($\pm$ 0.20)~$M_\odot$,
  $M_2$=0.36--0.43 ($\pm$ 0.10)~$M_\odot$,
  $a$=1.51--1.61~$R_{\odot}$,
  $i$=65$^\circ$--75$^\circ$ ($i=65^\circ$ corresponds to the largest masses and $a$).

  The additional step is to make use of a reasonable assumption that the secondary star fills its Roche lobe.
  The relative size of the donor star is therefore constrained by Roche geometry and the donor must obey
  the period-density relation for Roche lobe filling objects \citep{Warner}.
  Using any of the recently obtained empirical and theoretical mass-period relations we obtain the mass for
  the secondary 0.41--0.45 $M_\odot$ (see, for example, \citealt{Warner, Patterson05}), which is in a good
  agreement with the dynamical solution.

  During the Loiano observations, the emission lines exhibited double-peaked profiles with
  the averaged measured peak-to-peak velocity separation in \Halpha\ of $\approx740$ \kms.
  \citet{Smak1981} has shown that the distance between the peaks in the double-peaked profiles can be used
  to determine the velocity of the outer rim of the accretion disc $V_{out}$ which in turn depends on
  the mass $M_1$ and the radius of the disc.
  The largest radius of the accretion disc, should it be determined from observations, can further limit
  the system parameters.

  The outer parts of the disc are under the gravitational influence of the secondary star, which prevents the disc
  from growing larger than $R_{max}$. This largest radius is determined by tidal limitations. Tidal stress and
  viscous stress become comparable at this radius and truncate the disc (\citealt{Warner} and references therein).
  The largest radius can be estimated by the use of equation (2.61) from \citet{Warner},
  and for the mass ratio $q$ = 0.43 we obtain
   \begin{equation}
    \label{RmaxA}
      R_{max} = {0.6\,a / (1+q)} = 0.42\,a = 0.61\,R_{L1}
   \end{equation}
  \noindent where $R_{L1}$ is the distance of the inner Lagrangian point from the WD.

  Unfortunately, although the observed spiral-like pattern is an indirect manifestation of the large
  accretion disc in UX~UMa, its estimated size appears to be implausibly large.
  Based on the obtained system parameters and assuming the Keplerian velocity in
  the accretion disc, it gives an estimate of the outer radius of the disc to be $R_{out} \approx 0.80\,a$,
  that is even greater than the Roche lobe of the primary.


  Should this discrepancy be considered as serious? Generally speaking, yes. Tidal limitations are of
  fundamental nature whereas the peak separation of a double-peaked emission line is an accurate measure of
  the velocity of the outer edge of the accretion disc obeying perfect Keplerian rotation.

  Nevertheless, there are physical mechanisms which may alter the distance between the peaks. For example, spiral
  shocks disturb the orderly Keplerian motion in the outermost disk that may result in variability of the
  inter-peak distance, mostly towards smaller values \citep{Neustroev1998}, and this effect is clearly seen in
  the trailed spectra (Fig.~\ref{dopmaps2}, bottom half). Assuming that spiral waves could exist in the accretion
  disc of UX~UMa, we made an attempt to reduce their influence by taking the largest peak separation in \Halpha.
  This approach, however, does not guarantee that this extremal value of $V_{out}\sim390$ \kms\ is
  close enough to the true velocity of the outer edge of the accretion disc in case of strong shocks,
  and we indeed found the disc to be very large, almost filling the Roche lobe of the primary
  ($R_{out} / R_{L1} \sim 0.94$).

  Another reason why the distance between the peaks in the double-peaked emission lines can be reduced is
  a significant amount of emitting matter in the outermost accretion disc with a velocity component
  perpendicular to the disk plane. Evidences for the presence of winds in nonmagnetic CVs in a state of
  high mass accretion are unambiguous. Regarding to UX~UMa, \citet{Knigge1997} strongly suggested the presence
  of a relatively dense and slowly outflowing transition region between the disk photosphere and the
  fast-moving wind in the system. The presence of strong Balmer emission lines at mid-eclipse during our
  observations  of UX~UMa also implies that a part of the line-forming region extends out to distances from the WD.
  Unfortunately, we can say little about how strong the wind is in the outmost regions of the disc of UX~UMa,
  if it exists at all. Anyway, realization of any of these mechanisms could lead to underestimating of $V_{out}$.

  In conclusion, we have to note that our parameters are not as consistent with Doppler tomography as
  the parameters of \citet{Baptista95}: the emission from the secondary star overfills the bubble marking
  the Roche lobe on the tomograms, and also the stream trajectory is located higher on the maps than the
  brightest area of the region where we believe the bright-spot is (Fig.~\ref{alldopmaps}).
  As it was mentioned above, the Maximum Entropy Method applied during the Doppler tomography reconstruction
  can perhaps be partly responsible for the above effects by blurring/smoothing the compact emission sources.
  Otherwise, it may testify that we still have not constrained the radial velocity semi-amplitudes very
  reliably in UX~UMa. Moreover, it also puts in doubt the independent measurements of $K_2$ by
  \citet{VandePutte2003}.
  Thus, we are now in a situation, when on one side there is a photometric solution which better describes
  the Doppler maps but definitely cannot be correct, and on the other side is another, spectroscopic solution
  which we think is closer to the true parameters of UX~UMa but does not agree as well with Doppler tomography.
  One more set of the system parameters obtained by \citet{Smak1994sys}, is also inconsistent with the former two.
  Thus, there are still many problems with the parameters of the system, which is considered to be one of the best
  studied CVs. In order to solve them, new \emph{simultaneous} photometric and spectroscopic observations
  with high-time resolution will be very useful.

\subsection{UX~UMa as a SW Sextantis star}

  During the SAO observations UX~UMa showed some of the defining properties of the SW~Sex stars, such as
  single-peaked emission lines with transient absorption features at phase $\sim0.4-0.7$. In this
  section we summarize these characteristics and compare them with properties of the founding members of
  the SW~Sex sub-class:

\begin{enumerate}
  \item UX~UMa is the eclipsing system as the founding members of the SW~Sex sub-class. Its orbital
  period of 4.72~h is slightly above but very close to the 3 {--} 4.5 hr period interval.

  \item The emission lines in the blue wavelength range are single-peaked as in all the SW~Sex stars.

  \item Significant high excitation spectral features including \HeII\lam 4686 and Bowen blend emission
  are typical for the class.

  \item The transient absorption of the Balmer and \HeI\ emission lines reaches its maximum strength at phase
  $\sim0.58$. The relative intensity of this component increases with the line excitation level. In the \HeI\
  lines this absorption is a dominating component. The \HeII\ \lam 4686 emission line is perhaps not affected
  by this absorption. Note, that the phase 0.5 absorption in UX~UMa is a part of an absorption S-wave.
  To the best of our knowledge, this is not generally the case in other SW~Sex stars. However, we
  admit that visibility of such S-wave can be governed by some special physical conditions.

  \item The Doppler maps of the SW~Sex stars consistently share a common feature: the bulk of Balmer emission
  is concentrated in the (-Vx,-Vy) quadrant. In this area of UX~UMa's tomograms the unique dark spot is located.
  The dark spot is compact in the Balmer lines and deep and spacious in the \HeI\ lines.

  \item Many SW Sex stars exhibit high-velocity emission S-waves with maximum blueshift near phase $\sim0.5$.
  This emission is not seen in UX~UMa. On the opposite, we have observed the broad absorption depression around the
  red wings of the Balmer lines just prior to eclipse. This depression is certainly linked with the
  above-mentioned transient absorption.

  \item In the eclipsing SW~Sex stars the emission-line radial velocity curves show a delay of a 0.1 - 0.2
  orbital cycle relative to the motion of the WD. This effect in UX~UMa is weak, if exists. Note, however,
  that \citet{Froning2003} have detected a phase lag of $\simeq20^{\circ}$, compatible with the derived parameters
  of the radial velocity curves (Table~\ref{radvelfit}).
\end{enumerate}

Thus, many of the characteristics of UX~UMa, a prototype of the NL stars and the UX~UMa
sub-class, are in general accord with the SW~Sex systems. Nevertheless, all these features have been observed
only in 1999 and almost completely disappeared in 2008. Many symptoms indicate that the system was in different
states during those observations. \citet{Groot2004} proposed that the appearance of a NL system as a UX~UMa/RW~Tri
or SW~Sex star seems to be mainly governed by the mass-transfer rate from the secondary at the time of observation,
where the SW~Sex behaviour becomes more prominent with increasing mass-transfer rate. However, \citet{Ballouz2009}
found little evidence to suggest that the SW~Sex stars have higher accretion rates than other NLs above the period
gap within the same range of orbital periods. From our observations, we come to a rather contrary conclusion --
UX~UMa behaved as the SW~Sex stars when its mass-transfer rate was perhaps lower than usual.
Thus, this hypothesis certainly merits further study.

\section{Summary}
\label{SumSec}

  We have presented an analysis of time-resolved, medium resolution optical spectroscopic observations of
  the system in the blue (3920--5250~{\AA}) and red (6100--7200~{\AA}) wavelength ranges, that were obtained
  in April 1999 and March 2008 respectively.

  The appearance and behaviour of the spectra indicates that UX~UMa has been in different states during
  those observations. The blue spectra are very complex. They are dominated by strong and broad single-peaked
  emission lines of hydrogen. The high-excitation lines of \HeII\ $\lambda4686$ and the Bowen blend are quite
  strong as well. All the lines consist of a mixture of absorption and emission components. Using Doppler
  tomography we have identified four distinct components of the system: the non-uniform accretion disc,
  the secondary star,
  the bright spot from the gas stream/disc impact region, and the unique dark spot in the lower-left quadrant
  of the tomograms. In the red wavelength range, both the hydrogen (\Halpha) and neutral helium
  (\HeI\ $\lambda6678$ and \HeI\ $\lambda7065$) lines were observed in emission and both exhibited
  double-peaked profiles. Doppler tomography of these lines reveals spiral structure in the accretion disc,
  but in contrast to the blue wavelength range, there is no evidence for either the dark spot or
  the gas stream/disc impact region emission, while the emission from the secondary star is weak.

  During the SAO observations UX~UMa showed many of the defining properties of the SW~Sex stars. However,
  all these features have been observed only in 1999 and almost completely disappeared in 2008.

  We have measured, from the motion of the wings of the \Halpha\ and \Hbeta\ emission lines,
  the radial velocity semi-amplitude $K_1$ to be $113\pm11$ \kms. We have also restricted the radial velocity
  semi-amplitude of the secondary star to be $K_2 \ga 200$ \kms. We have used these results to derive
  the system parameters of UX~UMa and found that these estimates are inconsistent with previous values derived
  by means of analysis of WD eclipse features in the light curve in the different wavelength ranges.
  It may testify that we still have not constrained the radial velocity semi-amplitudes very reliably in UX~UMa.


\section*{Acknowledgments}

  The authors thank Natalia Neustroeva for help in preparation of the paper.
  We would like to acknowledge the anonymous referee whose comments have significantly improved this paper.
  VS thanks DFG for financial support (grant SFB/Transregio 7 "Gravitational Wave Astronomy"),
  RBRF (grant 09-02-97013-p-povolzh'e-a), and President's programme for support of leading science
  schools (partial support, grant NSh-4224.2008.2).

\bsp
\label{lastpage}
\end{document}